%% file: main.tex
\documentclass[twocolumn]{aastex631}

\usepackage{graphicx}
\usepackage{dcolumn}
\usepackage{bm}
\usepackage[nointegrals]{wasysym}
\usepackage{verbatim}
\usepackage{color}
\usepackage{amsmath}
\usepackage[utf8]{inputenc}
\usepackage[caption=false]{subfig}

\renewcommand{\bold}[1]{\boldsymbol{#1}}

\newcommand{\si}[1]{{\rm{#1}}}

\newcommand{\Planck}{{\it Planck}}

\newcommand{\Frank}[1]{\textcolor{black}{#1}}
\newcommand{\fk}[1]{\textcolor{black}{#1}}
\newcommand{\fkk}[1]{\textcolor{black}{#1}}

\newcommand{\DW}[1]{\textcolor{black}{#1}}
\newcommand{\blake}[1]{\textcolor{cyan}{}}

\newcommand{\bds}[1]{\textcolor{black}{#1}}

\newcommand{\NS}[1]{\textcolor{black}{#1}}

\newcommand{\abds}[1]{\textcolor{black}{#1}}

\newcommand{\unitvec}{\ensuremath{\hat{\boldsymbol{n}}}}

\newcommand{\Alens}{A_{\mathrm{lens}}}

\newcommand{\nullb}{C^{\text{null}}_L}

\newcommand{\mycomment}[1]{}
\newcommand{\LCDM}{$\Lambda$CDM}
\newcommand{\Lmax}{L_{\mathrm{max}}}

\newcommand{\albiasMiscalMV}{0.00432}
\newcommand{\albiasMiscalMVPol}{0.00419}
\newcommand{\albiasBeamMV}{0.00113}
\newcommand{\albiasBeamMVPol}{0.00110}
\newcommand{\albiasPolAngMV}{-0.00025}
\newcommand{\albiasPolAngMVPol}{0.00070}
\newcommand{\albiasLeakConstMV}{-0.00461}
\newcommand{\albiasLeakConstMVPol}{-0.00887}
\newcommand{\albiasLeakBeamMV}{-0.00021}
\newcommand{\albiasLeakBeamMVPol}{-0.00052}
\newcommand{\albiasPolEffMV}{-0.00896}
\newcommand{\albiasPolEffMVPol}{-0.01884}
\newcommand{\biasSigmaMiscalMV}{0.18 \sigma}
\newcommand{\biasSigmaMiscalMVPol}{0.09 \sigma}
\newcommand{\biasSigmaBeamMV}{0.05 \sigma}
\newcommand{\biasSigmaBeamMVPol}{0.02 \sigma}
\newcommand{\biasSigmaPolAngMV}{-0.01 \sigma}
\newcommand{\biasSigmaPolAngMVPol}{0.01 \sigma}
\newcommand{\biasSigmaLeakConstMV}{-0.19 \sigma}
\newcommand{\biasSigmaLeakConstMVPol}{-0.18 \sigma}
\newcommand{\biasSigmaLeakBeamMV}{-0.01 \sigma}
\newcommand{\biasSigmaLeakBeamMVPol}{-0.01 \sigma}
\newcommand{\biasSigmaPolEffMV}{-0.37 \sigma}
\newcommand{\biasSigmaPolEffMVPol}{-0.38 \sigma}

\newcommand{\uKam}{\mu\text{K-arcmin}}

\newcommand{\ba}{\begin{eqnarray}}
\newcommand{\ea}{\end{eqnarray}}

\usepackage{aas_macros}

\begin{document}

\title{The Atacama Cosmology Telescope: A Measurement of the DR6 CMB Lensing Power Spectrum and its Implications for Structure Growth}
 \shorttitle{ACT DR6 lensing power spectrum}
  \shortauthors{Qu, Sherwin, Madhavacheril, Han, Crowley et al.}








\input{authors_qu.tex}

\begin{abstract}
We present new measurements of cosmic microwave background (CMB) lensing over $9400\,\si{deg}^2$ of the sky. \abds{These lensing measurements are derived from the Atacama Cosmology Telescope (ACT) Data Release 6 (DR6) CMB dataset, which consists of five seasons of ACT CMB temperature and polarization observations.} We determine the amplitude of the CMB lensing power spectrum at $2.3\%$ precision ($43\sigma$ significance) using a novel pipeline that minimizes sensitivity to foregrounds and to noise properties. To ensure our results are robust, we analyze an extensive set of null tests, consistency tests, and systematic error estimates and employ a blinded analysis framework. Our CMB lensing power spectrum measurement provides constraints on the amplitude of cosmic structure that do not depend on \textit{Planck} or galaxy survey data, thus giving independent information about large-scale structure growth and potential tensions in structure measurements. The baseline spectrum is well fit by a lensing amplitude of $A_{\mathrm{lens}}=1.013\pm0.023$  relative to the \textit{Planck} 2018 CMB power spectra best-fit $\Lambda$CDM model {\NS{and $A_{\mathrm{lens}}=1.005\pm0.023$ relative to the $\text{ACT DR4} + \text{\textit{WMAP}}$ best-fit model.}} From our lensing power spectrum measurement, we derive constraints on the parameter combination $S^{\mathrm{CMBL}}_8 \equiv \sigma_8 \left({\Omega_m}/{0.3}\right)^{0.25}$ of $S^{\mathrm{CMBL}}_8= 0.818\pm0.022$ from ACT DR6 CMB lensing alone and $S^{\mathrm{CMBL}}_8= 0.813\pm0.018$ when combining ACT DR6 and \textit{Planck} \texttt{NPIPE} CMB lensing power spectra. These results are in excellent agreement with $\Lambda$CDM model constraints from \textit{Planck} or $\text{ACT DR4} + \text{\textit{WMAP}}$ CMB power spectrum measurements. Our lensing measurements from redshifts $z\sim0.5$--$5$ are thus fully consistent with $\Lambda$CDM structure growth predictions based on CMB anisotropies probing primarily $z\sim1100$. We find no evidence for a suppression of the amplitude of cosmic structure at low redshifts.





\end{abstract}





\section{Introduction}


\bds{The cosmic microwave background (CMB) is a unique backlight for illuminating the growth of structure in our Universe. As the CMB photons travel from the last-scattering surface to our telescopes, they are gravitationally deflected, or lensed, by large-scale structure along their paths. The resulting arcminute-scale lensing deflections distort the observed image of the CMB fluctuations, imprinting a distinctive non-Gaussian four-point correlation function (or trispectrum) in both the temperature and polarization anisotropies \citep{Lewis:2006fu,1987A&A...184....1B}}. A measurement of this lensing-induced four-point correlation function enables a direct determination of the power spectrum of the CMB lensing field; the CMB lensing power spectrum, in turn, probes the matter power spectrum projected along the line of sight, \abds{with the signal arising from a range of redshifts $z\sim 0.5$--$5$}.\footnote{\abds{See Appendix \ref{app.wkk} and Figure \ref{fig:1dkernel} for a more accurate characterization of the redshift origin of the CMB lensing signal we measure. While the mean redshift of the lensing signal is at $z\sim2$, the lensing redshift distribution is characterized by a peak at $z\sim1$ and a tail out to high $z$.}} Since most of the lensing signal originates from high redshifts and large scales, the signal is near-linear and simple to model, with complexities arising from baryonic feedback and highly non-linear evolution negligible at current levels of precision. 
Furthermore, the physics and redshift of the primordial CMB source are well understood, with the statistical properties of the unlensed source described accurately as a statistically isotropic Gaussian random field. These properties make CMB lensing a robust probe of cosmology, and, in particular, cosmic structure growth.

Measurements of the growth of cosmic structure \bds{can provide powerful insights into} new physics. For example, the comparison of low-redshift structure with primary CMB measurements constrains the sum of the neutrino masses, because massive neutrinos suppress the growth of structure in a characteristic way \citep{LESGOURGUES_2006}. Furthermore, high-precision tomographic measurements of structure growth at low redshifts allow us to test whether dark energy continues to be well described by a cosmological constant or whether there is any evidence for dynamical behaviour or even a breakdown of general relativity. 

\abds{A particularly powerful test of structure growth is the following: we can fit a $\Lambda$CDM model to CMB power spectrum measurements arising (mostly}\footnote{Note that, while CMB power spectrum measurements primarily probe structure at $z\sim 1100$, they also have a degree of sensitivity to lower redshift structure, e.g., due to gravitational lensing effects on the CMB power spectra.}) \abds{from $z\sim 1100$, predict the amplitude of density fluctuations at low redshifts assuming standard growth, and compare this with direct, high-precision measurements at low redshift. Intriguingly, for some recent low-redshift observations, it is not clear that this test has been passed: several recent lensing and galaxy surveys have found a lower value of $S_8 \equiv \sigma_8 \left( \Omega_m / 0.3 \right)^{0.5}$ than predicted by extrapolating the \textit{Planck} CMB power spectrum measurements to low redshifts in $\Lambda$CDM \citep{10.1093/mnras/stt601,KiDS:2020suj,Heymans_2021,Krolewski:2021yqy,Philcox:2021kcw,PhysRevD.105.023520,Loureiro_2022,HSCY3Real,HSCY3Fourier}.}\footnote{We note that the best-constrained weak-lensing parameter $S_8 \equiv \sigma_8 \left( \Omega_m / 0.3 \right)^{0.5}$ has a slightly different exponent than $S^\mathrm{CMBL}_8 \equiv \sigma_8 \left( \Omega_m / 0.3 \right)^{0.25}$, the best constrained parameter for CMB lensing. These different definitions of $S_8$ reflect the different degeneracy directions in the $\sigma_8$--$\Omega_m$ plane due to galaxy lensing and CMB lensing being sensitive to different redshift ranges and scales.} These discrepancies are generally referred to as the ``$S_8$ tension''. CMB lensing measurements that do not rely on either \textit{Planck}{\footnote{\textit{Planck} also did not find a low value of $S_8$ from CMB lensing \cite{Planck:2018,Carron_2022}.} or galaxy survey data have the potential to provide independent insights into this tension.\footnote{\NS{For example, if ACT were to obtain a lower lensing amplitude, in tension with that predicted from the measurements of the primordial CMB anisotropies, this could indicate new physics at high redshifts and on large scales (or unaccounted-for systematic effects in either data). On the other hand, if ACT lensing were entirely consistent with CMB anisotropies but inconsistent with other lensing measurements, this could imply either systematics in the measurements or new physics that only affects very low redshifts and/or small scales.}} 

With the advent of low-noise, high-resolution CMB telescopes such as the Atacama Cosmology Telescope (ACT), the South Pole Telescope (SPT), and the \textit{Planck} satellite, 
CMB lensing has progressed rapidly from first detections to high-precision measurements. First direct evidence of CMB lensing came from cross-correlation measurements with  \textit{Wilkinson Microwave Anisotropy Probe (WMAP)} data~\citep{Smith2007}; ACT reported the first CMB lensing power spectrum detection and the first constraints on cosmological parameters from lensing spectra, including evidence of dark energy from the CMB alone~\citep{Das2011, sherwin2011}. Since then, lensing power spectrum measurements have been made by multiple groups, with important advances made by the SPT, POLARBEAR  and BICEP/Keck teams \abds{as well as ACT}~\citep{van_Engelen_2012,polarbear,Story_2015,BICEP2016,PhysRevD.95.123529,Omori_2017,Wu_2019,Bianchini_2020}.  The \textit{Planck} team has made key contributions to CMB lensing over the past decade and has made the highest precision measurement of the lensing power spectrum prior to this work, with a $40\sigma$ significance\footnote{\abds{Throughout this work, the significance of a lensing power spectrum measurement is defined as the ratio of the best-fit lensing amplitude $\Alens$ to the error on this quantity.}} measurement presented in their official 2018 release~\citep{Planck:2018} and a $42\sigma$ measurement demonstrated with the \texttt{NPIPE} data~\citep{Carron_2022}. \abds{With \textit{Planck} lensing and now separately with the measurements presented in this paper, CMB lensing measurements have achieved precision that is competitive with any galaxy weak lensing measurement. 
CMB lensing is thus one of our most powerful modern probes of the growth of cosmic structure.}

\bds{The goal of our work is to perform a new measurement of the CMB lensing power spectrum with state-of-the-art precision. This lensing spectrum will allow us to perform a stringent test of our cosmological model, comparing our lensing measurements from redshifts $z\sim0.5$--$5$ with flat-$\Lambda$CDM \footnote{Unless otherwise stated we will refer to $\Lambda$CDM as an abbreviation to flat-$\Lambda$CDM.} structure growth predictions based on CMB power spectra probing primarily $z\sim1100$. Our lensing power spectrum will also constrain key parameters such as the sum of neutrino masses, the Hubble parameter, and the curvature of the Universe, as explored in our companion paper \citep{dr6-lensing-cosmo}.}  

\section{Summary of Key Results}\label{summary}

In this paper, we present CMB lensing measurements using data taken by ACT between 2017 and 2021. \fk{This is part of the ACT collaboration's Data Release 6 (DR6), as described in detail in Section \ref{data}}. Section \ref{sec:sim} discusses the simulations used to calculate lensing biases and covariances. In Section \ref{sec:methods}, we describe our pipeline used to measure the CMB lensing spectrum. We verify our measurements with a series of map-level and power-spectrum-level null tests summarised in Section \ref{sec:null} and we quantify our systematic error estimates in Section \ref{sec.systematics}. Our main CMB lensing power spectrum results are presented in Section \ref{sec:results}; \abds{readers interested primarily in the cosmological implications of our work, rather than how we perform our analysis, may wish to skip to this Section}. We discuss our results in Section \ref{sec.discussion} and conclude in Section \ref{sec:conclusion}. This paper is part of a larger set of ACT DR6 papers, and is accompanied by two others: \citet{dr6-lensing-cosmo} presents the released DR6 CMB lensing mass map, and explores the consequences for cosmology from the combination and comparison of our measurements with external data; \citet{dr6-lensing-fgs} investigates the levels of foreground biases -- arguably the most significant potential source of systematic errors -- and ensures these are well-controlled in our analysis.

We briefly summarize the key results of our work in the following paragraphs. Of course, for a detailed discussion, we encourage the reader to consult the appropriate section of the paper.

\begin{figure*}
  \includegraphics[width=0.85\paperwidth]{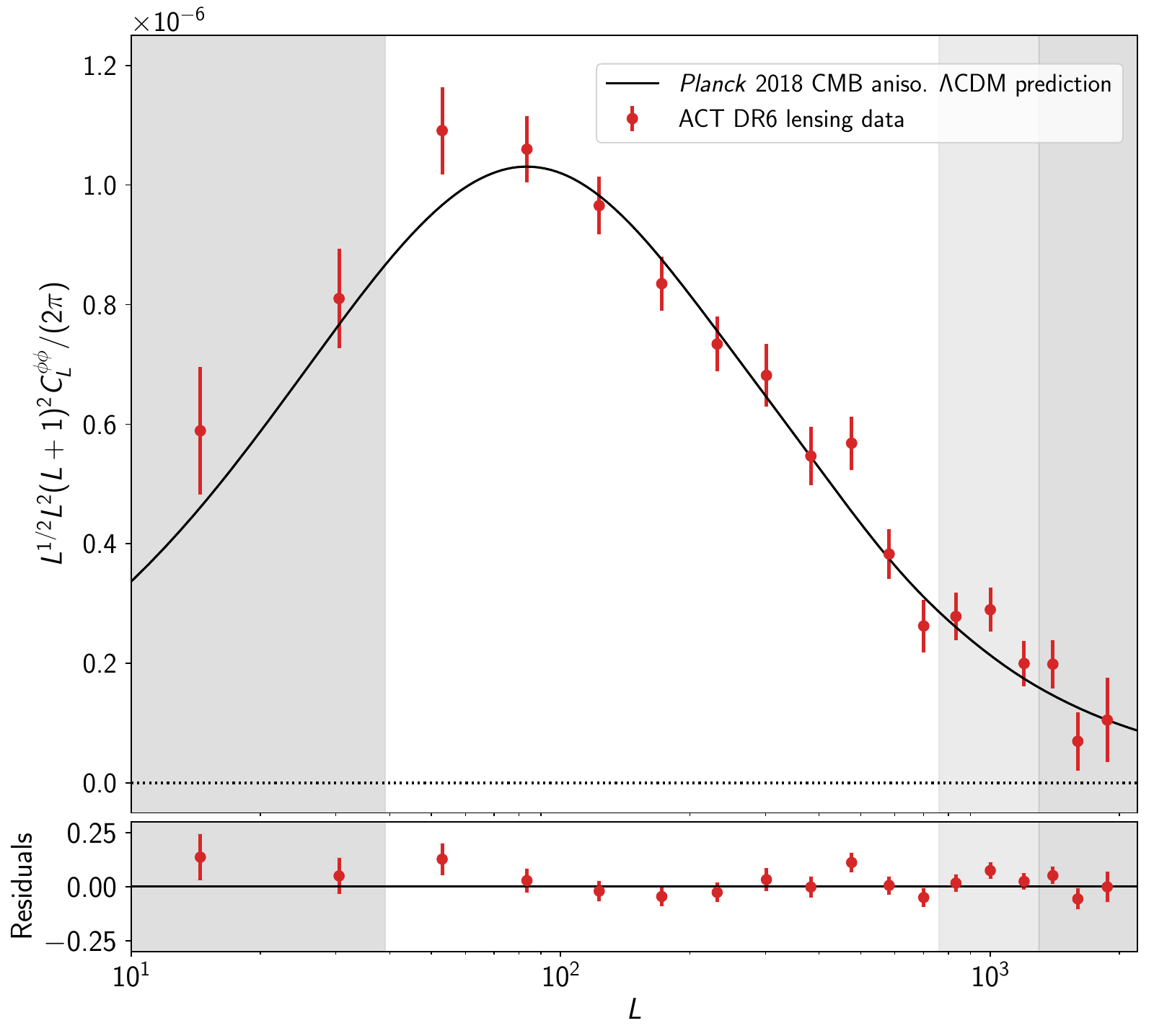}
  \caption{
  The upper panel shows in red the ACT DR6 lensing potential power spectrum bandpowers for our baseline (combined temperature and polarization) analysis. \bds{The bandpowers within the shaded regions are excluded in our baseline analysis that only analyzes the conservative range of lensing multipoles $40 < L < 763$, although we also include scales up to $L_{\mathrm{max}}=1300$, indicated with a lighter shading, in our extended-range analysis.} {\abds{We find good agreement with the $\Lambda$CDM theoretical predictions based on either the \textit{Planck} 2018 or ACT DR4 + \textit{WMAP} CMB power spectra best-fit cosmology; the solid line shows the \textit{Planck} 2018 prediction, which we emphasize does not arise from a fit to our data. Residuals of our measurement with respect to the \textit{Planck} prediction are shown in the lower panel of the figure. Our ACT lensing data fits a model based on a best-fit rescaling of the \textit{Planck} prediction with a lensing amplitude of $A_{\mathrm{lens}}=1.013\pm0.023$, and one based on a rescaling of the ACT DR4 + \textit{WMAP} prediction with $A_{\mathrm{lens}}=1.005\pm0.023$.}}}
  \label{Fig.results_clkk}
\end{figure*}

\begin{itemize}
    \item We reconstruct lensing and lensing power spectra from $9400\,\si{deg}^2$ of temperature and polarization data. Our measurements are performed with a new cross-correlation-based curved-sky lensing power spectrum pipeline that is optimized for ground-based observations with complex noise.

    \item An extensive suite of null tests and instrument systematic estimates shows no significant evidence for any systematic bias in our measurement. These tests form a key part of our blinded analysis framework, which was adopted to avoid confirmation bias in our work. Foregrounds appear well mitigated by our baseline profile-hardening approach, and we find good consistency of our baseline results with spectra determined using other foreground-mitigation methods.

    \item We measure the amplitude of the CMB lensing power spectrum at state-of-the-art 2.3\% precision, corresponding to a measurement signal-to-noise ratio of $43\sigma$. 
    This signal-to-noise ratio independently matches the 42$\sigma$ achieved in the latest {\it Planck} lensing analysis \abds{and is competitive with the precision achieved in any galaxy weak lensing analysis.}
    Our lensing power spectrum measurement is shown in Figure \ref{Fig.results_clkk}.

\begin{figure*}[ht!]
    \centering
    \includegraphics[width=0.7\paperwidth]{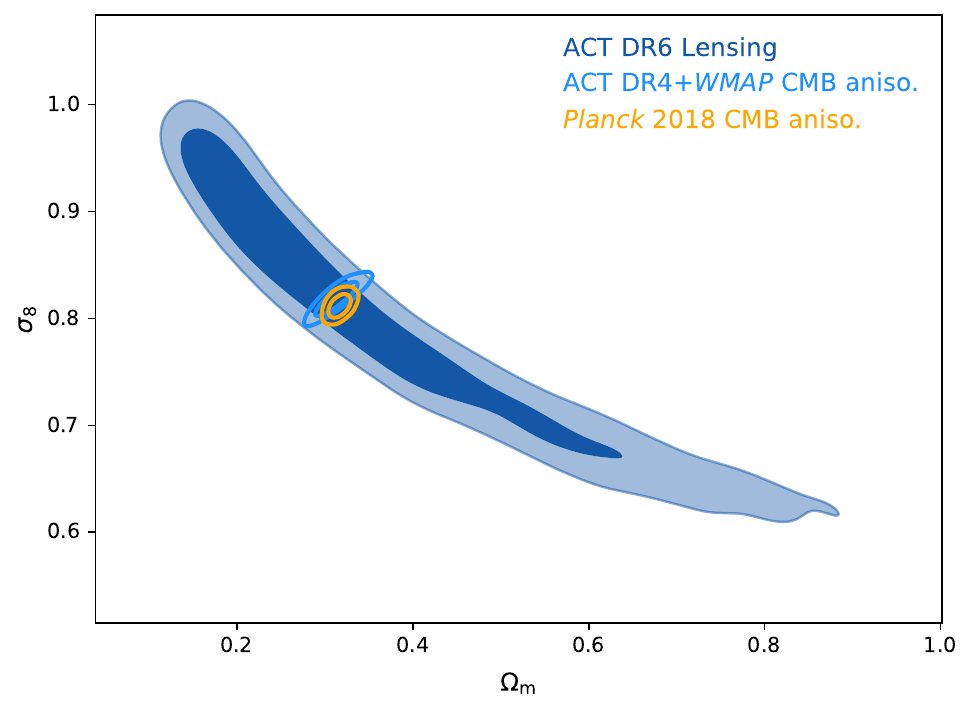}
    \caption{Constraints in the $\sigma_8$--$\Omega_m$ plane from our baseline ACT DR6 lensing power spectrum measurement (blue). These can be compared with the predictions from standard $\Lambda$CDM structure growth and \textit{Planck} or ACT DR4 + \textit{WMAP} CMB power spectra (orange open and blue open contours, respectively). In all cases, 68\% and 95\% contours are shown. Our results are in excellent agreement with \textit{Planck} (or ACT DR4 + \textit{WMAP}) and $\Lambda$CDM structure growth. The parameter combination measured best by CMB lensing alone $S^{\mathrm{CMBL}}_8 \equiv \sigma_8 \left({\Omega_m}/{0.3}\right)^{0.25}$ is measured to be $S^{\mathrm{CMBL}}_8=0.818\pm0.022$ and the individual parameters are constrained to  $\Omega_m=0.355\pm0.178$ and $\sigma_8=0.814\pm0.099$.}
    \label{fig:sig8om_nobao}
\end{figure*}

    \item The lensing power spectrum is well fit by a \LCDM~  cosmology and, in particular, by the \textit{Planck} 2018 CMB power spectrum model. Fitting a lensing amplitude that rescales the lensing power spectrum from this model, we obtain a constraint on this amplitude of $A_{\mathrm{lens}}=1.013\pm0.023$. If we fit instead to the best-fit model from ACT DR4 + \textit{WMAP} power spectra, we obtain a lensing amplitude of \Frank{$A_{\mathrm{lens}}=1.005\pm0.023$}.

    \item From our measurement of the DR6 lensing power spectrum alone, we measure the best-constrained parameter combination\footnote{The degeneracy between $\sigma_8$ and $\Omega_m$ prevents strong constraints on either of these parameters individually, and indeed (although we report them in the figure caption) any such constraints derived will depend strongly on the prior ranges; in order to break the degeneracy between these two parameters we combine with BAO as shown in the companion paper \citep{dr6-lensing-cosmo}.} $S^{\mathrm{CMBL}}_8 \equiv \sigma_8 \left({\Omega_m}/{0.3}\right)^{0.25}$ as $S^{\mathrm{CMBL}}_8= 0.818\pm0.022$. This key result is illustrated in Figure \ref{fig:sig8om_nobao}.
    
    \item We combine ACT DR6 and \textit{Planck} 2018 CMB lensing power spectrum observations, accounting for the appropriate covariances between the two measurements. For this combined dataset, we obtain a constraint of $S^{\mathrm{CMBL}}_8= 0.813\pm0.018$.
    
    \item All our results are fully consistent with expectations from \textit{Planck} 2018 or ACT DR4 + \textit{WMAP} CMB power spectra measurements and standard $\Lambda$CDM structure growth. \bds{This is an impressive success for the standard model of cosmology: with no additional free parameters, we find that a $\Lambda$CDM model fit to CMB power spectra probing (primarily) $z \sim 1100$ correctly predicts cosmic structure growth (and lensing) down to $z\sim 0.5$--$5$ at $2\%$ precision. }
    
    \item \bds{We find no evidence for tensions in structure growth and we do not see a suppression of the amplitude of cosmic structure at the redshifts and scales we probe ($z\sim 0.5$--$5$ on near-linear scales). This has implications for models \abds{of new physics} that seek to explain the $S_8$ tension: such models cannot strongly affect linear scales and redshifts $z\sim0.5$--$5$ or above, although new physics affecting primarily  small scales or low redshifts might evade our constraints.} 

\end{itemize}

\section{CMB Data}\label{data}

ACT was a six-meter aplanatic Gregorian telescope located in the Atacama Desert in Chile. The Advanced ACTPol (AdvACT) receivers fitted to the telescope were equipped with arrays of superconducting transition-edge-sensor bolometers, sensitive to both temperature and polarization at frequencies of 30, 40, 97, 149 ,and 225\,$\si{GHz}$\footnote{In the following, we denote them f030, f040, f090, f150 and f220.} \citep{fowler2007optical,thornton2016atacama}. This analysis focuses on data collected from 2017 to 2021 covering two frequency bands f090 (77--112\,GHz), f150 (124--172\,GHz). The observations were made using three dichroic detector modules, known as polarization arrays (PA), with PA4 observing
in the f150 (PA4 f150) and f220 (PA4 f220) bands; PA5 in the f090 (PA5 f090) and
f150 (PA5 150) bands, and PA6 in the f090 (PA6 f090) and f150 (PA6 f150) bands.
 We will refer to these data and the resulting maps as DR6; although further refinements and improvements of the DR6 data and sky maps can be expected before they are finalized and released, extensive testing has shown that the current versions are already suitable for the lensing analysis presented in this paper. For arrays PA4--6, we use the DR6 night-time data and the f090 and f150 bands
only. Although including additional datasets in our pipeline is straightforward, this choice was made because daytime data require more extensive efforts to ensure instrumental systematics (such as beam variation) are well controlled and because including the f220 band adds analysis complexity while not significantly improving our lensing signal-to-noise ratio.  We, therefore, defer the analysis of the daytime and f220 data to future work.

\subsection{Maps}
The maps were made with the same methodology as \abds{in} \cite{Aiola_2020}; they will be described in full detail in \cite{dr6-maps}. To summarize briefly, maximum-likelihood maps are built at $0.5^\prime$ resolution using 756 days of data observed in the period 2017-05-10 to 2021-06-18. Samples contaminated by detector glitches or the presence of the Sun or Moon in the telescope's far sidelobes are cut, but scan-synchronous pickup, like ground pickup, is left in the data since it is easier to characterize in map space.


The maps of each array-frequency band are made separately, for a total of five  array-band combinations. For each of these, we split the data into three categories due to differences in systematics and scanning patterns: \textit{night}, \textit{day-deep} and \textit{day-wide}. Of these categories, night makes up 2/3 of the statistical power and, as previously stated, is the only dataset considered in this analysis. 

Each set of night-time data is  split into eight subsets with independent instrument \abds{and atmospheric} noise noise. 
\bds{These data-split maps are useful for characterizing the noise properties with map differences and for applying} the \abds{cross-correlation-based estimator} described in Section~\ref{sec:reconstruction}.  
In total, for this lensing analysis we use 40 separate night \abds{split maps} in the f090 and f150 bands. 

The data used in this analysis initially cover approximately $19\,000\,\si{deg}^2$ before Galactic cuts are applied and have a total 
\abds{inverse variance}
of 0.55/nK$^2$ for the night-time data. 
Figure~\ref{fig:coverage} shows the sky coverage and full-survey depth of ACT DR6 night-time observations. 

 \begin{figure*}
  \includegraphics[width=\textwidth]{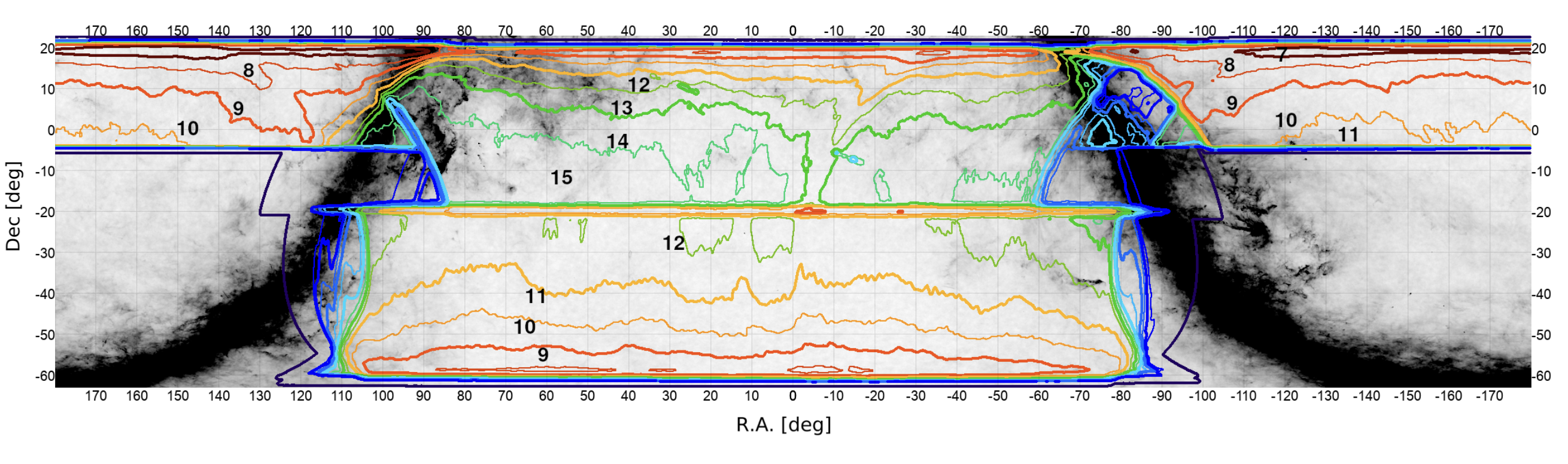}
  \caption{Sky coverage and full survey depth of the ACT DR6 night-time observations in equatorial coordinates.  Here the $x$-axis gives right ascension and the $y$-axis indicates declination. The background grayscale map corresponds to the \textit{Planck} $353\si{GHz}$ intensity. Coloured lines are the depth contours with the numbers corresponding to the noise levels in $\uKam$ units. \fk{The odd and even numbers have different thickness to help distinguish contours with similar colour.}}
  \label{fig:coverage}
\end{figure*}

 \begin{figure*}
 \centering
\hspace*{-0.8cm}\includegraphics[width=1.11\textwidth]
{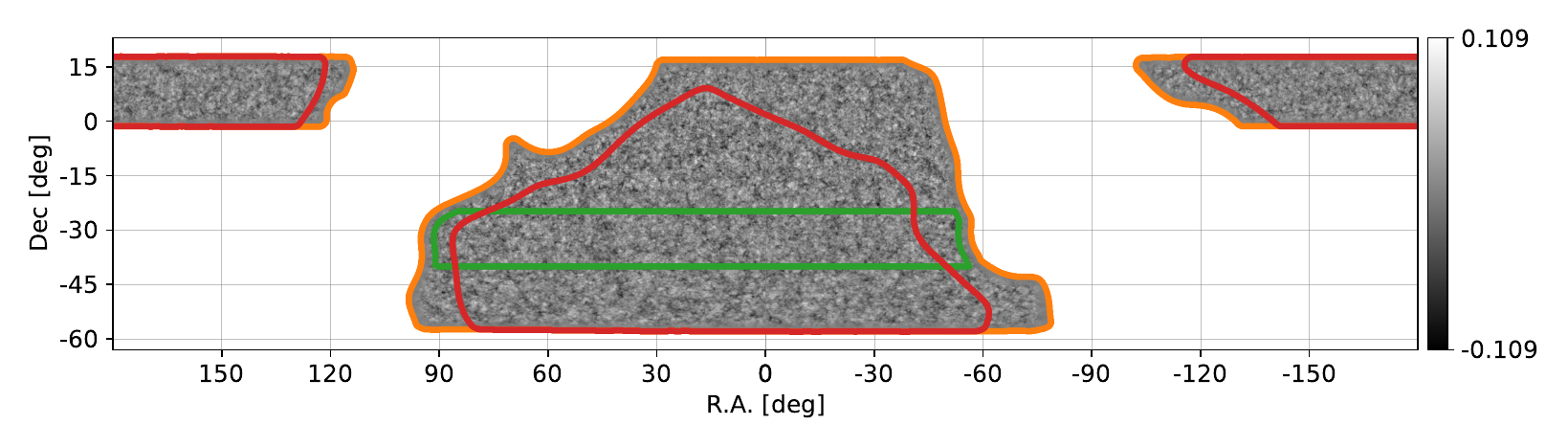}
  \caption{ The DR6 lensing convergence map filtered with a signal-over-noise (Wiener) filter in order to highlight the signal-dominated scales. 
      The coordinate system is the same as in Figure~\ref{fig:coverage}. We note that stretching features in the lower part of the map are due to cylindrical projection. This map is produced using the cross-correlation-based estimator described in Section~\ref{sec:qe}, which avoids using data with the same noise realization, and provides a high-fidelity mapping of the dark matter distribution over $23\,\%$ of the sky. The gray-scale has white corresponding to regions with high matter density and dark to under-dense regions. Our companion paper, \citet{dr6-lensing-cosmo}, describes this lensing map in detail. We show the fiducial analysis mask in orange as well as two additional masks used for consistency tests as described in Sections~\ref{sec:maskxlink} and~\ref{mask40}.} 
  \label{fig:lensing_map}
\end{figure*}

In addition to the maps of the CMB sky, ancillary products are produced by the map-making algorithm. One such set of products is the ``inverse-variance maps'' denoted by $\bold{h}$, which provide the per-pixel inverse noise \abds{variance} of the individual array-frequencies.

\subsection{Beams}\label{subsection:beams}
The instrumental beams are determined from dedicated observations of Uranus and Saturn. The beam estimation closely follows the method used for ACT DR4  \citep{Lungu2022}. In short, the main beams are modelled as azimuthally symmetric and estimated for each observing season from Uranus observations. 
\abds{An additional correction that broadens the beam is determined from point source profiles; this correction is then included in the beam.}
Polarized sidelobes are estimated from Saturn observations and removed during the map-making process. Just as the five observing seasons that make up the DR6 dataset are jointly mapped into eight disjoint splits of the data, the per-season beams are also combined into eight per-split beams using a weighted average that reflects the statistical contribution of each season to the final maps (determined within the footprint of the nominal mask used for this lensing analysis). One notable improvement over the DR4 beam pipeline is the way the frequency dependence of the beam is handled. We now compute, using a self-consistent and Bayesian approach, the scale-dependent colour corrections that convert the beams from describing the response to the approximate Rayleigh--Jeans spectrum of Uranus to one describing the response to the CMB blackbody spectrum. The formalism will be described in a forthcoming paper~\citep{dr6-beams}. The CMB colour correction is below $1\%$ for the relatively low angular multipole limit $\ell_{\mathrm{max}}$ used in this paper.


The planet observations are also used to quantify the temperature-to-polarization leakage of the instrument. The procedure again follows the description in \cite{Lungu2022}. To summarize, Stokes $Q$ and $U$ maps of Uranus are constructed for each detector array and interpreted as an estimate of the instantaneous temperature-to-polarization leakage. After rotating the $Q$ and $U$ maps to the north pole of the standard spherical coordinate system an azimuthally symmetric model is fitted to the maps. The resulting model is then converted to a one-dimensional leakage beam in harmonic space: $B^{T \rightarrow E}_{\ell}$ and $B^{T \rightarrow B}_{\ell}$, which relates the Stokes $I$ sky signal to leakage in the $E$- or $B$-mode linear polarization field.

\subsection{Calibration and transfer function}\label{subsection:cal}
Our filter-free, maximum-likelihood map-making should ideally be
unbiased, but that requires having the correct model for the
data. In practice,  subtle model errors bias the
result. The following two main sources of bias have been identified \citep{2022arXiv221002243N}.
\begin{enumerate}
\item Sub-pixel error: the real CMB sky has infinite resolution while our nominal maps are made at $0.5^\prime$ resolution. While we could have expected this only to affect the smallest angular scales, the coupling of this model error with down-weighting of the data to mitigate effects of atmospheric noise leads to a deficit of power on the largest scales of the maps.
\item Detector gain calibration: inconsistent detector gains can also cause a lack of power in our maps at f090 and f150 on large angular scales.  This inconsistency arises due to errors in gain calibration at the time-ordered-data (TOD) processing stage. The current DR6  maps\footnote{This analysis uses the first science-grade version of the ACT DR6 maps, labeled \texttt{dr6.01}. Since these maps were generated, we have made some refinements to the map-making that improve the large-scale transfer function and polarization noise levels, and include data taken in 2022. We expect to use a second version of the maps for further science analyses and for the DR6 public \abds{CMB} data release.} use a preliminary calibration procedure; alternative calibration procedures are currently being investigated to mitigate this effect.
\end{enumerate} 

To assess the impact of the loss of power at large angular scales on the lensing power spectrum, a multipole-dependent transfer function $t^{T}_\ell$ is calculated at each frequency by taking the ratio of the corresponding ACT CMB temperature bandpowers $C^{\text{ACT}  \times \text{ACT} }_\ell$ and the ACT--\textit{Planck} (\texttt{NPIPE}) temperature cross-correlation bandpowers ${C^{ \text{ACT}  \times \text{P}}_\ell}$:
\ba
    t^{T}_\ell=\frac{C^{\text{ACT}  \times \text{ACT} }_\ell}{C^{ \text{ACT}  \times \text{P}}_\ell}.
\ea
Here $C^{\text{ACT}  \times \text{ACT} }_\ell$ is a noise-free cross-spectrum between data splits and ${C^{ \text{ACT}  \times \text{P}}_\ell}$ is computed by cross-correlating with the \textit{Planck} map which is nearest in frequency.

A logistic function with three free parameters is fit to the above $t^{T}_\ell$. We then divide the temperature maps in harmonic space by the resulting curve, $T_{\ell{m}}\rightarrow T_{\ell{m}}/t^{T}_\ell$, in order to deconvolve the transfer function. Due to the modest sensitivity of our lensing estimator to low CMB multipoles, deconvolving this transfer function results in only a negligible change in 
the lensing power spectrum \abds{amplitude}, $\Delta{A}_\text{lens}=0.004$ (corresponding to less than $0.2\sigma$). Therefore, we have negligible sensitivity to the details of the transfer function.

We determine calibration factors $c_{\textrm{A}_\textrm{f}}$ at each \abds{array-frequency combination} $\textrm{A}_\textrm{f}$ of ACT relative to \textit{Planck} by minimizing differences between the ACT temperature power spectra, $C^{\text{ACT}  \times \text{ACT} ,\textrm{A}_\textrm{f}}_\ell$, and the cross-spectrum with \textit{Planck}, $C^{\text{ACT}  \times \text{P},\textrm{A}_\textrm{f}}_\ell$, at intermediate multipoles. 
In these DR6 maps, the transfer functions approach unity as $\ell$ increases, and eventually \abds{plateau at this value} for $\ell > 800$ and $\ell>1250$ at f090 and f150, respectively; we therefore use the multipoles $800$--$1200$ at f090 and $1250$--$1800$ at f150 to determine the calibration factors $c_{\textrm{A}_\textrm{f}}$ by minimizing the following $\chi^{2}$:
\ba
\chi^{2}(c_{\textrm{A}_\textrm{f}})= \sum_{\ell_b = \ell^{\rm min}_b}^{ \ell^{\rm max}_b}\sum_{\ell^{'}_b = \ell^{\rm min}}^{ \ell^{\rm max}_b}    {\Delta_{\ell_b}(c_{\textrm{A}_\textrm{f}})}\left[\Sigma^{\textrm{A}_\textrm{f}}\right]^{-1}_{\ell_b, \ell_b' } {\Delta_{\ell_b'}(c_{\textrm{A}_\textrm{f}})},
\ea
where the sum is over bandpowers. Here, the difference bandpowers are given by
\begin{equation}
    \Delta_{\ell_b}=c_{\textrm{A}_\textrm{f}}C^{ \rm{ACT} \times \rm{ACT}, \textrm{A}_\textrm{f}}_{\ell_b} -  C^{ \rm{ACT} \times \rm{P}, \textrm{A}_\textrm{f}}_{\ell_b}, 
\end{equation}
and $\Sigma^{\textrm{A}_\textrm{f}}_{\ell_b, \ell_b' }$ is their covariance matrix computed \abds{analytically, using noise power spectra measured from data,} at $c_{\textrm{A}_\textrm{f}}=1$. 

The errors we achieve on the calibration factors are small enough that they can be neglected in our lensing analysis; see Appendix~\ref{app:calandbeam}  for details.

\subsection{Self-calibration of polarization efficiencies}\label{sec.polareff}
Polarization efficiencies scale the true polarization signal on the sky to the signal component in the observed polarization maps.
Assuming incorrect polarization efficiencies in the sky maps leads to biases in the lensing reconstruction amplitude because our quadratic lensing estimator uses up to two powers of the mis-normalized polarization maps; for example, polarization-only quadratic lensing estimators will be biased by the square of the efficiency error $(p^{\textrm{A}_\textrm{f}}_{\mathrm{eff}})^2$.

However, the normalization of the estimator involves dividing the unnormalized estimator, which is quadratic in CMB maps, by fiducial $\Lambda$CDM $C_{\ell}$s. If these fiducial $\Lambda$CDM spectra are rescaled by the same two powers of the efficiency error then the estimator will again become unbiased. In other words, as long as we ensure that the amplitude of the spectra used in the normalization is scaled to be consistent with the amplitude of spectra of the data, our estimator will reconstruct lensing without any bias. The physical explanation of this observation is that \abds{lensing} does not affect the amplitude of the CMB correlations, only their shapes.

To ensure an unbiased polarization lensing estimator, even though the ACT blinding policy in Section \ref{sec:Blind} does not yet allow either a direct comparison of polarization power spectra of ACT and \textit{Planck} or a detailed comparison of the ACT power spectra with respect to $\Lambda$CDM, we employed a simple efficiency self-calibration procedure, which aims to ensure amplitude consistency between fiducial spectra and map spectra. 
The procedure is explained in detail in Appendix \ref{app:polfit}. In short, we fit for a single amplitude scaling $p_{\rm eff}^{{\textrm{A}_\textrm{f}}}$ \blake{fix} between our data polarization power spectra and the fiducial model \abds{power spectra} assumed for the normalization of the estimator. 
We then simply correct the polarization data maps by this \abds{amplitude} scaling parameter to ensure an unbiased lensing measurement. We verify in Appendix \ref{app:syst.poleff} that the uncertainties in this \abds{correction for} the polarization efficiencies are negligible for our analysis.

\subsection{Point-source subtraction}\label{subsection:srcs}
Point-source-subtracted maps are made using a two-step process. First, we run a matched filter on a version of the DR5 ACT+\textit{Planck} maps \citep{Naess_2020} updated to use the new data in DR6, and we register objects detected at greater than $4\sigma$ in a catalog for each frequency band. The object fluxes are then fit individually in each split map using forced photometry at the catalog positions and subtracted from the map. This is done to take into account the strong variability of the quasars that make up the majority of our point source sample. Due to our variable map depth, this procedure results in a subtraction threshold that varies from $4$--$7\,\si{mJy}$ in the f090 band, and $5$--$10\,\si{mJy}$ in the f150 band. An extra map processing step to reduce the effect of point-source residuals not accounted for in the map-making step is described in Section \ref{sec.compact_object}, below.

\subsection{Cluster template subtraction}\label{sec.template_subtract}
Our baseline analysis mitigates biases related to the thermal Sunyaev--Zeldovich (tSZ) effect by subtracting models for the tSZ contribution due to galaxy clusters. We use the \textsc{Nemo}\footnote{\url{https://nemo-sz.readthedocs.io/}} software, which performs a matched-filter search for clusters via their tSZ signal (see \citealt{2021sz} for details). We model the cluster signal using the Universal Pressure Profile (UPP) described by \citet{arnaud10} and construct a set of 15 filters with different angular sizes by varying the mass and redshift of the cluster model.
We construct cluster tSZ model maps for both ACT frequencies by placing beam-convolved UPP-model clusters with an angular size corresponding to that of the maximal signal-to-noise detection across all 15 filter scales as reported by \textsc{Nemo}, for all clusters detected with signal-to-noise ratio (SNR) greater than $5$ on the ACT footprint. This model image is then subtracted from the single-frequency ACT data before coadding. Further details about point-source and cluster template subtraction can be found in \citet{dr6-lensing-fgs}.

\section{Simulations}\label{sec:sim}

Our pipeline requires ensembles of noise and signal simulations. Because ACT is a ground-based telescope, its dominant noise component is slowly varying, large-scale microwave emission by precipitable water vapour in the atmosphere \citep{Errard2015,morris2021}. When combined with the ACT scanning strategy, the atmospheric noise produces several nontrivial noise properties in the ACT DR6 maps. These include steep, red, and spatially varying noise power spectra, spatially varying stripy noise patterns, and correlations between frequency bands \citep{dr6-noise}. 

Simulating the complicated ACT DR6 noise necessitated the development of novel map-based noise models, as described in \cite{dr6-noise}. In our main analysis, we utilize noise simulations drawn from that work's ``isotropic wavelet" noise model \footnote{\abds{We use the \texttt{mnms} (Map-based Noise ModelS) code available at  \url{https://github.
com/simonsobs/mnms.}}}. This model builds empirical noise covariance matrices by performing a wavelet decomposition on differences of ACT map splits.  It is designed to target the spatially varying noise power spectra, which makes it an attractive choice for our lensing reconstruction pipeline,  which, in the large-lens limit, approximates a measurement of the spatially-varying CMB power spectrum \citep{PhysRevD.85.043016,Prince_2018}. 
In Appendix \ref{app:crossnoise} we show that our cross-correlation-based lensing estimator (Section \ref{sec:qe}) is robust to the choice of noise model, producing consistent results when the isotropic wavelet model is replaced with one of the other noise models from \cite{dr6-noise} (the “tiled” or “directional wavelet” models); unlike the isotropic wavelet model, these additionally model the stripy correlated noise features present in the ACT noise maps. \abds{We also emphasize} that since the cross-correlation-based estimator is immune to noise bias and hence insensitive to assumptions of the noise modelling, accurate noise simulations are only required in our pipeline for estimation of the lensing power spectrum's covariance matrix; in contrast, for bias calculation steps, accurate noise simulations are not needed.
 
We then generate full-sky simulations of the lensed CMB \citep{PhysRevD.71.083008} and Gaussian foregrounds  (obtained from the average of foreground power spectra in the \citealt{websky} and \citealt{Sehgal_2010} simulations) at a resolution of $0.5^\prime$ and apply a taper mask at the edge with cosine apodization of width $10^\prime$. We apply the corresponding pixel window function to this CMB signal in Fourier space and then downgrade this map to $1^\prime$ resolution. We add this signal simulation to the noise simulation described above. The full simulation power spectra\abds{, including noise power,} were found to match those of the data to within $3\%$.\footnote{\NS{Note that our blinding policy allows us to compare noise-biased TT power spectra above $\ell=500$ to fiducial noise-biased power spectra.} We also note that, at this level of agreement, our bias subtraction methods such as RDN0 (see Appendix \ref{app.rdn0}) are expected to perform well. We also note that, since these simulations are not used to estimate foreground biases, we may approximate them safely as Gaussian.} For each array-frequency, we generate 800 such simulated sky maps that are used to calculate multiplicative and additive Monte-Carlo (MC) biases \abds{as well as} the covariance matrix (see Section \ref{sec.covmat}). 

We also generate a set of noiseless CMB simulations used to estimate the the mean-field correction and the RDN0 bias (see Section \ref{app.rdn0}) and two sets of noiseless CMB simulations with different CMB signals but with a common lensing field used to estimate the $N_1$ bias (see Section \ref{app.n1}). In Section \ref{sec.covplanck} we also make use of 480 FFP10 CMB simulations \citep{ffpsims} to obtain an accurate estimate of the covariance between ACT DR6 lensing and \textit{Planck} \texttt{NPIPE} lensing.

\section{Pipeline and Methodology}\label{sec:methods}
This section explains the reconstruction of the CMB lensing map and the associated CMB lensing power spectrum, starting from the observed sky maps.

\subsection{Downgrading}
The sky maps are produced at a resolution of $0.5^\prime$, but because our lensing reconstruction uses a maximum CMB multipole of $\ell_\mathrm{max}=3000$, a downgraded pixel resolution of $1^\prime$ is sufficient for the unbiased recovery of the lensing power spectrum and reduces computation time. Therefore,  we downgrade the CMB data maps by block-averaging neighbouring CMB pixels. Similarly, the inverse-variance maps are downgraded by summing the contiguous full-resolution inverse-variance values.

\subsection{Compact-object treatment}\label{sec.compact_object}

The sky maps are further processed to reduce the effect of point sources not accounted for in the map-making step. As described in Section \ref{subsection:srcs}, we work with maps in which point sources above a threshold of roughly 4--10\,mJy (corresponding to an SNR threshold of $4\sigma$) have been fit and subtracted at the map level. However, very bright and/or extended sources may still have residuals in these maps. To address this, we prepare a catalog of 1779 objects for masking with holes of radius $6^\prime$: these include especially bright sources that require a specialized point-source treatment in the map-maker \citep[see][]{Aiola_2020,Naess_2020}, extended sources with $\rm{SNR}>10$ identified through cross-matching with external catalogs, all point sources with $\rm{SNR}>70$ at f150 and an additional list of locations with residuals from point-source subtraction 
that were found by visual inspection. We include an additional set of 14 objects for masking with holes of radius $10^\prime$: these are regions of diffuse or extended positive emission identified by eye in matched-filtered co-adds of ACT maps. They include nebulae, Galactic dust knots, radio lobes and large nearby galaxies. We subsequently inpaint these holes using a constrained Gaussian realisation with a Gaussian field consistent with the CMB signal and noise of the CMB fields and matching the boundary conditions at the hole's edges \citep{Bucher_2012,1911.05717}. This step is required to prevent sharp discontinuities in the sky map that can introduce spurious features in the lensing reconstruction. The total compact-source area inpainted corresponds to a sky fraction of $0.147\,\%$. Further, more detailed discussion of compact object treatment can be found in \citep{dr6-lensing-fgs}.


\subsection{Real-space mask}\label{sec.mask}

To exclude regions of bright Galactic emission and regions of the ACT survey with very high noise, we prepare edge-apodized binary masks over the observation footprint as follows. We start with Galactic-emission masks based on $353\,\si{GHz}$ emission from \Planck\ PR2,\footnote{\texttt{HFI\_Mask\_GalPlane-apo0\_2048\_R2.00.fits}} rotating and reprojecting these to our \textit{Plate-Carr\'{e}e} cylindrical (CAR) pixelization in Equatorial coordinates. We use a \bds{Galactic mask that leaves (in this initial step) 60\% of the full sky as our baseline; we use a more conservative mask retaining initially 40\% of the full sky for a consistency test described in Section \ref{mask40}}. From here on we denote \fk{the masks constructed using these Galactic masks as $60\%$ and $40\%$ masks}\bds{Galactic masks}\blake{or do you mean the full mask constructed using this galactic mask?}. We additionally apply a mask that removes any regions with root-mean-square map noise larger than $70\,\mu\si{K}$-arcmin in any of our input f090 and f150 maps; this removes very noisy regions at the edges of our observed sky area. \bds{Regions with clearly visible Galactic dust clouds and knots are additionally masked, by hand, with appropriately-sized circular holes.}\footnote{Null tests, such as the Galactic mask null tests in Section \ref{mask40} and the consistency between temperature and polarization lensing bandpowers in Section \ref{sec.nullpol}, show that we are insensitive to details of the treatment of Galactic knots.} After identifying these spurious features in either match-filtered maps or lensing reconstructions themselves, masking them removes a further sky fraction of $f_\mathrm{sky}= 0.00138$. The resulting final mask is then adjusted  to round sharp corners. We finally apodize the mask with a cosine-squared edge roll-off of total width of $3\,\text{deg.}$ The total usable area after masking is $9400\,\si{deg}^2$, which corresponds to a sky fraction of $f_\mathrm{sky}= 0.23$.

\subsection{Pixel window deconvolution}
The block averaging operation used to downgrade the sky maps from $0.5^\prime$ to $1^\prime$ convolves the downgraded map with a top-hat function that needs to be deconvolved.\footnote{\abds{Of course, even without downgrading, a pixel window function is present, although it has less impact on the scales of interest.}}
\blake{What is this sentence fragment doing? ``In this case, this is just to account for the pixelization'' }We do this by transforming the temperature and polarization maps $X$ to Fourier space,\footnote{In this paper, we distinguish between Fourier space, obtained from a 2D Fourier transform of the cylindrically projected CAR maps, and harmonic space, which is shorthand for spherical harmonic space} giving $\text{FFT}(X)$, and dividing by the $\mathrm{sinc}(f_x)$ and $\mathrm{sinc}(f_y)$ functions, where $f_x$ and $f_y$ are the dimensionless wavenumbers\footnote{$f_x$ and $f_y$ range from 0 to 0.5 and are generated using the \texttt{numpy} routine \texttt{numpy.fft.rfftfreq}. }
\begin{equation}
    X^{\text{pixel-deconvolved}}=\mathrm{IFFT}\Big[\frac{\mathrm{FFT}(X)}{\mathrm{sinc}(f_x)\mathrm{sinc}(f_y)}\Big],
\end{equation}
where $\text{IFFT}$ denotes the inverse (discrete) Fourier transform.  For simplicity, $X$ without a superscript used in the subsequent sections will refer to the pixel-window-deconvolved maps unless otherwise stated.

\subsection{Fourier-space mask}\label{sec.fourier}
Contamination by ground, magnetic, and other types of pick up in the data due to the scanning of the ACT telescope manifests as excess power at constant declination stripes in the sky maps and thus can be localised in Fourier space.
We mask Fourier modes with $|\ell_x|<90$ and $|\ell_y|<50$ to remove this contamination as in \citet{Louis_2017,Choi_2020}. This masking is carried out both in the data and in our realistic CMB simulations. We demonstrate in Appendix \ref{sec:verification} that this Fourier-mode masking reduces the recovered lensing signal by around $10\%$; we account for this well-understood effect with a multiplicative bias correction obtained from simulations. 

\subsection{Co-addition and noise model}
In the following section, we describe the method we use to combine the individual array-frequency to form the final sky maps used for the lensing measurement.

We first define for each array-frequency's data the map-based coadd map $\boldsymbol{c}$, an unbiased estimate of the sky signal, by taking the inverse-variance-weighted average of the eight split maps $\boldsymbol{m}_i$:
\begin{equation}\label{eq.ivarc}
\boldsymbol{c}=\frac{\sum^{i=7}_{i=0}\mathrm{\bold{h}}_i*\boldsymbol{m}_i}{\sum^{i=7}_{i=0}\mathrm{\bold{h}}_i} .
\end{equation}
Note that in the above equation, the multiplication ($\ast$) and division denote element-wise operations. These coadd maps provide our best estimate of the sky signal for each array, and are used for noise estimation as explained below.

As we will describe in Section \ref{sec:reconstruction}, the cross-correlation-based estimator we use requires the construction of four sky maps $\boldsymbol{d}$ with independent noise. We construct these maps $\boldsymbol{d}$ in the same manner as Equation~\eqref{eq.ivarc}, coadding together split $j$ and $j+4$ with $j\in\{0,1,2,3\}$.

\subsubsection{Inverse-variance coaddition of the array-frequencies}

We combine the different coadded data maps $\bold{d}_{\mathrm{A}_\mathrm{f}}$ with array-frequencies  $\mathrm{A}_\mathrm{f}\in\{\text{PA4 f150}, \text{PA5 f090}, \text{PA5 f150}, \text{PA6 f090}, \text{PA6 f150}\}$ into single CMB fields $M_{\ell{m}}^{X}$, with \abds{$X\in(T,E,B)$} 
on which lensing reconstruction is performed.  The coadding of the maps is done in spherical-harmonic space,\footnote{\fk{The harmonic-space coadding we perform here does not fully account for spatial inhomogeneities in the noise, as opposed to the coadd method presented in \cite{Naess_2020}.   However, this is justified because all array-frequencies have similar spatial noise variations as they are observed with the same scanning pattern. Hence the spatial part should approximately factor out.}}
\begin{equation}
    M_{\ell{m}}=\sum_{\mathrm{A}_\mathrm{f}} w^{\mathrm{A}_\mathrm{f}}_\ell d^{{\mathrm{A}_\mathrm{f}}}_{\ell{m}}\left(B^{\mathrm{A}_\mathrm{f}}_\ell\right)^{-1},
\end{equation}
where 
\begin{equation}\label{eq.weight}
    w_\ell^{\mathrm{A}_\mathrm{f}}=\frac{\left(N^{\mathrm{A}_\mathrm{f}}_\ell\right)^{-1}\left(B^{\mathrm{A}_\mathrm{f}}_\ell\right)^2}{\sum_{(\mathrm{A}_\mathrm{f})}\left(N^{\mathrm{A}_\mathrm{f}}_\ell\right)^{-1}\left(B^{\mathrm{A}_\mathrm{f}}_\ell\right)^2}
\end{equation}
are the normalized inverse-variance weights in harmonic space. These weights, giving the relative contributions of each array-frequency, are shown in Figure~\ref{fig.weights} and are constructed to sum to unity at each multipole $\ell$. Note that a deconvolution of the harmonic beam transfer functions $B^{\mathrm{A}_\mathrm{f}}_\ell$ is performed for each array-frequency\footnote{\fk{We use the same beam for temperature and polarization and neglect $T\rightarrow P$ leakage beams. The latter is justified in Section \ref{app:syst.t2pleakage}, where we show that including $T\rightarrow P$ has a small impact on the lensing bandpowers (a shift of less than $0.1\%$).}}. 
The noise power spectra $N^{\mathrm{A}_\mathrm{f}}_\ell$ are obtained from the beam-deconvolved noise maps of the individual sky maps with the following prescription.

We construct a noise-only map, $\boldsymbol{n}_i$ by subtracting the pixel-wise coadd $\boldsymbol{c}$ of each map\footnote{For simplicity, we suppress the subscripts indicating the array-frequency $(A_f)$.} from the individual  data splits $\boldsymbol{m}_i$; this noise-only map is given by:
\begin{equation}
    \boldsymbol{n}_i=\boldsymbol{m}_i-\boldsymbol{c}.
\end{equation}
We then transform the real-space noise-only maps $\boldsymbol{n}_i$ into spherical-harmonic space $n^{(i)}_{\ell{m}}$ and use these to compute the noise power spectra used for the weights in Equation~\eqref{eq.weight}. Since we have $k=8$ splits, we can reduce statistical variance by finding the average of these noise spectra\footnote{The factors of $1/[k(k-1)]$ are explained as follows: the $1/(k-1)$ factor converts the null noise power per split to an estimate of the coadd noise power; this is $(k-1)$ since the coadd map enters into $n$ removing 1 degree of freedom. The additional $1/k$ averages over 8 independent realizations. Refer to \citep{dr6-noise} for a detailed discussion.}:
\begin{equation}
    N_\ell=\frac{1}{w_2}\frac{1}{k(k-1)}\frac{1}{2\ell+1}\sum^{k}_i\sum^{\ell}_{m=-\ell} {n^{(i)}_{\ell{m}}}n^{(i)*}_{\ell{m}},
\end{equation}
where $w_2=\int d^2\hat{\bold{n}} \, M^2(\hat{\bold{n}})/(4\pi)$ is the average value of the second power of the mask $M(\hat{\bold{n}})$, which corrects for the missing sky fraction due to the application of the analysis mask, as described in Section \ref{sec.mask}. The resulting noise power is further smoothed over by applying a linear binning\footnote{We checked that the resulting coadded map is stable to different choices of binning $\Delta\ell$ as long as the resultant $N_\ell$ are smooth.} of $\Delta\ell=14$ .

\begin{figure}
  \centering
  \includegraphics[width=\linewidth]{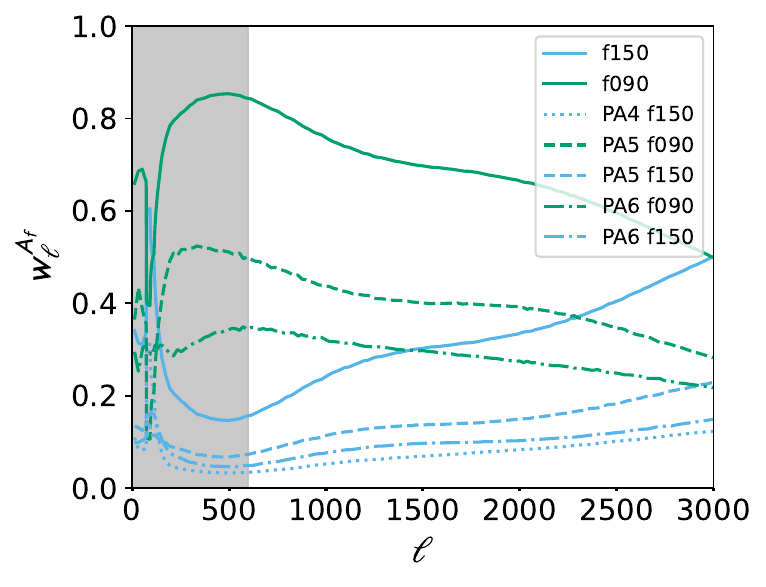}
  \caption{Maps from detector array-frequencies are combined using a weighted average in harmonic space to form a coadded sky map. This figure shows the weights applied to the map from each array-frequency as a function of multipole; the weights sum to unity at each multipole. The white regions show the CMB scales used in our baseline analysis, namely $600\leq\ell\leq3000$. The combined f090 and f150 weights are shown in solid green and blue  respectively. We note that the PA4 f150 array map has the smallest weight (shown as dotted light blue), less than $10\%$, while PA5 f090 map provides the largest contributions to our coadded map (with the weight shown in dashed green).
  \label{fig.weights}}
\end{figure}
The same coadding operation is performed on simulations containing lensed sky maps and noise maps. The resulting suite of coadded CMB simulations is used throughout our baseline analysis.

\subsubsection{Internal linear combination coaddition}
As an alternative to our baseline approach of combining only the ACT maps in harmonic space, we also explore a frequency cleaning approach which includes high-frequency data from \textit{Planck} ($353\,\si{GHz}$ and $545\,\si{GHz}$). This approach is described in detail in \cite{dr6-lensing-fgs} but, to summarise, we produce harmonic-space constrained internal linear combinations (ILC) of the ACT and high-frequency \textit{Planck} maps that minimise the variance of the output maps while also approximately deprojecting the cosmic infrared background (CIB).  Comparisons of the consistency of this approach against the baseline method are described in Section \ref{sec:cibdeprojection} and provide a useful test of our methods for mitigating foreground biases.

\subsection{Filtering}\label{sec.filtering}

Optimal quadratic lensing reconstruction requires as inputs Wiener-filtered $X=T,E$ and $B$ CMB multipoles and inverse-variance-filtered maps (the latter can be obtained from the former by dividing the Wiener-filtered multipoles by the fiducial lensed power spectra ${C}^{\text{fid}}$ before projecting back to maps). The filtering step is important because \abds{an optimal analysis of the observed CMB sky requires} both the downweighting of noise and the removal of masked areas} \citep{PhysRevD.83.043005}.

We write the temperature $T$ and polarization $_{\pm2}P\equiv{Q}\pm{i}U$ (beam- and pixel-deconvolved) data maps as
\begin{equation}
    \begin{pmatrix}
        T\\_2P\\_{-2}P
    \end{pmatrix}=\mathcal{Y}\begin{pmatrix}
        T_{\ell{m}}\\E_{\ell{m}}\\B_{\ell{m}}
    \end{pmatrix}+\mathrm{noise},
\end{equation}
where the matrix $\mathcal{Y}$ contains the spin-weighted spherical harmonic functions to convert the spherical harmonics $T_{\ell{m}}, E_{\ell{m}}$ and $B_{\ell{m}}$ to real-space maps over the unmasked region. The real-space covariance matrix of the data maps is
\begin{equation}
C = \mathcal{Y} \mathbb{C}^{\text{fid}} \mathcal{Y}^\dagger + C_\text{noise} ,
\end{equation}
where $\mathbb{C}^{\text{fid}}$ is the matrix of our fiducial lensed CMB spectra with elements
\begin{equation}
\label{eq:lensedfilters}
    [\mathbb{C}^{\text{fid}}]_{\ell{m},\ell^{\prime}{m^{\prime}}}=\delta_{\ell\ell^\prime}\delta_{mm^\prime}\begin{pmatrix}C^{TT}_\ell&C^{TE}_\ell&0\\C^{TE}_\ell&C^{EE}_\ell&0\\0&0&C^{BB}_\ell \end{pmatrix} ,
\end{equation}
and $C_\text{noise}$ is the real-space noise covariance matrix. The Wiener-filtered multipoles are then obtained as
\begin{equation}
{X}^{\text{WF}}_{\ell{m}} = \mathbb{C}^{\text{fid}} \mathcal{Y}^\dagger C^{-1} \begin{pmatrix} T \\ {}_2 P \\ {}_{-2} P \end{pmatrix} .
\label{eq:exactWF}
\end{equation}

For our main analysis, we employ an approximate form of the Wiener filter that follows from the rearrangement 
\begin{align}
\mathbb{C}^{\text{fid}} \mathcal{Y}^\dagger C^{-1} &= \left(\left(\mathbb{C}^{\text{fid}}\right)^{-1} + \mathbb{N}^{-1}\right)^{-1} \mathcal{Y}^\dagger C^{-1} \nonumber \\
&= \left(\left(\mathbb{C}^{\text{fid}}\right)^{-1} + \mathbb{N}^{-1}\right)^{-1} \mathbb{N}^{-1} \mathcal{Y}^\dagger \left(\mathcal{Y}\mathcal{Y}^\dagger\right)^{-1} \nonumber \\
&= \mathbb{C}^{\text{fid}} \left(\mathbb{C}^{\text{fid}} + \mathbb{N}\right)^{-1} \mathcal{Y}^\dagger \left(\mathcal{Y}\mathcal{Y}^\dagger\right)^{-1} ,
\end{align}
where $\mathbb{N}^{-1} \equiv \mathcal{Y}^\dagger C_{\text{noise}}^{-1} \mathcal{Y}$. The operation $\mathcal{Y}^\dagger \left(\mathcal{Y}\mathcal{Y}^\dagger\right)^{-1}$ takes the (pseudo-)spherical transform of the masked maps, with $\mathcal{Y}\mathcal{Y}^\dagger = \text{diag}(1,2,2)\delta^{(2)}(\unitvec-{\unitvec}')$. Our approximate form of the Wiener filter takes the form
\begin{equation}\label{wfmultipole}
{X}^{\text{WF}}_{\ell{m}}\approx \mathbb{C}^{\text{fid}}\mathcal{F}\mathcal{Y}^\dagger \left(\mathcal{Y}\mathcal{Y}^\dagger\right)^{-1} \begin{pmatrix}
        T\\_2P\\_{-2}P
    \end{pmatrix},
\end{equation}
%
%
where $\mathcal{F}$ is the filtering operation applied to the temperature and polarization spherical harmonics. The filters used are diagonal in harmonic space such that each component of $X\in{T,E,B}$ is filtered separately by  $F^X_\ell=1/(C^{XX}_\ell+N^{XX}_\ell)$.

The above diagonal filtering neglects small amounts of mode mixing due to masking, does not account for noise inhomogeneities over the map, and also ignores cross-correlation in $C^{TE}_\ell$. However, it has the advantage of allowing the temperature and polarization map to be filtered independently and is a good approximation on scales for which the CMB fields are signal dominated, and in situations when the noise level is close to homogeneous, as is the case for ACT DR6.\footnote{The sky maps used have only a factor of two variation on the spatial dependence of the depth after the cuts done in Fig. \ref{fig:coverage}.} 
This method is also significantly faster than using the more optimal filter in Equation~\eqref{eq:exactWF}, which requires evaluation of the inverse of the covariance matrix $C$ with, for example, conjugate-gradient methods. Therefore, for the main analysis, we employ this diagonal filter. 


The inverse-variance filtered maps
\begin{equation}
\bar{\bold{X}}(\unitvec) = C^{-1} \begin{pmatrix} T \\ {}_2 P \\ {}_{-2}P \end{pmatrix}
\end{equation}
are related to the Wiener-filtered multipoles ${X}^{\text{WF}}_{\ell{m}}$ in Equation~\eqref{wfmultipole} via
\begin{align}
    \bar{\bold{X}}(\unitvec)&=(\mathcal{Y}\mathcal{Y}^\dagger)^{-1}\mathcal{Y}(\mathbb{C}^{\text{fid}})^{-1}{X}^{\text{WF}}_{\ell{m}}\nonumber\\&=\mathrm{diag}\left(1,\frac{1}{2},\frac{1}{2}\right)\mathcal{Y}(\mathbb{C}^{\text{fid}})^{-1}{X}^{\text{WF}}_{\ell{m}}\nonumber\\&=\mathrm{diag}\left(1,\frac{1}{2},\frac{1}{2}\right)\mathcal{Y}{\bar{X}}_{\ell{m}},
\end{align}
where, in the last line, $\bar{X}_{\ell{m}} = (\mathbb{C}^{\text{fid}})^{-1}{X}^{\text{WF}}_{\ell{m}}$.

\subsection{Lensing reconstruction}\label{sec:reconstruction}
In this section, we describe the methodology used to estimate CMB lensing using the quadratic estimator (QE). Our baseline methodology closely follows the pipeline used in \citet{Planck:2018}, albeit with key improvements in areas such as foreground mitigation (using a profile-hardened estimator that is more robust to extragalactic foregrounds, see Section \ref{sec:biashardening}) and immunity to noise modeling (using the more robust cross-correlation-based estimator described in Section \ref{sec.crossqe}).

\subsubsection{Standard Quadratic Estimator}\label{sec:qe}
A fixed realization of gravitational lenses imprints preferred directions into the CMB, thereby breaking the statistical isotropy of the unlensed CMB. Mathematically, the breaking of statistical isotropy corresponds to the introduction of new correlations between different, formerly independent modes of the CMB sky, with the correlations proportional to the lensing potential $\phi_{LM}$. Adopting the usual convention of using $L$ and $M$ to refer to lensing multipoles and $\ell$ and $m$ to CMB multipoles, we may write the new, lensing-induced correlation between two different CMB modes ${X}_{\ell_1m_1}$ and ${Y}_{\ell_2m_2}$ as follows :
\begin{multline}
    \langle{X}_{\ell_1m_1}{Y}_{\ell_2m_2}\rangle_{\text{CMB}}=\sum_{LM}(-1)^M\begin{pmatrix}{\ell_1}&\ell_2&L\\ m_1&m_2&-M\end{pmatrix} \\ \times f^{XY}_{\ell_1\ell_2L}\phi_{LM}.
\end{multline}
The average $\langle\quad \rangle_{\text{CMB}}$ is taken over CMB realizations with a fixed lensing potential $\phi$. Here the fields ${X}_{\ell{m}},{Y}_{\ell{m}}\in\{T_{\ell{m}},E_{\ell{m}},B_{\ell{m}}\}$ and the bracketed term is a Wigner $3j$ symbol. The response functions $f^{XY}_{\ell\ell^\prime{L}}$ for the different quadratic pairs $XY$ can be found in \cite{Okamoto2003} and are linear functions of the CMB power spectra (the lensed spectra are used to cancel a higher-order correction \citealt{Lewis_2011}). 

The correlation between different modes induced by lensing motivates the use of quadratic combinations of the lensed temperature and polarization maps to reconstruct the lensing field. Pairs of Wiener-filtered maps, $X^{\text{WF}}$, and inverse-variance-filtered maps, $\bar{X}$, are provided as inputs to a quadratic estimator that reconstructs an un-normalized, minimum-variance (MV) estimate of the spin-$1$ component of the real-space lensing displacement field:
%
\begin{equation}\label{real_space}
    {}_1 \hat{d}(\unitvec)=-\sum_{s=0,\pm2}{_{-s}\bar{X}}(\unitvec)[\eth_sX^{\text{WF}}](\unitvec).
\end{equation}
 Here, $\eth$ is the spin-raising operator acting on spin spherical harmonics and the pre-subscript $s$ denotes the spin of the field. The gradients of the Wiener-filtered maps are given explicitly by
\begin{widetext}
\begin{align}
[\eth_0 X^{\text{WF}}](\unitvec)&\equiv\sum_{\ell{m}}\sqrt{\ell(\ell+1)}T^{\text{WF}}_{\ell{m}}{_1}Y_{\ell{m}}(\unitvec),\nonumber\\
[\eth_{-2}X^{\text{WF}}](\unitvec)&\equiv-\sum_{\ell{m}}\sqrt{(\ell+2)(\ell-1)}[E^{\text{WF}}_{\ell{m}}-iB^{\text{WF}}_{\ell{m}}]{_{-1}}Y_{\ell{m}}(\unitvec),\nonumber\\
[\eth_{2}X^{\text{WF}}](\unitvec)&\equiv-\sum_{\ell{m}}\sqrt{(\ell-2)(\ell+3)}[E^{\text{WF}}_{\ell{m}}+iB^{\text{WF}}_{\ell{m}}]{_3}Y_{\ell{m}}(\unitvec).
\label{eq:WF}
\end{align}
\end{widetext}

The displacement field can be decomposed into the gradient $\phi$ and curl $\Omega$ components by expanding in spin-weighted spherical harmonics:
\begin{equation}\label{real_harmonic}
   {}_{\pm1}\hat{d}(\unitvec)=\mp\sum_{LM}\Bigg(\frac{\bar{\phi}_{LM}\pm{i}\bar{\Omega}_{LM}}{\sqrt{L(L+1)}}\Bigg){_{\pm1}}Y_{LM}(\unitvec).
\end{equation}
%
Hence, by taking spin-$\pm 1$ spherical-harmonic transforms of ${}_{\pm1}\hat{d}(\unitvec)$, where ${}_{-1}\hat{d} = {}_1 \hat{d}^\ast$,
and taking linear combinations of the resulting coefficients, we can isolate the gradient and curl components. 
The gradient component $\phi_{LM}$ contains the information about lensing that is the focus of our analysis. \footnote{\fk{Here we adopt the notation of using the overbar to refer to  unnormalized quantities.}} \blake{do we define un-normalized phi vs normalized phi with and without bars somewhere? should do so when it's mentioned first} The curl $\Omega_{LM}$ is expected to be zero (up to small post-Born corrections; e.g.,~\citealt{Pratten_2016} and references therein) and can therefore serve as a useful null test, as discussed in Section \ref{sec:curl}.




Even in the absence of lensing, other sources of statistical anisotropy in the sky maps, such as masking or noise inhomogeneities, can affect the naive lensing estimator. One can correct such effects by subtracting the lensing estimator's response to such non-lensing statistical anisotropies, which is commonly referred to as the mean-field $\langle\bar{\phi}_{LM}\rangle$. 
We estimate this mean-field signal by averaging the reconstructions produced by the naive lensing estimator from 180 noiseless\footnote{The reason we do not include instrumental noise here is that we use the cross-correlation-based estimator, presented in Section \ref{sec.crossqe}, which cancels the noise contribution to the mean-field.} simulations, each with independent CMB and lensing potential realizations. 
This averaging ensures that only the response to spurious, non-lensing statistical anisotropy remains (as the masking 
is the same in all simulations, whereas CMB and lensing fluctuations average to zero).  Subtracting this mean-field leads us to the following lensing estimator:
\begin{equation}
    \bar{\phi}_{LM}\rightarrow\bar{\phi}_{LM}-\langle\bar{\phi}_{LM}\rangle.
\end{equation}

The temperature-only $(s=0)$ and polarization-only $(s=\pm2)$ estimators\footnote{Note that the $s=0$ estimator includes part of the standard Hu and Okamoto $TE$ estimator (with $E$ on the gradient leg) through the Wiener filter, and the $s=2$ includes part of the usual $TE$ and $TB$ estimators. When obtaining temperature-only estimators, we therefore also set the  input $E$-fields to zero. } in Equation~\eqref{real_space} are combined at the field level \footnote{As opposed to the alternative of combining at the lensing power spectrum level.} to produce the full un-normalized MV estimator.

Expanding the Wiener-filtered fields in terms of the inverse-variance-filtered multipoles $\bar{X}_{\ell m}$, and extracting the gradient part, approximately recovers the usual estimators $\bar{\phi}^{XY}_{LM}$ of \cite{Okamoto2003}, where $XY\in\{TT,TE,ET,EE,EB,BE\}$. 
More specifically, the MV estimator presented here is approximately equivalent\footnote{Our implementation corresponds to the SQE estimator from \cite{2101.12193}, which is slightly sub-optimal compared to \cite{Okamoto2003}.}
to combining the individual estimators $\bar{\phi}^{XY}_{LM}$
 with a weighting given by the inverse of their respective normalization $({\mathcal{R}^{XY}_L})^{-1}$:
\begin{equation}\label{eq:mv.estimator}
    \hat{\phi}^{{\mathrm{MV}}}_{LM}=({\mathcal{R}^{\mathrm{MV}}_L})^{-1}\sum_{XY}\bar{\phi}^{{XY}}_{LM}.
\end{equation}
%

Here, $({\mathcal{R}^{\mathrm{MV}}_L})^{-1}$ is the MV estimator normalization that ensures our reconstructed lensing field is unbiased; by construction, it is defined via $\phi_{LM}=({\mathcal{R}^{XY}_L})^{-1}\langle\bar{\phi}^{XY}_{LM}\rangle_{\text{CMB}}$. The normalization is given explicitly by
\begin{equation}
({\mathcal{R}^{\mathrm{MV}}_L})^{-1}=\frac{1}{\sum_{XY}{(\mathcal{R}^{XY}_L})}.
\end{equation}

\fk{In the notation adopted here, the unnormalized estimator $\bar{\phi}_{LM}$ is related to the normalized estimator $\hat{\phi}_{LM}$ via the normalization $\mathcal{R}^{-1}_L$ as $\hat{\phi}_{LM}=\mathcal{R}^{-1}_L\bar{\phi}_{LM}$. }

\abds{To first approximation, this normalization is calculated analytically with curved-sky expressions from \cite{Okamoto2003}. We generally use fiducial lensed spectra in this calculation (as well as the filtering of Eq.~\ref{eq:lensedfilters}), which reduces the higher-order $N^{(2)}$ bias to sub-percent levels; however, for the $TT$ estimator, we use the lensed temperature-gradient power spectrum $C^{T\nabla{T}}_\ell$ to further improve the fidelity of the reconstruction \citep{Lewis_2011}.} This analytic, isotropic normalization is fairly accurate, but it does not account for effects induced by Fourier-space filtering and sky masking. Therefore, we additionally apply a multiplicative Monte-Carlo (MC) correction $\Delta{A}^{\text{MC,mul}}_L$ to all lensing estimators, so that $\hat{\phi}_{LM}\rightarrow \Delta{A}^{\text{MC,mul}}_L \hat{\phi}_{LM}$. 
This correction is obtained by first cross-correlating reconstructions from simulations with the true lensing map; we then divide the average of the input simulation power spectrum by the result, i.e.,
\begin{equation}
    \Delta{A}^{\text{MC,mul}}_L=\frac{\langle{C}^{{\phi}\phi}_L\rangle}{\langle{C}^{\hat{\phi}\phi}_L\rangle}.
\end{equation}
In practice, this multiplicative MC correction is computed after binning both spectra into bandpowers.

An explanation of the origin of the multiplicative MC correction is provided in Appendix \ref{sec:verification}: it is found to be primarily a consequence of the Fourier-space filtering.

Having obtained our estimate of the lensing map in harmonic space, $\hat{\phi}_{LM} $, we can compute a naive, biased estimate of the lensing power spectrum. Using two instances of the lensing map estimates $\hat{\phi}_{LM}^{AB}$ and $\hat{\phi}_{LM}^{CD}$, this power spectrum is given by
\begin{equation}
    \hat{C}^{\hat{\phi}\hat{\phi}}_L(\hat{\phi}_{LM}^{AB},\hat{\phi}_{LM}^{CD})\equiv\frac{1}{w_4(2L+1)}\sum^{L}_{M=-L}\hat{\phi}_{LM}^{AB}(\hat{\phi}_{LM}^{CD})^*,
\end{equation}
where $w_4=\int d^2 \unitvec\, M^4(\unitvec)/(4\pi)$, the average value of the fourth power of the mask $M(\bold{\hat{n}})$, corrects for the missing sky fraction due to the application of the analysis mask. In Equation~\eqref{crossqe}, below, we will introduce a new version of these spectra that ensures that only \emph{different} splits of the data are used in order to avoid any noise contribution. This will allow us to obtain an estimate of the lensing power spectrum that is not biased by any mischaracterization of the noise in our CMB observations.

Nevertheless, biases arising from CMB and lensing signals still need to be removed from the naive lensing power spectrum estimator. We discuss the subtraction of these biases in Section \ref{sec.biassub}.

\subsubsection{{Profile hardening for foreground mitigation}}\label{sec:biashardening}
Extragalactic foreground contamination from Sunyaev--Zel'dovich clusters, the cosmic infrared background, and radio sources can affect the quadratic estimator and hence produce large biases in the recovered lensing power spectrum if unaccounted for. For our baseline analysis, we use a geometric approach to mitigating foregrounds and make use of bias-hardened estimators \citep{Namikawa2013,Osborne_2014,PhysRevD.102.063517}. As with lensing, other sources of statistical anisotropy in the map such as point sources and tSZ clusters can be related to a response function $f^s_{\ell_1\ell_2L}$ and a field $s_{LM}$ describing the anisotropic spatial dependence. Bias-hardened estimators work by reconstructing simultaneously both lensing and non-lensing statistical anisotropies and subtracting the latter, with a scaling to ensure the resulting estimator has no remaining response to non-lensing anisotropies. Explicitly, the bias-hardened $TT$ part of the lensing estimator is given by
\begin{equation}
    \hat{\phi}^{TT,\mathrm{BH}}_{{LM}}=\frac{\hat{\phi}^{TT}_{{LM}}-(\mathcal{R}_L^{TT})^{-1}\mathcal{R}^{\phi,s}_{{L}}\hat{s}_{{LM}}}{1-(\mathcal{R}^{\phi,s}_{{L}})^2(\mathcal{R}^{TT}_{{L}})^{-1}(\mathcal{R}^{s}_{{L}})^{-1} },
\end{equation}
\fk{
where $\mathcal{R}^{\phi,s}_{{L}}$ is the cross-response between the lensing field $\hat{\phi}^{TT}_{{LM}}$ and the source  field $\hat{s}_{LM}$ and $(\mathcal{R}^{s}_{{L}})^{-1}$ is the normalization for the source estimator.}

In our case, we optimise the response to the presence of tSZ cluster ``sources'', and as shown in \cite{sailer2023}, this estimator is also effective in reducing the effect of point sources by a factor of around five. The cross-response function of this \abds{tSZ-profile-hardened estimator} is given by
%
%
%
\fk{\begin{equation}
    \mathcal{R}^{\phi,\mathrm{tSZ}}_{{L}}=\frac{1}{2L+1}\sum_{\ell\ell^{\prime}}\frac{f^{\phi}_{\ell{L}\ell^\prime}f^{tSZ}_{\ell{L}\ell^\prime}}{2C^{\mathrm{total}}_\ell{C_{\ell^{\prime}}^{\mathrm{total}}}} ,
\end{equation}}
\fk{where $C^{\mathrm{total}}_\ell=C^{TT}_\ell+N^{TT}_\ell$ is the total temperature power spectrum including instrumental noise and $f^{tSZ}_{\ell{L}\ell^\prime}$ is the response function to tSZ sources. This response function requires a model for cluster profiles; we estimate an effective profile from the square root of the smoothed tSZ angular power spectrum (which is dominated by the one-halo term) obtained from a \textsc{websky} simulation~\citep{PhysRevD.102.063517}.}

In the formalism presented here, the appropriately normalized MV estimator, with the temperature estimator part `hardened' against tSZ, is obtained by first subtracting the standard temperature lensing estimator from the MV estimator and then adding back the profile-hardened temperature estimator, i.e.,
\begin{multline}
    \hat{\phi}^{\mathrm{MV},BH}_{LM}=({\mathcal{R}^{\mathrm{MV}}_L})^{-1}\left[\frac{\hat{\phi}^{\mathrm{MV}}_{LM}}{({\mathcal{R}^{\mathrm{MV}}_L})^{-1}}-\frac{\hat{\phi}^{TT}_{LM}}{({\mathcal{R}^{TT}_L})^{-1}} \right. \\
    \left. +\frac{\hat{\phi}^{TT,\text{BH}}_{LM}}{({\mathcal{R}^{TT}_L})^{-1}}\right].
\end{multline}

Both the investigation of foreground mitigation in \cite{dr6-lensing-fgs}, summarized in this paper in \ref{foreground}, and the foreground null tests discussed in Section \ref{sec:null} show that this \abds{baseline method can control the foreground biases on the lensing amplitude $\Alens$ to levels below $0.2\sigma$, where $\sigma$ is the statistical error on this quantity.}

\subsubsection{Cross-correlation-based quadratic estimator}\label{sec.crossqe}

The lensing power spectrum constructed using the standard QE  is  sensitive to assumptions made in simulating and modelling the instrument noise used for calculating the lensing power-spectrum biases. This is despite the use of realization-dependent methods, as described in Appendix \ref{app.rdn0} (which discusses power-spectrum bias subtraction). Hence, in practice, we construct our lensing power spectrum using lensing maps $\hat{\phi}^{(ij),XY}_{LM}$ reconstructed from different data splits, indexed by $i$ and $j$, which have independent noise.
Using the shorthand notation of $\mathrm{QE}(X^A,Y^B)$ 
for the quadratic estimator (see Eq. \ref{real_space}) operating on two sky maps $X^A$ and $Y^B$, $\hat{\phi}^{(ij),XY}_{LM}$ is defined as 
\begin{equation}
    \hat{\phi}^{(ij),XY}_{LM}=\frac{1}{2}[\mathrm{QE}(X^i,Y^j)+\mathrm{QE}(X^j,Y^i)].
\end{equation}
Note that this is symmetric under interchange of the splits.

\abds{We use this cross-correlation-based estimator from \cite{Madhavacheril2021} with independent data splits to ensure our analysis is immune to instrumental and atmospheric noise effects in the mean-field and $N_0$ (Gaussian) biases (introduced below in Section~\ref{sec.biassub}). This makes our analysis highly robust to potential inaccuracies in simulating the complex atmospheric and instrumental noise in the ACT data.}

The coadded, standard lensing estimator, equivalent to Equation~\eqref{eq:mv.estimator}, which uses all the map-split combinations, is given by
\begin{equation}
    \hat{\phi}^{XY}_{LM}=\frac{1}{4^2}\sum_{ij}\hat{\phi}^{(ij),XY}_{LM}.
\end{equation}
\abds{The corresponding estimate of the power spectrum from $XY$ and $UV$ standard QEs is then
\begin{equation}
    C^{\hat{\phi}\hat{\phi}}_L[XY,UV] = \frac{1}{4^4} \sum_{ijkl} \hat{C}^{\hat{\phi}\hat{\phi}}_L(\hat{\phi}^{(ij),{XY}}_{LM},\hat{\phi}^{(kl),{UV}}_{LM}) .
\end{equation}
This is modified by removing any terms where the same split is repeated to give the cross-correlation-based estimator:
\begin{equation}
    C^{\hat{\phi}\hat{\phi},\times}_L[XY,UV] = \frac{1}{4!} \sum_{i\neq j \neq k \neq l} \hat{C}^{\hat{\phi}\hat{\phi}}_L(\hat{\phi}^{(ij),{XY}}_{LM},\hat{\phi}^{(kl),{UV}}_{LM}) . \label{eq:PScrossslow}
\end{equation}
In this way, only lensing maps constructed from CMB maps with independent noise are included, so noise mis-modelling does not affect the mean-field estimation, and any cross-powers between lensing maps that repeat splits (and hence contribute to the Gaussian noise bias) are discarded.}

We can accelerate the computation of Equation~\eqref{eq:PScrossslow} following~\citet{Madhavacheril2021}.
We introduce the following auxiliary estimators using different combinations of splits:
\begin{align}
       \hat{\phi}^{\times,{XY}}_{LM}&=\hat{\phi}^{XY}_{LM}-\frac{1}{16}\sum^{4}_{i=1}\hat{\phi}^{(ii),XY}_{LM}, \\ \hat{\phi}^{(i),{XY}}_{LM}&=\frac{1}{4}\sum_{j=1}^4{\hat{\phi}^{(ij),XY}}_{LM},\\    \hat{\phi}^{(i)\times,{XY}}_{LM}&=\hat{\phi}^{(i),XY}_{LM}-\frac{1}{4}\hat{\phi}^{(ii),XY}_{LM},
\end{align}
%
in terms of which the cross-correlation-based estimator may be written as
\begin{align}\label{crossqe}
    C^{\hat{\phi}\hat{\phi},\times}_L[XY,UV]=\frac{1}{4!}\Big[&256\hat{C}^{\hat{\phi}\hat{\phi}}_L(\hat{\phi}^{\times,{XY}}_{LM},\hat{\phi}^{\times,{UV}}_{LM})\nonumber \\&-64\sum^{4}_{i=1}\hat{C}^{\hat{\phi}\hat{\phi}}_L(\hat{\phi}^{(i)\times,{XY}}_{LM},\hat{\phi}^{(i)\times,{UV}}_{LM})\nonumber \\&+4\sum_{i\leq{j}}\hat{C}^{\hat{\phi}\hat{\phi}}_L(\hat{\phi}^{(ij),{XY}}_{LM},\hat{\phi}^{(ij),{UV}}_{LM})\Big].
\end{align}


Finally, the baseline lensing map we produce, which again avoids repeating the same data splits in the estimator, is given by
\begin{equation}
    \hat{\phi}^{XY}_{LM}=\frac{1}{6}\sum_{i<j}\hat{\phi}_{LM}^{(ij),XY}.
\label{eq:lensingmap}
\end{equation}
The resulting lensing map is shown in CAR projection in Figure~\ref{fig:lensing_map}, with the map filtered to highlight the signal-dominated scales.

\subsection{Bias Subtraction}
 \label{sec.biassub}

Naive lensing power spectrum estimators based on the auto-correlation of a reconstructed map are known to be biased due to both reconstruction noise and higher-order lensing terms. This is also true for the cross-correlation-based lensing power spectrum in Equation~\eqref{crossqe}, despite its insensitivity to noise. To obtain an unbiased lensing power spectrum from the naive lensing power spectrum estimator, we must subtract the well-known lensing power spectrum biases: the $N_0$ and $N_1$ biases as well as a small additive MC bias. The bias-subtracted lensing power spectrum is thus given by
\begin{equation}\label{eq.sub}
    \hat{C}^{\phi\phi,\times}_L=C^{\hat{\phi}\hat{\phi},\times}_L-\Delta{C^{\text{Gauss}}_{L}}-\Delta{C^{N_1}_{L}}-\Delta{C}^{\mathrm{MC}}_L .
\end{equation}

These biases can be understood in more detail as follows. The $N_0$ or Gaussian bias, $\Delta{C^{\text{Gauss}}_{L}}$, is effectively a lensing reconstruction noise bias. Equivalently, since the lensing power spectrum can be measured by computing the connected part of the four-point correlation function of the CMB, $\Delta{C^{\text{Gauss}}_{L}}$ can be understood as the disconnected part that must be subtracted off the full four-point function; these disconnected contractions are produced by Gaussian fluctuations present even in the absence of lensing. The $N_0$ bias is calculated using the now-standard realization-dependent $N_0$ algorithm introduced in \cite{Namikawa2013,planck2013}. This algorithm, which combines simulation and data maps in specific combinations to isolate the different contractions of the bias, is described in detail in Appendix \ref{app.rdn0}. The use of a realization-dependent $N_0$ bias reduces correlations between different lensing bandpowers and also makes the bias computation insensitive to inaccuracies in the simulations.

The $N_1$ bias subtracts contributions from  ``accidental'' correlations of lensing modes that are not targeted by the quadratic estimator 
(\abds{see \citealt{PhysRevD.67.123507} for details;} the nomenclature arises because the $N_1$ bias is first order in $C_L^{\phi \phi}$, unlike the $N_0$ bias, which is zeroth order in the lensing spectrum). The $N_1$ bias is computed using the standard procedure introduced in~\citet{Story_2015}, and described in Appendix \ref{app.n1}. 

Finally, we absorb any additional residuals arising from non-idealities, such as the effects of masking, in a small additive MC bias $\Delta{C}^{\mathrm{MC}}_L$ that is calculated with simulations. We describe the computation of this MC bias in detail in Appendix \ref{app.mcbias}.



\fk{The unbiased lensing spectrum, scaled by $L^2(L+1)^2/4$, is binned in bandpowers with uniform weighting in $L$.} \fk{Details regarding the bins and ranges adopted in our analysis can be found in Section~\ref{sec:binning}.}

\begin{figure*}
 \centering
  \includegraphics[width=0.8\paperwidth]{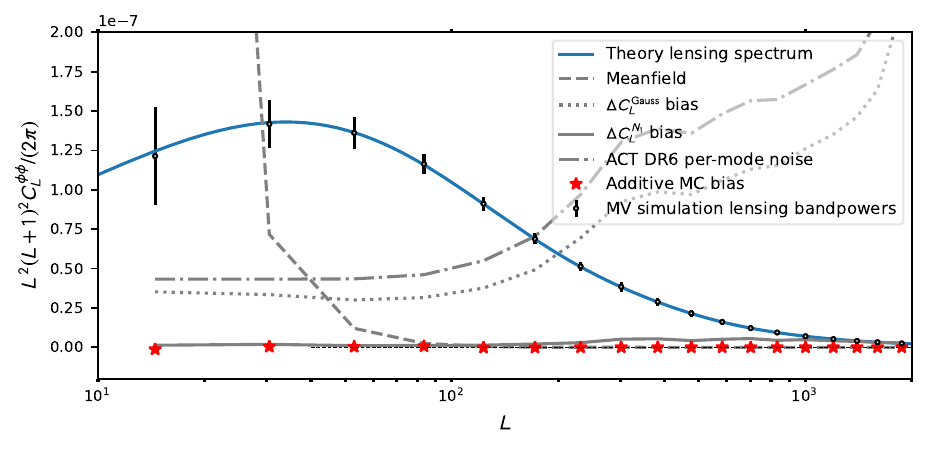}
  \caption{Summary of the different biases subtracted in our lensing power spectrum measurement; \abds{this figure also serves as a test of our pipeline}. The dotted line shows the large Gaussian or $N_0$ bias as a function of multipole $L$; note that this is smaller than the Gaussian bias of the standard QE estimator, as noise does not contribute to the bias for the cross-correlation-based estimator we are using. The effective reconstruction noise for our measurement, which is a more accurate reflection of the noise in our lensing map than the $N_0$ bias, is indicated by the dot-dashed line.
  Similarly, the $N_1$ bias is given by the solid grey line. The power spectrum of the mean-field is shown with a dashed grey line. This term becomes larger than the lensing signal at large scales, $L\sim20$, and the inability to estimate it using simulations with sufficient accuracy is partially the reason why we set the lower limit $L_\mathrm{min}=40$ for cosmological interpretation.
  The open black circles show the reconstructed bandpowers for the mean of 480 simulations with all biases except the MC bias subtracted; these bandpowers are measured by passing realistic sky simulations that closely match the theory spectra (shown in blue) through the pipeline. The error bars on these points represent the errors on a single realisation. An important test of our pipeline is that the simulated residual from an average over many simulations, known as the MC bias, is small; the MC bias is shown as red stars in the plot and is indeed nearly negligible over a wide range of scales. All curves shown are for the minimum-variance (i.e., combined temperature and polarization) estimator.}
  \label{fig:pipeline_ver}
\end{figure*}

To illustrate the sizes of the different bias terms subtracted, we plot them all as a function of scale in Figure~\ref{fig:pipeline_ver}. The fact that the additive MC bias is small is an important test of our pipeline and indicates that it is functioning well.  The procedures laid out above constitute our core full-sky lensing pipeline, which enables the unbiased recovery of the lensing power spectrum after debiasing.



\subsection{Normalization: {dependence on cosmology}}\label{sub:norm.corr.summary}

Prior to normalization, the quadratic lensing estimator probes not just the lensing potential $\phi$; it is instead sensitive to a combination ${\phi}_{L,M} \times \mathcal{R}_{L}\rvert_{C_\ell^{\textrm{CMB}}}$, where the response $\mathcal{R}_{L}\rvert_{C_\ell^{\textrm{CMB}}}$ is a function of the true CMB two-point power spectra. Applying the normalization factor $\mathcal{R}^{-1}_{L}\rvert_{C_\ell^{\textrm{CMB,fid}}}$, where $C_\ell^{\textrm{CMB, fid}}$ are the fiducial CMB power spectra assumed in the lensing reconstruction, attempts to divide out this CMB power spectrum dependence and provide an unbiased lensing map. If the power spectra describing the data are equal to the fiducial CMB power spectra (i.e., $C_\ell^{\textrm{CMB}}=C_\ell^{\textrm{CMB, fid}}$), the estimated lensing map is indeed unbiased. Otherwise, the estimated lensing potential is biased by a factor $\mathcal{R}^{-1}_{L}\rvert_{C_\ell^{\textrm{CMB,fid}}} / \mathcal{R}^{-1}_{L}\rvert_{C_\ell^{\textrm{CMB}}}$. 


In early CMB lensing analyses, it was assumed that the CMB power spectra were determined much more precisely than the \abds{lensing field}, so that any uncertainty in the CMB two-point function and in the normalization could be neglected; however, with current high-precision lensing measurements, the impact of CMB power spectrum uncertainty must be considered. We use as our fiducial CMB power spectra the standard $\Lambda$CDM model from \Planck~2015 TTTEEE cosmology with an updated $\tau$ prior as in~\cite{Calabrese2017}. In Appendix~\ref{app:norm.corr.details}, we describe in detail our tests of the sensitivity of our lensing power spectrum measurements to this assumption;  we summarize the conclusions \NS{below}. 


\abds{We analytically compare the amplitude of the lensing power spectrum $\Alens$ when changing the fiducial CMB power spectra described above to the best-fit model CMB power spectra for an independent dataset, namely ACT DR4+{\it{WMAP}} \citep{Aiola_2020}; \abds{we account for the impact of calibration and polarization efficiency characterization in this comparison.}  Doing this we find a change in $\Alens$ of only $0.23\sigma$, comfortably subdominant to our statistical uncertainty.} An important reason why this change is so small is that our pre-processing procedures, which involve calibration and polarization efficiency corrections relative to the \textit{Planck} spectra, drive the amplitudes of the spectra in our data closer to our \NS{original} fiducial model.  This result reassures us that the CMB power spectra are sufficiently well measured, by independent experiments, not to degrade our uncertainties on the lensing power spectrum significantly. 

Nevertheless, we additionally account for uncertainty in the CMB power spectra in our cosmological inference from the lensing measurements alone (i.e., when not also including CMB anisotropy measurements) by adding to the covariance matrix a small correction calculated numerically from an ensemble of cosmological models sampled from a joint ACT DR4+\textit{Planck} chain (see \cite{Aiola_2020} for details). This results in a small increase in our errors (\abds{by approximately $3\%$ for the lensing spectrum bandpower error bars}), although the changes to the cosmological parameter constraints obtained are nearly negligible.\footnote{\abds{The error on $S_8^{\mathrm{CMBL}}\equiv \sigma_8 (\Omega_m/0.3)^{0.25}$ determined from ACT DR6 CMB lensing alone increases from $0.021$ to $0.022$ when we include the additional term in our covariance matrix.}}

\subsection{Covariance Matrix} \label{sec.covmat}
We obtain the band-power covariance matrix from $N_s=792$ simulations. We do not subtract the computationally expensive realization-dependent RDN0 from all the simulations when evaluating the covariance matrix. Instead, we use an approximate, faster version, referred to as the semi-analytic $N_0$, which we describe briefly below in Section~\ref{subsection:diagonal}.

To account for the fact that the inverse of the above covariance matrix is not an unbiased estimate of the inverse covariance matrix, we rescale the estimated inverse covariance matrix by the Hartlap factor \citep{Hartlap_2006}:
\begin{equation}
    \alpha_\mathrm{cov}=\frac{N_s-N_{\text{bins}}-2}{N_s-1},
\end{equation}
where $N_{\text{bins}}$ is the number of bandpowers.

\subsubsection{Semi-analytic $N_0$}\label{subsection:diagonal}

The realization-dependent $N_0$ algorithm (see Equation~\ref{eq:rdn0}) used to estimate the lensing potential power spectrum is computationally expensive since it involves averaging hundreds of realisations of spectra obtained from different combinations of data and simulations. For covariance matrix computation, which requires the estimation of many simulated lensing spectra to produce the covariance matrix, we adopt a semi-analytical approximation to this Gaussian bias term, referred to as semi-analytic  RDN0. This approximation ignores any off-diagonal terms \abds{involving two different modes $\langle X_{\ell m} Y^\ast_{\ell' m'}\rangle$} when calculating RDN0. The use of the faster semi-analytic RDN0 provides a very good approximation to the covariance matrix obtained using the full realization-dependent $N_0$, with both algorithms similarly reducing correlations between different bandpowers.\footnote{Not including this semi-analytic $N_0$ can lead to correlations of order $20\%$ between neighbouring bandpowers.} We stress that this approximate semi-analytic $N_0$ is only used in the covariance computation and is not employed to debias our data. Further details of the calculation of the semi-analytic RDN0 bias correction are presented in Apppendix \ref{app:dumb}.



\subsubsection{Covariance verification}

We verify that 792 simulations are sufficient to obtain converged results for our covariance matrix as follows. We compute two additional estimates of the covariance matrix from subsets containing 398 simulations each and verify that our results are stable: even when using covariances obtained from only 398 simulations, we obtain the same lensing amplitude parameter, $A_{\text{lens}}$, to within $0.1\sigma$.
In addition, the fact that our null-test suite passes, and in particular the fact that our noise-only null tests in Section~\ref{sec. noise_only} (containing no signal) generally pass, provides further evidence that our covariance estimate describes the statistics of the data well. 

We verify the assumption that our bandpowers are distributed according to a Gaussian in Appendix~\ref{app:gaussianity}.

\subsubsection{Covariance matrix results and correlation between bandpowers}
\begin{figure}
  \includegraphics[width=\linewidth]{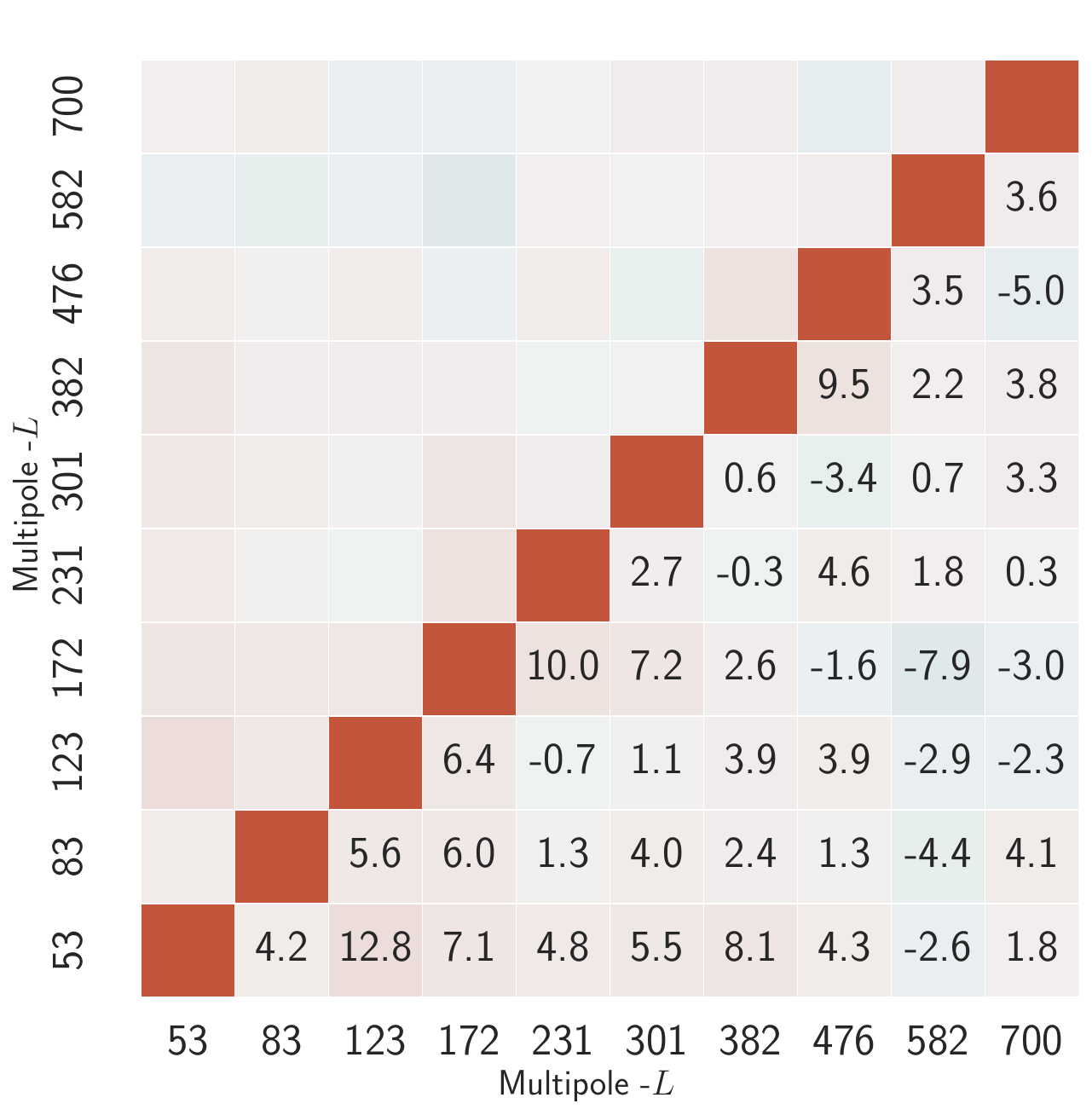}
  \caption{Size of the off-diagonal correlations for our lensing power spectrum bandpower covariance matrix. The covariance matrix is estimated from 792 simulated lensing spectrum measurements. The number in each lower-diagonal element of the matrix shows the correlation coefficient between the relevant bandpowers, expressed as a percentage, for the bins used in our analysis range. It can be seen that the off-diagonal correlations do not exceed the $15\%$ level. The band centers are shown along the axes of the matrix.}
  \label{Fig.correlation}
\end{figure}

The correlation matrix for our lensing power spectrum bandpowers, obtained using a set of 792 simulations, can be seen in Figure~\ref{Fig.correlation}. We find that correlations between different bandpowers are small, with off-diagonal correlations typically below $10\%$.

\section{Null and consistency tests}\label{sec:null}
We now summarize the set of tests we use to assess the robustness of our lensing measurement and the quality of the data we use.  \fk{We first introduce the baseline and extended multipole ranges used in our analysis, and describe how the null tests we have performed guided these choices.} In Section \ref{goodness_fit}, we describe how we compute the $\chi^2$ and probability to exceed (PTE\footnote{The PTE is the probability of obtaining a higher $\chi^2$ than what we actually obtain, given a distribution with the same number of degrees of freedom.}) to characterize passing and failing null tests. In Section \ref{sec:Blind} we describe our blinding procedure, the criteria used to determine readiness for unblinding, and the unblinding process itself. We then describe in detail the map-level null tests in Section \ref{sec:map_level_null} and bandpower-level null tests in Section \ref{sec:bandpower_null}. Section \ref{sec.chi_dist} provides a summary of the distribution of the combined map- and bandpower-level null tests. \abds{Finally, while we aim to present the most powerful null tests in the main text, a discussion of additional null tests performed can be found in Appendix \ref{app.null}.} 

\subsection{Selection of baseline and extended multipole range}\label{sec:binning}

\fk{For our baseline analysis, we use the lensing multipoles $40<L< 763$ with the following non-overlapping bin edges for $N_{\text{bins}}=10$ bins at $[40,66,101,145,199,264,339,426,526,638,763]$.}
The baseline multipole range $40<L<763$ was decided prior to unblinding.
This range is informed by both the results of the null tests and the simulated foreground estimates. The scales below $L=40$ are removed due to large fluctuations at low $L$ observed in a small number of null tests; these scales are difficult to measure robustly since the simulated mean-field becomes significantly larger than the signal, although the cross-based estimator relaxes simulation accuracy requirements on the statistical properties of the noise. 
The $L_\mathrm{max}$ limit is motivated by the results of the foreground tests on simulations performed in \cite{dr6-lensing-fgs}, where at $L_{\text{max}}=763$ the magnitude of fractional biases in the fit of the lensing amplitude is still less than $0.2\sigma$ ($0.5\%$) although biases rise when including smaller scales. This upper range is rather conservative, and hence we also provide an analysis with an extended cosmology range up to $L_{\text{max}}=1300$, although we note that this extended range was not determined before unblinding and that instrumental systematics have only been rigorously tested for the baseline range. (We also note that the null-test PTEs and simulated foregrounds biases still appear acceptable in the extended range, although, again, we caution that we only carefully examined the extended-range null tests after we had unblinded.) 

\subsection{Calculation of goodness of fit}\label{goodness_fit}

In any null test, we construct a set of null bandpowers $\bold{d}^{\text{null}}$, which (after appropriate bias subtraction) should be statistically consistent with zero. For map-level null tests, $\bold{d}^{\text{null}}$ are the bandpowers obtained by performing lensing power spectrum estimation on CMB maps differenced to null the signal, while for bandpower-level null tests they are given by differences of reconstructed lensing power spectra, $\Delta{C}^{\hat{\phi}\hat{\phi}}_L$.
We test consistency of the null bandpowers with zero by calculating the $\chi^2$ with respect to null:
\begin{equation}
\chi^2=(\bold{d}^{\mathrm{null}}){}^\intercal \mathbb{C}^{-1}\bold{d}^{\mathrm{null}}.
\end{equation}
The relevant covariance matrix $\mathbb{C}$ for each null test  is estimated by performing the exact same analysis on 792 simulations, ensuring that all correlations between the different datasets being nulled are correctly captured. The PTE is calculated from the $\chi^2$ with $10$ degrees of freedom as we have 10 bandpowers in the baseline range. (We also consider and compute PTEs for our extended scale range, which has $13$ degrees of freedom.)

\subsection{Blinding procedure}\label{sec:Blind}

We adopt a blinding policy that is intended to be a reasonable compromise between reducing the effect of confirmation bias and improving our ability to discover and diagnose issues with the data and the pipeline efficiently. We define an initial blinded phase after which, when pre-defined criteria are met, we unblind the data.

In the initial blinded phase, as part of our blinding policy, we agree in advance to abide by the following rules.
\begin{enumerate}
\item We do not fit any cosmological parameters or lensing amplitudes to the lensing power spectrum bandpowers. While we allow debiased lensing power spectra to be plotted, we do not allow them to be compared or plotted either against any theoretical predictions or against bandpowers from any previous CMB lensing analyses, including the \textit{Planck} analyses. In this way, we are blind to the amplitude of lensing at the precision needed to inform the $S_8$ tension and constrain neutrino masses, but we can still rapidly identify any catastrophic problems with our data -- although none were found.
\item During the blinded phase, we also allow unprocessed lensing power spectra without debiasing to be plotted against theory curves or \textit{Planck} bandpowers for $L>200$. The justification for this is that the unprocessed spectra are dominated on small scales by the Gaussian $N_0$ bias and hence not informative for cosmology, although they allow for useful checks of bias subtraction and noise levels. When analysing bandpowers of individual array-frequency reconstructions, we allow unprocessed spectra to be plotted over all multipoles, because individual array-frequency lensing spectra are typically noise-bias dominated on all scales.\footnote{Such comparisons allow for a small number of order-of-magnitude sanity checks of intermediate results from different array-frequencies.}  
\end{enumerate}

We calculate PTE values of bandpowers in our map-level null tests (see Section~\ref{sec:map_level_null}) and for differences of bandpowers in our consistency tests (Section \ref{sec:bandpower_null}) during the blinded phase. For the power spectra of the CMB maps themselves (as opposed to those of lensing reconstructions), we follow a blinding policy that will be described in an upcoming ACT DR6 CMB power spectrum paper.

After unblinding, all these restrictions are lifted and we proceed to the derivation of cosmological parameters. We require the following criteria to be satisfied before unblinding.
\begin{enumerate}
    \item All baseline analysis choices made in running our pipeline, such as the range of CMB angular scales used, are frozen.
    
    \item No individual null-test PTE should lie outside the range $0.001 < {\rm PTE} < 0.999$.
    
    \item The distribution of PTEs for different null tests should be consistent with a uniform distribution (verified via a Kolmogorov--Smirnov test, with the caveat that this neglects correlations). 

    \item The number of null and consistency tests that fall outside the range $0.01 < {\rm PTE} < 0.99$ should not be significantly inconsistent with the expectations from random fluctuations.
    
    \item The comparison of the sum of $\chi^2$ for several different types of tests against expectations from simulations should 
    fall within $2\sigma$ of the simulation distributions.
\end{enumerate}

The PTE ranges we accept are motivated by the fact that we calculate $\mathcal{O}(100)$ PTEs but not $\mathcal{O}(1000)$.

\subsubsection{Post-unblinding change}\label{sec:post-unblinding}

\abds{As described in Section~\ref{sec.template_subtract}, our baseline analysis models bright galaxy clusters and subtracts them from maps. However, this procedure was introduced after unblinding. Before unblinding, bright galaxy clusters were masked and inpainted, similar to our treatment of compact objects described in Section~\ref{sec.compact_object}. This minor modification to the analysis, which had only a small effect on the results, was not}
prompted by any of the post-unblinding results we obtained, but rather from concerns arising in an entirely different project focused on cluster mass calibration. In the course of this project, a series of tests for the inpainting of cluster locations were performed using \textsc{websky}  simulations \citep{websky} and Sehgal  simulations \citep{Sehgal_2010}. We discovered that in simulations, our inpainting algorithm can be unstable, as it is heavily dependent on the assumptions of the underlying noise, on the map pre-processing and on inpainting-specific hyperparameters; small inpainting artifacts at the inpainted cluster locations can correlate easily with the true lensing field, leading to significant biases to the lensing results in simulations. Although the same kind of stability tests performed on data show no indication of issues related to inpainting (likely due to the actual noise properties and processing in the data not producing any significant instabilities), concerns about the instability of inpainting on simulations motivated us to switch, for our baseline analysis, to the cluster template  subtraction described in Section \ref{sec.template_subtract} as an alternative method for the treatment of clusters. 
 
 Model subtraction shows excellent stability in the simulations, with no biases found, and foreground studies show that an equivalent level of foreground mitigation is achieved with this method, even when the template cluster profile differs somewhat from the exact profile in the simulations \citep{dr6-lensing-fgs}. We, therefore, expect lensing results obtained using the template subtraction method to be more accurate. Fortunately, changing from cluster inpainting to cluster template subtraction only causes a small change to the relevant $S^\mathrm{CMBL}_8$ parameter: 
$S^\mathrm{CMBL}_8$ decreases by only $0.15\sigma$, as shown later in Figure \ref{Fig.compare}; the inferred lensing amplitude increases by $0.75\sigma$ (the shifts differ in sign due to minor differences in the scale dependences of the lensing amplitude parameter and $S^\mathrm{CMBL}_8$). The small shift in $S^\mathrm{CMBL}_8$ that results from our change in methodology does not significantly affect any of the conclusions drawn from our analysis.

\subsection{Map-level null tests}\label{sec:map_level_null}

\begin{table}
    \centering
    \caption{Summary of the map-level null tests described in Section \ref{sec:map_level_null}. For each test, we show the $\chi^2$ and associated PTE values for the baseline range. \blake{Why PTEs in brackets?}}

    \label{table:null} 
    \begin{tabular}{c c c}
     \hline\hline 
     Map level null test & $\chi^2$ & (PTE) \\ [0.5ex] 
     \hline
     PA4 f150 noise-only  & 8.5 & (0.58) \\ 
     PA5 f090 noise-only  & 6.4  & (0.77)\\ 
     PA5 f150 noise-only  & 11 & (0.35) \\ 
     PA6 f090 noise-only  & 10 & (0.49) \\ 
     PA6 f150 noise-only  & 14 & (0.17) \\ 
     Coadded noise & 21.2 & (0.02) \\
     $\text{PA4 f150}-\text{PA5 f090}$ & 23 & (0.01) \\
     $\text{PA4 f150}-\text{PA5 f150}$ & 19.5 & (0.03) \\
     $\text{PA4 f150}-\text{PA6 f090}$ & 13.7 & (0.19) \\
     $\text{PA4 f150}-\text{PA6 f150}$ & 19.0 & (0.04) \\
     $\text{PA5 f090}-\text{PA5 f150}$ & 5.0 & (0.89) \\
     $\text{PA5 f090}-\text{PA6 f090}$ & 7.5 & (0.68) \\
     $\text{PA5 f090}-\text{PA6 f150}$ & 18.0 & (0.06) \\
     $\text{PA5 f150}-\text{PA6 f090}$ & 12.3 & (0.27) \\
     $\text{PA5 f150}-\text{PA6 f150}$ & 8.2 & (0.61) \\
     $\text{PA6 f090}-\text{PA6 f150}$ & 9.7 & (0.46) \\
     $f090-f150$ MV & 7.6 & (0.67)  \\
     $f090-f150$ TT & 5.7 & (0.84)  \\
     $(f090-f150)\times{\phi}$ MV & 8.2 & (0.61)  \\
    $(f090-f150)\times{\phi}$ TT & 4.3 & (0.93)  \\
     Time-split difference & 18.6 & (0.05)  \\
     \hline
    \end{tabular}
\end{table}

This subsection describes null tests in which we apply the full lensing power spectrum estimation pipeline to maps that are expected to contain no signal in the absence of systematic effects. In all cases except for the curl reconstruction in Section \ref{sec:curl}, this typically involves differencing two variants of the sky maps at the map level (hence nulling the signal) and then proceeding to obtain debiased lensing power spectra from these null maps. \fk{To adhere closely to the baseline lensing analysis, we always prepare four signal-differenced maps and make use of the cross-correlation-based estimator.}

We describe each of our map-level null tests in more detail in the sections below.

\subsubsection{Curl}\label{sec:curl}
The lensing deflection field $\bold{d}$ can be decomposed into gradient and curl parts based on the potentials $\phi$ and $\Omega$, respectively, i.e., in terms of components $d_i= \nabla_i \phi + \epsilon_i{}^j \nabla_j \Omega$, where $\Omega$ is the divergence-free or ``curl'' component $\Omega$ of the deflection field and $\phi$ is again the lensing potential. (Here, $\epsilon_{ij}$ is the alternating tensor on the unit sphere.)
The curl $\Omega$ is expected to be zero at leading order and therefore negligible at ACT DR6 reconstruction noise levels (although a small curl component induced by post-Born and higher-order effects may be detectable in future surveys; \citealt{Pratten_2016}). However, systematic effects do not necessarily respect a pure gradient-like symmetry and hence could induce a non-zero curl-like signal. An estimate of this curl field can thus provide a convenient diagnostic for systematic errors that can mimic lensing. Furthermore, curl reconstruction also provides an excellent test of our simulations, our pipeline, and our covariance estimation. 

\begin{figure}
  \includegraphics[width=\linewidth]{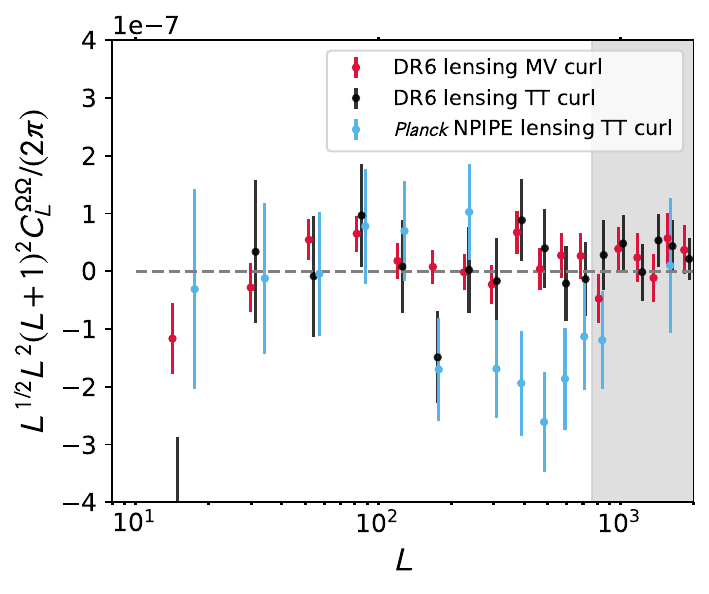}
  \caption{Power spectrum of the reconstructed curl mode of the lensing deflection field. Since the cosmological lensing field is irrotational, a measurement of the curl component can serve as a valuable null test for several systematic errors. Results of this curl null test are shown for our baseline, coadded dataset for the MV estimator (red) and TT estimator (black). Neither show any evidence for systematic contamination, with a PTE with respect to zero of 0.37 and 0.75, respectively. This can be contrasted with the \textit{Planck} \texttt{NPIPE} TT curl bandpowers shown in blue, which exhibit a significant deviation from zero in the range $264\leq{L}\leq901$. Our results provide further evidence that the negative curl power seen in the \textit{Planck} TT reconstruction is not a real cosmological signal \citep{Planck:2018}.} 
  \label{Fig.base_comb}
\end{figure}

We obtain a reconstruction of this curl field in the same manner as described in Section \ref{sec:qe}, by taking linear combinations of the spin-1 spherical harmonic transform of the deflection field. The bias estimation steps are then repeated in the same way as for the lensing estimator. The result for this null test is shown in Figure~\ref{Fig.base_comb} for the MV coadded result {\footnote{A note on the y-scaling used in the plots: For the null test plots, we scale our bandpowers by an factor of $\sqrt{L}$ with the visual purpose of enhancing the smaller scales with aids with identifying potential issues on the small scales that we probe with significant SNR. For Fig.\ref{fig:pipeline_ver} and Fig. \ref{Fig.results_compilation} we adopt the scaling of $L^2(L+1)^2/(2\pi)$ used by other CMB lensing measurements in the literature for easier comparison.}}, which is the curl equivalent of our baseline lensing spectrum. This test has a PTE of 0.37, in good agreement with null. We also show curl null test results for the temperature-only (TT) version of our estimator in Figure~\ref{Fig.base_comb}.

 
 The consistency of our curl measurement with zero provides further evidence of the robustness of our lensing measurement. Intriguingly, the curl null test was not passed for the TT estimator in \textit{Planck}, and instead (despite valiant efforts to explain it) a $4.1\sigma$ deviation\footnote{Note that the significance falls to $2.9\sigma$ after accounting for ``look-elsewhere'' effects.} from zero has remained, located in the range $264<L<901$ \cite{Planck:2018}; see Figure~\ref{Fig.base_comb}. Our result \abds{provides} further evidence that this non-zero curl is not physical in origin.

For completeness, we also compute curl tests associated with all other null tests described in the subsequent sections; we summarise the results and figures in Appendix \ref{app.null}. These results also show that there is no evidence of curl modes found even in subsets of our data.

\subsubsection{Noise-only null tests: Individual array-frequency split differences}\label{sec. noise_only}
We can test our pipeline, verify our covariance matrices, and assess the modelling of the noise for each array-frequency by differencing splits $\bold{m}_i$ of the data with equal weighting, and hence cancelling the signal, to form null maps $X^{i,\mathrm{null}}=\bold{m}_i-\bold{m}_{i+4}$. (There are various combinations from which this null map could be formed; we choose to difference split $i$ and split $i+4$, where $i\in\{0,1,2,3\}$.)
The resulting four signal-nulled maps are passed through the cross-correlation-based estimator.
We perform lensing reconstruction on these null maps with isotropic filtering. The power spectra used in this filter are obtained by averaging the power spectra of 80 simulations of lensed CMB with noise realizations consistent with the inverse-variance-weighted noise of the eight splits.
For these tests we thus use the filter appropriate to the coadded noise of the individual array-frequency instead of the baseline coadd filter, since otherwise the high noise in the individual array-frequencies leads to less sensitive null tests. Only the coadd noise null test discussed in Appendix~\ref{app.coadd_array} uses the baseline weights.
The normalization is computed with the same filters and applied to the resulting null spectrum. 
Because we are using the cross-correlation-based estimator and the signal is assumed absent, we do not need to estimate the mean-field, or the RDN0 and $N_1$ biases \abds{(which should all be zero)}; therefore, the simulations are used solely to estimate the covariance matrix.
The summary results for this category of tests, written in terms of the sum of the $\chi^2$ for all the array-frequencies, are shown in Figure~\ref{Fig.noise_only_hist_test} of Section \ref{sec.chi_dist}. 
These tests show no evidence of a discrepancy between different splits of the data map; this fact also confirms that our noise simulations provide accurate estimates of the covariance matrix. An additional noise-only null test can be obtained by coadding all the individual noise-only null maps; this stringent null test is shown in Appendix \ref{app.null}.

\subsubsection{Map-level frequency-difference test}\label{sec:maplevelfreq}
We prepare frequency-differenced null maps by subtracting the beam-deconvolved f150 split maps from the f090 split maps. The resulting difference maps are passed into the lensing reconstruction pipeline with the filters, normalization, \abds{and bias-hardening procedure} the same as used for the baseline reconstruction, which combines f150 and f090. \fk{This filter choice weights different scales in the null maps in the same way as for our baseline lensing measurement, which ensures that null-test results can be directly compared with our baseline lensing results.}
The null lensing power spectrum $\nullb$ is given schematically by
\begin{multline}
    \nullb=\langle \mathrm{QE}(T^{90}-T^{150},T^{90}-T^{150}) \\ \times \mathrm{QE}(T^{90}-T^{150},T^{90}-T^{150})\rangle.
\end{multline}

 This measurement is a rigorous test for our mitigation of foregrounds: the effect of foregrounds such as CIB and tSZ is expected to be quite different in these two frequency channels (with f090 more sensitive to tSZ and less to CIB) so we do not expect full cancellation of foregrounds in the difference maps. In particular, this null test targets the residual foreground-only trispectrum of the lensing maps; we compare our results with the levels expected from simulations in \cite{dr6-lensing-fgs}. In addition, this map-level null test is also sensitive to beam-related differences between the two frequency channels. 

\begin{figure}
  \includegraphics[width=\linewidth]{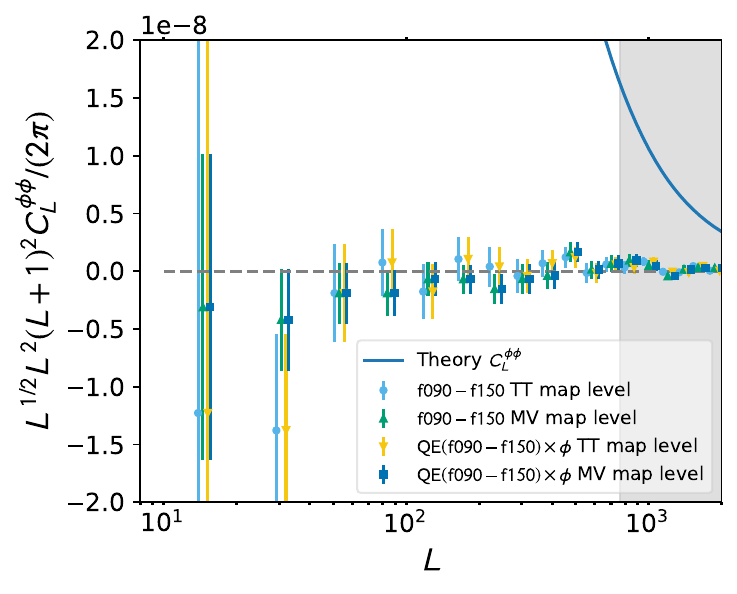}
  \caption{Lensing null tests based on frequency differences, which are a valuable diagnostic for insufficient foreground bias mitigation as well as for instrument systematics. The light-blue and green points show a lensing power spectrum measurement from a $\text{f090}-\text{f150}$  difference map, for TT and MV, respectively. Yellow and blue points show a cross-correlation of the null lensing map, made from the $\text{f090}-\text{f150}$ difference maps, with the baseline lensing map, for TT and MV, respectively. These two type of null tests are sensitive to different foreground bias terms. All tests are consistent with zero and thus provide no evidence for any foreground bias (or other systematic bias) in our measurement.}
  \label{Fig.90150MVdiff}
\end{figure}

As shown in Figure~\ref{Fig.90150MVdiff}, these null tests are consistent with zero, with PTEs of 0.67 and 0.84 for MV and TT respectively; no evidence for un-mitigated foreground contamination is found.

\subsubsection{Frequency-nulled map$~\times~\hat{\phi}^{\text{MV}}$  }
To perform an additional, similarly powerful, test of foregrounds, we cross-correlate the null reconstruction from the frequency-difference maps, 
  obtained as in the previous null test, with the baseline reconstruction  $\hat{\phi}^{\text{MV}}$; i.e., schematically, we compute:
\begin{equation}
    C_L^{\mathrm{null}} =\left<\mathrm{QE}(T^{90}-T^{150},T^{90}-T^{150}) \times \hat{\phi}^\mathrm{MV}\right>.
\end{equation}

This measurement is sensitive mainly to the foreground bispectrum\footnote{\abds{See \cite{dr6-lensing-fgs} for an explanation of the foreground bispectrum and trispectrum terms.}} involving two powers of foreground residuals and one power of the true convergence field. To a lesser extent, given the small residual foreground biases remaining in $\hat{\phi}^{\mathrm{MV}}$, 
the test is also sensitive to a foreground trispectrum contribution.
The null test results in Figure~\ref{Fig.90150MVdiff} show good consistency with zero with a passing PTE of 0.61 and 0.93 for MV and TT, respectively.

Since the foreground bias probed by this test is the dominant one on large scales, the consistency of this test with null is a particularly powerful test of foreground mitigation in our analysis.

\subsubsection{Array-frequency differences}\label{sec:arraydiff}

We test for consistency between the data obtained from the different instrument array-frequencies by taking differences between single array-frequency maps. Since we have five array-frequencies we obtain 10 possible combinations of such null maps. We pass these signal-nulled maps through the pipeline and use a filter that consists of the average power spectra of the two array-frequencies making up the difference map. We find no evidence of inconsistency between the different array-frequencies except for a marginal failure for the difference between PA4 f150 and PA5 f090 (with a PTE of $0.01$), which we discuss further in Section~\ref{sec.chi_dist} and argue is not concerning. The histogram for the $\chi^2$ values of all such tests is summarised in Figure~\ref{Fig.instrument_only_hist} of Section \ref{sec.chi_dist}. These tests show that there is good inter-array consistency at the four-point level.

In Appendix \ref{timediff_map} we perform an additional, related test: we measure the lensing power spectrum from null maps obtained by differencing CMB maps made from 2017--2018 observations with maps from 2018--2021 observations. The passing PTE of 0.05 provides no significant evidence of inconsistency between the two periods.

\subsection{Bandpower-level consistency tests}\label{sec:bandpower_null}

This section describes tests that aim to assess whether lensing spectrum bandpowers from variations of our analysis, or sub-sets of our data, are consistent with each other. For each variation or sub-set, we subtract the resulting debiased lensing power spectrum from our baseline debiased lensing power spectrum; \abds{both spectra are obtained with our standard methodology described in Section \ref{sec:reconstruction}.} We obtain a covariance matrix for this difference by repeating this analysis (with semi-analytic debiasing described in Section \ref{subsection:diagonal}) on simulations. We then use the nulled bandpower vector and its covariance matrix to check for consistency with zero.  We summarise the results of these tests in Table \ref{table:nullbandpower}; in this table, we also utilize the statistic $\Delta\Alens$ to quantify the magnitude of any potential bias to the lensing amplitude produced by the departure of the null-test bandpowers from zero, i.e.,
\begin{equation}
    \Delta\Alens = \frac{\sum_{bb^\prime} \hat{C}^{\mathrm{null}}_{L_b} \, {\mathbb{C}}^{-1}_{bb^\prime} \, {\hat{C}^{\phi\phi}_{L_b}}}{\sum_{bb^\prime} \hat{C}^{\phi\phi}_{L_b}\,{\mathbb{C}}^{-1}_{bb^\prime}\,{\hat{C}^{\phi\phi}_{L_b}}}.
\label{eq.al.bias}
\end{equation}
Here, $\hat{C}^{\phi\phi}_{L_b}$ is the baseline lensing power spectrum and ${\mathbb{C}}_{bb^\prime}$ is the baseline covariance matrix. \fkk{The $\Delta\Alens$ results are summarised in Fig. \ref{Fig.nullalens}. } 

\fk{For all of the null tests discussed in subsequent sections, we present plots that show the lensing bandpowers in the upper panel, with the baseline analysis in red boxes. Additionally, we include a sub-panel showing differences of bandpowers divided by the baseline MV errors $\sigma_L$.}

\subsubsection{Temperature-polarization consistency}\label{sec.nullpol}

We compare our baseline minimum variance (MV$\times$MV) analysis against the polarization only measurement (MVPOL$\times$MVPOL) and the temperature-only measurement (TT$\times$TT). We additionally compare TT$\times$TT against MVPOL$\times$MVPOL. The lensing bandpowers and the null bandpowers from differencing the polarization combinations can be seen in Figure~\ref{Fig.polcomb_consistent}. The corresponding curl is shown in  Appendix \ref{app:null tests curl}. As can be seen in these plots, the null tests are consistent with zero with PTEs of 0.34 ($\mathrm{TT}-\mathrm{MV}$), 0.73 ($\mathrm{MVPOL}-\mathrm{MV}$) and 0.69 ($\mathrm{TT}-\mathrm{MVPOL}$).

 \begin{figure}
  \includegraphics[width=\linewidth]{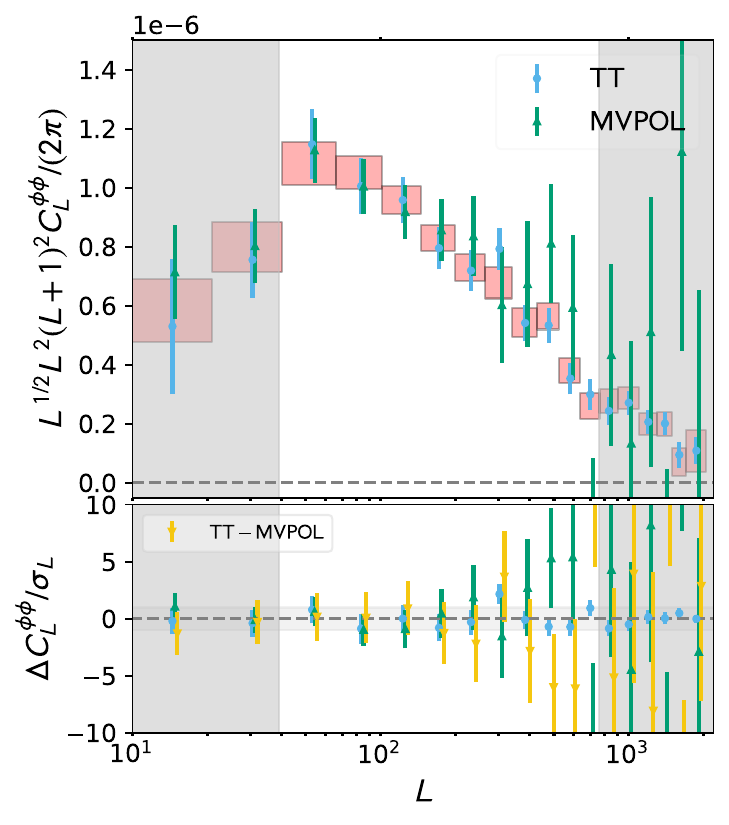}
  \caption{ACT DR6 lensing convergence bandpowers from the minimum-variance combination of temperature and polarization (our baseline, denoted as MV) in red, from temperature only (TT) in blue, and from polarization only (MVPOL) in green. \fk{The bottom panel shows the differences between the TT and MV spectra (blue), MVPOL and MV (green) and TT and MVPOL (yellow). These are consistent with null. \abds{Note that in the bottom panel, the difference results are divided by the error in the baseline bandpowers $\sigma_L$; we emphasize that this baseline error is not the same as the error in the difference.} }
  } 
  \label{Fig.polcomb_consistent}
\end{figure}

\begin{table}
    \centering
    \caption{Summary of the bandpower-level null tests described in Section \ref{sec:bandpower_null}. For each test, we show the $\chi^2$ and associated PTE values of the difference bandpowers as well as the shift in $\Alens$, in the form $\Delta{A^{\mathrm{lens}}}\pm\sigma({\Delta{A^{\mathrm{lens}}}})$. Where not indicated in the description of the test, the reported values are \abds{computed} with respect to the baseline MV reconstruction.} \blake{formatting} 
    \begin{tabular}{c c c c} 
     \toprule
     Bandpower null test & $\chi^2$ & (PTE) & $\Delta{A^{\mathrm{lens}}}$ \\ [0.5ex] 
     \hline
     $600<\ell_{\mathrm{CMB}}<2000$ & 2.9 & (0.98) & $-0.015\pm0.023$ \\ 
     $600<\ell_{\mathrm{CMB}}<2500$ & 9.6 & (0.48) & $-0.019\pm0.012$ \\ 
     $800<\ell_{\mathrm{CMB}}<3000$ & 10.9 & (0.37) & $0.01\pm0.01$ \\ 
     $1500<\ell_{\mathrm{CMB}}<3000$ & 4.4 & (0.93) & $-0.02\pm0.03$ \\ 
    $40\%$ mask  & 7.2 & (0.71) & $0.01\pm0.02$ \\
    Aggr. ground pick up  & 14.8 & (0.14) & $0.01\pm0.01$ \\ 
    Poor cross-linking reg.  &4.1 &(0.94) &$-0.06\pm0.06$ \\ 
    MV $\text{f090}-\text{f150}$  & 9.1 & (0.52) & $-0.002\pm0.04$ \\
    TT $\text{f090}-\text{f150}$  & 16.6 & (0.08) & $-0.05\pm0.06$ \\
    CIB deprojection   & 15.6 & (0.11) & $-0.02\pm0.02$ \\
    TT shear  & 13.5 & (0.20) & $0.01\pm0.05$ \\
    $\text{TT}-\text{MV}$    & 11.2 & (0.34) & $-0.004\pm0.03$ \\ 
    $\text{MVPOL}-\text{MV}$   & 6.9 & (0.73) & $0.06\pm0.06$ \\ 
    $\text{TT}-\text{MVPOL}$   & 7.4 & (0.69) & $-0.06\pm0.07$ \\ 
    $\text{South}-\text{North patch}$  & 4.77 & (0.91) & $0.04\pm0.05$ \\ 
    $\text{Time-split 1}-\text{2}$  & 11.4 & (0.33) & $0.003\pm0.036$ \\ 
    Time-split 1  & 8.1 & (0.62) & $-0.04\pm0.04$ \\ 
    Time-split 2  & 11.2 & (0.33) & $-0.04\pm0.04$ \\ 
    $\text{PA4 f150} - \text{PA5 f090}$  & 9.1 & (0.52) & $0.0\pm0.1$ \\
    $\text{PA4 f150} - \text{PA5 f150}$  & 7.0 & (0.73) & $0.1\pm0.2$ \\
    $\text{PA4 f150} - \text{PA6 f090}$  & 9.1 & (0.52) & $0.0\pm0.2$ \\
    $\text{PA4 f150} - \text{PA6 f150}$  & 20.1 & (0.03) & $0.11\pm0.2$ \\
    $\text{PA5 f090} - \text{PA5 f150}$  & 5.8 & (0.83) & $0.13\pm0.2$ \\
    $\text{PA5 f090} - \text{PA6 f090}$  & 9.5 & (0.49) & $0.02\pm0.05$ \\
    $\text{PA5 f090} - \text{PA6 f150}$  & 19.6 & (0.08) & $0.02\pm0.04$ \\
    $\text{PA5 f150} - \text{PA6 f090}$  & 10.4 & (0.41) & $-0.03\pm0.07$ \\
    $\text{PA5 f150} - \text{PA6 f150}$  & 16.6 & (0.08) & $0.1\pm0.2$ \\
    $\text{PA6 f090} - \text{PA6 f150}$  & 17.2 & (0.07) & $0.07\pm0.2$ \\
    $\text{PWV high} - \text{low}$   & 5.0 & (0.89) & $0.02\pm0.04$ \\
     \hline
    \end{tabular}
    \label{table:nullbandpower} 
\end{table}

 \begin{figure}
  \includegraphics[width=\linewidth]{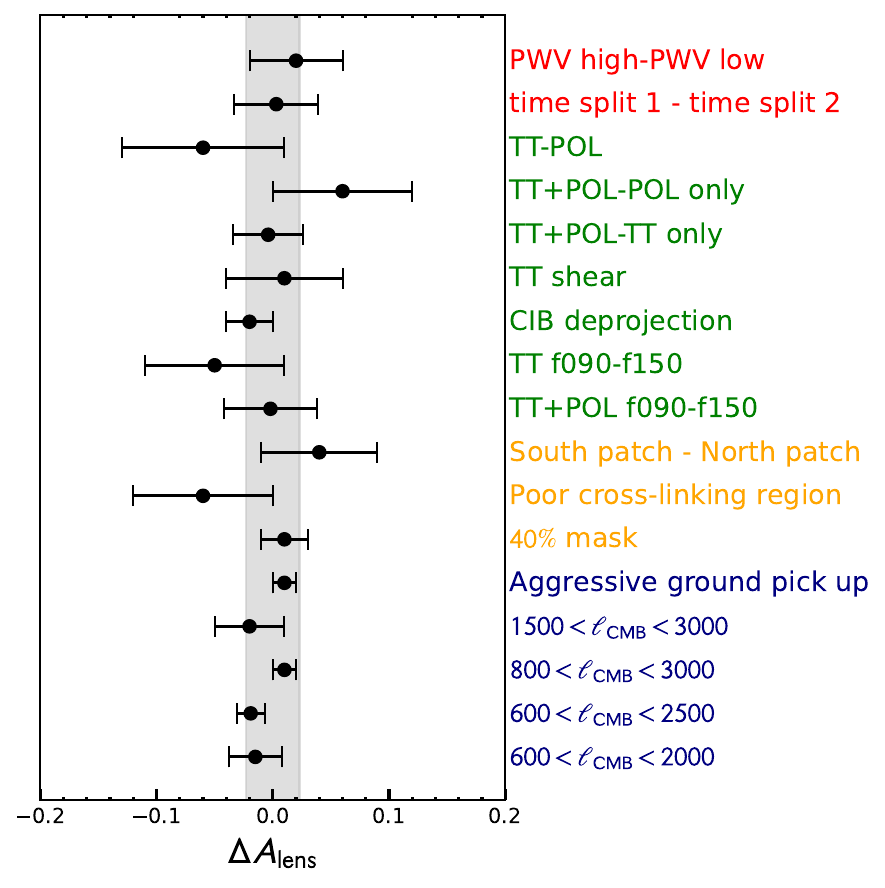}
  \caption{\fkk{Shift in $A_\mathrm{lens}$ for the lensing bandpower null tests described in Table \ref{table:nullbandpower}. These shifts are color coded as follows: the blue labels stand for scale consistency tests, orange for isotropy-related tests, green for polarization- and frequency-combination tests, and red for instrument-related tests. The nulled spectra are all consistent with producing zero shift in $A_\mathrm{lens}$; the grey band shows the 1$\sigma$ errors of our baseline lensing amplitude measurements.}}
  \label{Fig.nullalens}
\end{figure}

\subsubsection{Bandpower-level frequency-difference test}\label{sec:bandfreqtest}

\begin{figure}
  \includegraphics[width=\linewidth]{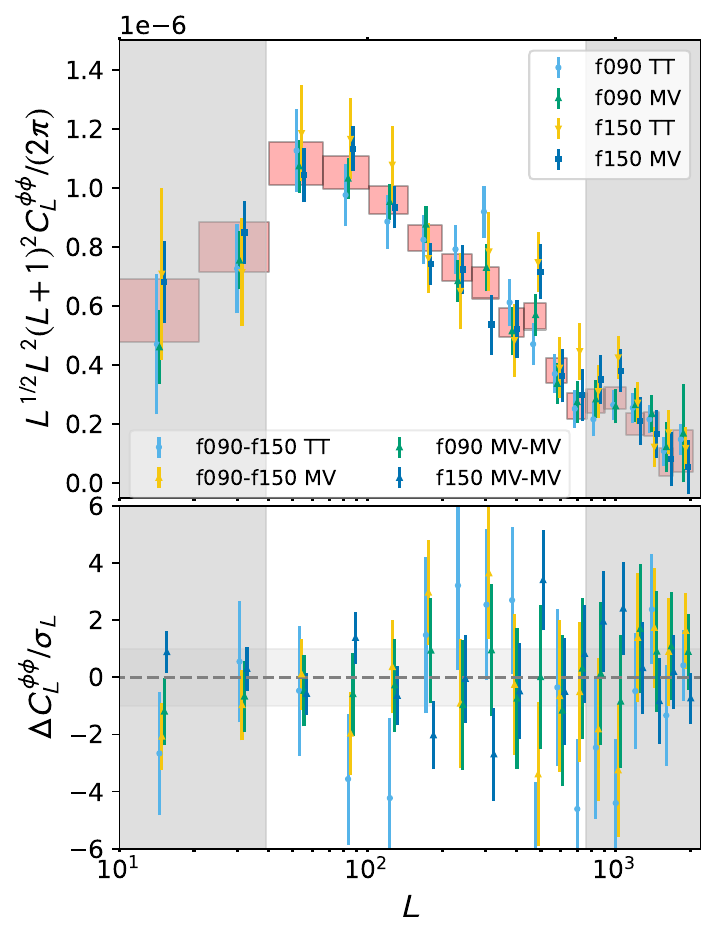}
  \caption{Lensing power spectra obtained from single-frequency maps with the TT or MV estimator, compared again with the baseline analysis (red boxes), which coadds f090 and f150 data and uses the MV estimator. As seen in the bottom panel, the single-frequency analyses are consistent with each other and with the baseline MV.}
  \label{Fig.freqbandpowers}
\end{figure}

We compare the lensing power spectrum derived from \abds{f090 and f150 data alone} to our baseline analysis in Figure \ref{Fig.freqbandpowers}.

The null bandpowers are formed by taking the difference
\begin{equation}
    C^{\text{null}}_L=C^{\hat{\phi}\hat{\phi},90\,\si{GHz}}_L-C^{\hat{\phi}\hat{\phi},150\,\si{GHz}}_L,
\end{equation}
where $C^{\hat{\phi}\hat{\phi},90\,\si{GHz}}_L$ is the lensing spectrum reconstructed with the f090 data only, i.e., PA5 f090 and PA6 f090, and $C^{\hat{\phi}\hat{\phi},90\si{GHz}}_L$ is obtained by reconstructing the data at f150 only (from the PA4 f150, PA5 f150 and PA6 f150 array-frequencies). For the co-addition of the data we use the same noise weights \abds{(up to normalization)} as in the baseline analysis. \fk{For the reconstruction we use the same filters used for the baseline analysis.} This null test is sensitive to all foreground contributions \abds{(including both bispectrum and trispectrum terms)}. 
 However, compared to the map-level frequency-difference null test above, this measurement has larger errors, since the lensed CMB is not nulled at the map level. Our results in Figure~\ref{Fig.freqbandpowers} show good agreement of the lensing reconstruction obtained from different frequencies. The curl is also shown in appendix \ref{app:null tests curl}. 

\subsubsection{Consistency with CIB-deprojection analysis}\label{sec:cibdeprojection}
The companion paper \cite{dr6-lensing-fgs} finds that a CIB-deprojected version of the analysis shows similar performance to our baseline analysis in mitigating foreground biases to a negligible level without incurring a large signal-to-noise penalty. We, therefore, perform a consistency check between this alternative, multifrequency-based foreground mitigation method and the geometry-based profile hardening method that is our baseline.

\fk{\cite{dr6-lensing-fgs} describe the production of CIB-deprojected temperature maps by performing a harmonic-space constrained internal linear combination (hILC)}
of the DR6 coadded temperature map and the high-frequency data from \textit{Planck} at $353\,\si{GHz}$ and $545\,\si{GHz}$. The high-frequency \textit{Planck} channels are chosen because the CIB is much brighter and the primary CMB information is subdominant at high frequencies; these high-frequency maps are hence valuable foreground monitors that can be used while still keeping our analysis largely independent of CMB measurements from \textit{Planck}.

Performing the hILC requires the use of the total auto- and cross-spectra for all the input maps; these are measured directly from the data, and are smoothed with  a Savitzky--Golay filter (\citealt{1964AnaCh..36.1627S}; window length 301 and polynomial order 2), to reduce ``ILC bias" (see, e.g.,~\citealt{delabrouille09}) arising from fluctuations in the spectrum measurements. 
We also generate 600 realizations of these maps, using the \textit{Planck} \texttt{NPIPE} noise simulations provided by \citet{plancknpipe}; these are used for the $N_0$ subtraction, mean-field correction and covariance matrix estimation. The deprojected temperature maps are then used in our lensing reconstruction and lensing power spectrum estimation (along with the same polarization data as is used in the baseline analysis and same cross-correlation-based estimator with profile hardening).

\begin{figure}
  \includegraphics[width=\linewidth]{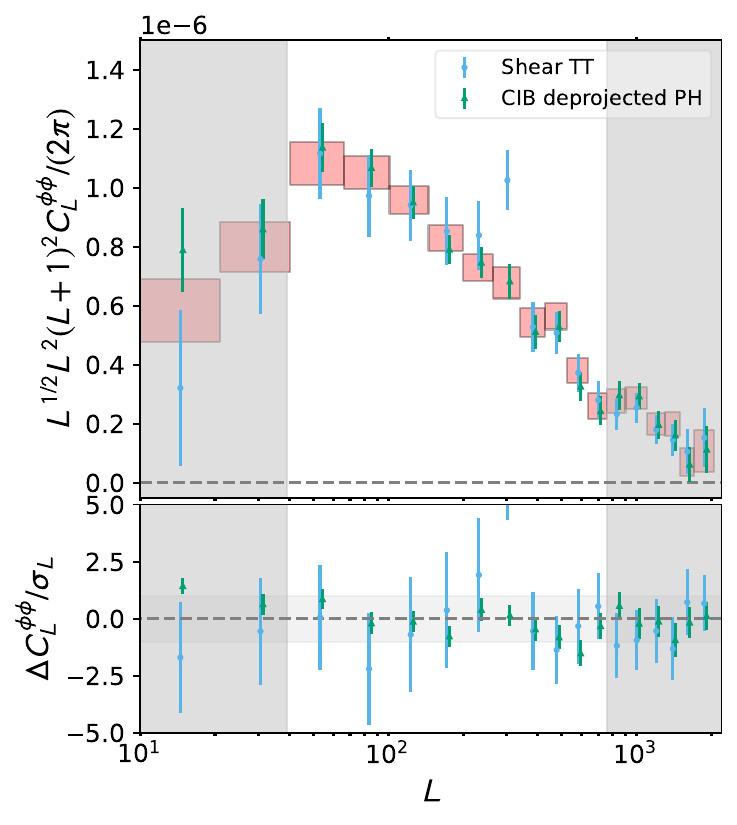}
  \caption{Alternative measurements of lensing with different foreground mitigation techniques; the difference with our baseline analysis provides an additional null test for foreground contamination. Green points show lensing power-spectrum bandpowers obtained using foreground mitigation based on multifrequency CIB deprojection, described in Section \ref{sec:cibdeprojection}, alongside profile hardening. Blue points show a measurement of the lensing power spectrum with the foreground-insensitive shear estimator, described in Section \ref{sec:shear}. These can be compared with the baseline analysis using just profile hardening, shown with red boxes. The differences of these alternative measurements with our baseline are shown in the bottom panel, with errors re-computed for this difference measurement. The measurements of lensing with different foreground mitigation methods are in good consistency with our baseline analysis, suggesting that foregrounds in our measurement have been mitigated successfully. While one bandpower in the shear-estimator measurement appears high, we note that such a sharp feature is not generally a signature of foreground contamination in lensing \citep{dr6-lensing-fgs}; since the overall shear PTE $=20\%$ is acceptable, we ascribe this point to a random fluctuation.}
  \label{fig:cibdiff}
\end{figure}

As seen in Figure~\ref{fig:cibdiff}, the results show that the bandpowers are consistent with the baseline analysis with a PTE of 0.11. This implies that CIB contamination is not significant in our lensing analysis. Given the similarity of the spectral energy distributions of the CIB and Galactic dust, these results also provide evidence against significant biases from Galactic dust contamination. At the same time, since CIB deprojection increases the amount of tSZ in our maps \citep{PhysRevD.103.103510,kusiak2023enhancing}, the stability of our results also suggests that tSZ is mitigated well.

\subsubsection{Shear estimator}\label{sec:shear}
 We can obtain alternative temperature-only lensing bandpower measurements using the shear-only estimator  \citep{Schaan_2019,qu_shear}. Shear estimators are a class of geometric methods that suppress extragalactic foreground contamination while making only minimal assumptions about foreground properties, at the cost of only a moderate decrease in signal-to-noise. The shear estimator decomposes the standard quadratic estimator into a monopole and a quadrupole part and discards the monopole part; the motivation for this is that foreground mode-couplings tend to be spherically symmetric, as argued in \cite{Schaan_2019}.

We verify that we obtain consistent lensing results with the full-sky shear estimator \citep{qu_shear}; the spectrum difference $\Delta{C_L^{\phi\phi}}$ is shown in Figure~\ref{fig:cibdiff} (blue points). With a PTE of 0.20, this test shows no significant discrepancies with the baseline and provides further evidence that the impact of extragalactic foreground biases is controlled within our levels of uncertainty.

\subsubsection{Array differences}\label{array_bandpower}
\begin{figure}
  \includegraphics[width=\linewidth]{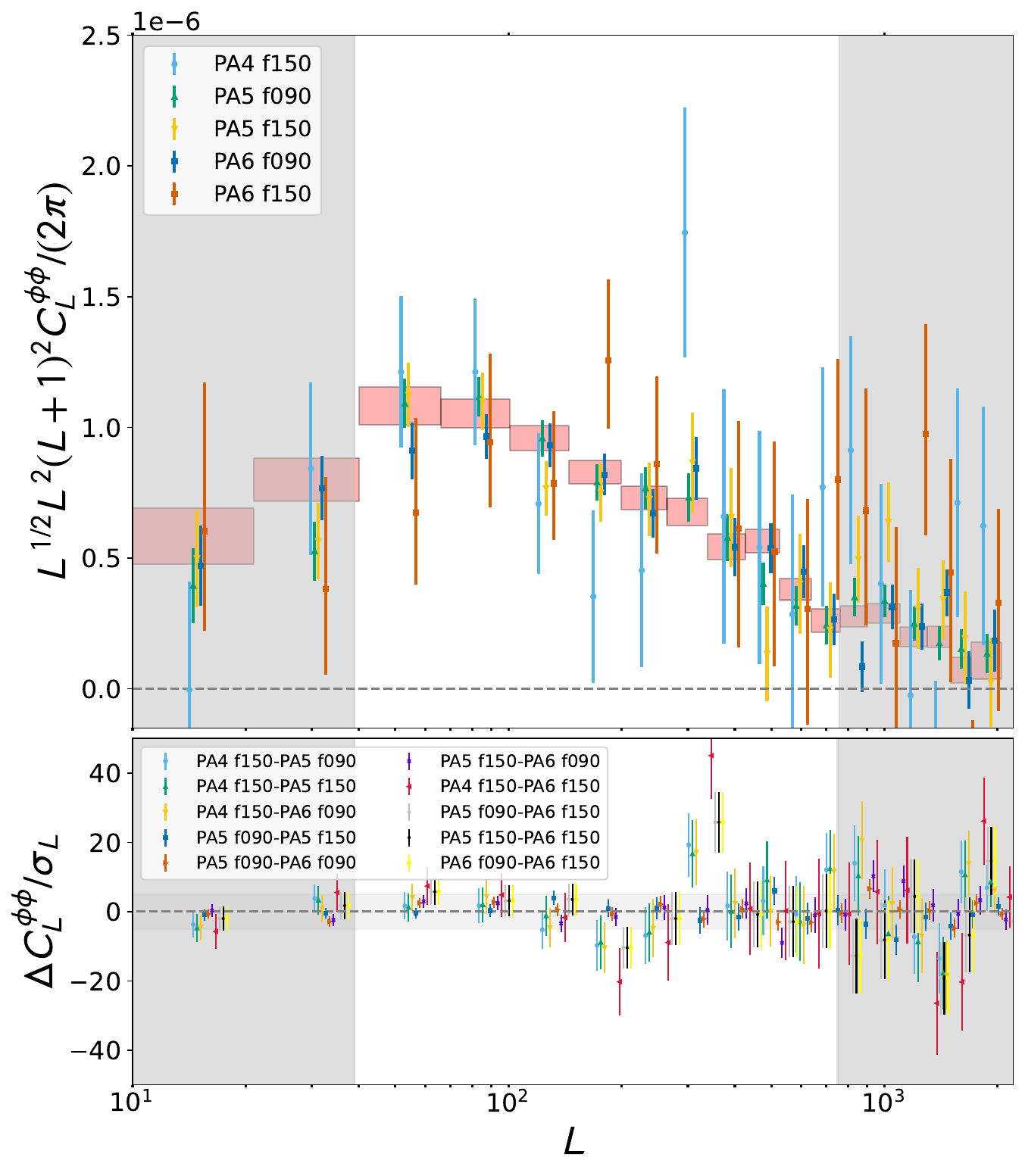}
  \caption{Lensing power spectra obtained from single telescope array-frequency maps, compared again with the baseline analysis (red boxes), with spectrum differences in the bottom panel. \abds{All results are for the MV estimator.} We conclude that the single-array analyses are consistent with our baseline measurement. \blake{legend text size}}
  \label{Fig.arraydiffband}
\end{figure}

In addition to testing the consistency of the different array-frequencies at the map level, we further test their consistency by comparing the lensing bandpowers obtained from each array. 
For the filtering operation, we use a filter with a noise level consistent with the given array instead of the coadded baseline noise level; this choice was made in order to increase the signal-to-noise for each single-array spectrum (which is not always high) and increase the sensitivity of the test. This bandpower null test provides a broader assessment of the presence of possible multiplicative biases that could create inconsistencies among the different array-frequencies. The results in Table \ref{table:nullbandpower} and Figure~\ref{Fig.arraydiffband} show there is no evidence of such effects in our data. 

\subsubsection{Cross-linking}\label{sec:maskxlink}
A pixel of the maps is said to be well cross-linked when it has been observed by different scans that are approximately orthogonally oriented at the location of the pixel on the sky. 
The DR6 scan strategy produces adequate cross-linking over the survey area except for a narrow region around $\text{Dec}=-35^{\circ}$. The poorly cross-linked region has significantly more correlated noise than the rest of the patch.

\begin{figure}
  \includegraphics[width=\linewidth]{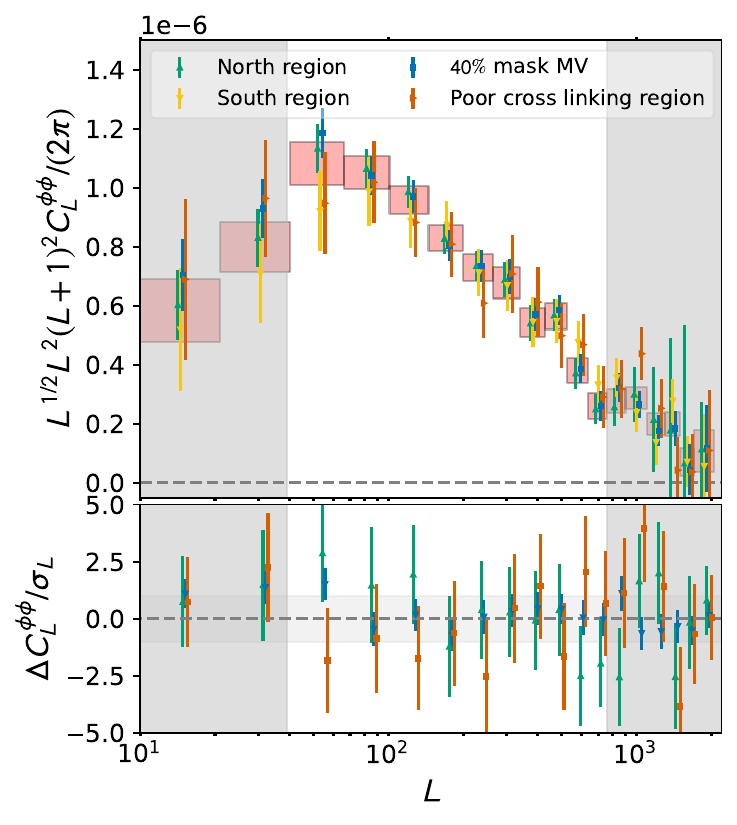}
  \caption{ Lensing power spectra obtained with different analysis masks, i.e., from different sky regions. These can be compared with our baseline analysis using our standard sky mask (red boxes). All results are for the MV estimator. Selected differences of the resulting lensing spectra are shown in the bottom panel. We find good stability of our results to variations in the sky region used, with the difference spectra consistent with zero.
}
  \label{Fig.galbandpowers}
\end{figure}

We isolate a region with poor cross-linking, with the footprint borders shown in green in Figure~\ref{fig:lensing_map}. We compare the bandpowers obtained from this region \fk{(using a filter consistent with the coadd noise of this small noisy patch)} against those from our baseline analysis in Figure~\ref{Fig.galbandpowers}  and find no statistically significant difference in the bandpowers obtained from both regions, with a null $\text{PTE}=0.94$.

\subsubsection{Multipole-range variation}
\label{sec:multipole}
We compare our baseline MV reconstruction, which uses the CMB with multipoles in the range $600 < \ell < 3000$, against reconstructed lensing spectra obtained from different CMB scale ranges: 
$500 < \ell < 3000$,
$600 < \ell < 3000$,
$800 < \ell < 3000$,
$1000 < \ell < 3000$,
$1500 < \ell < 3000$,
$300 < \ell < 2500$, and
$300 < \ell < 2000$.
This multipole range variation tests the following:
(i) the consistency of lensing spectra when including a more or less extended range of CMB modes and, more specifically, (ii) the impact of extragalactic foregrounds such as CIB and tSZ, which should increase as we increase the maximum multipole; and (iii) the impact of any ground pickup, transfer function, or Galactic foreground systematics, which should worsen when decreasing the minimum multipole.

\begin{figure*}
  \includegraphics[width=\textwidth]{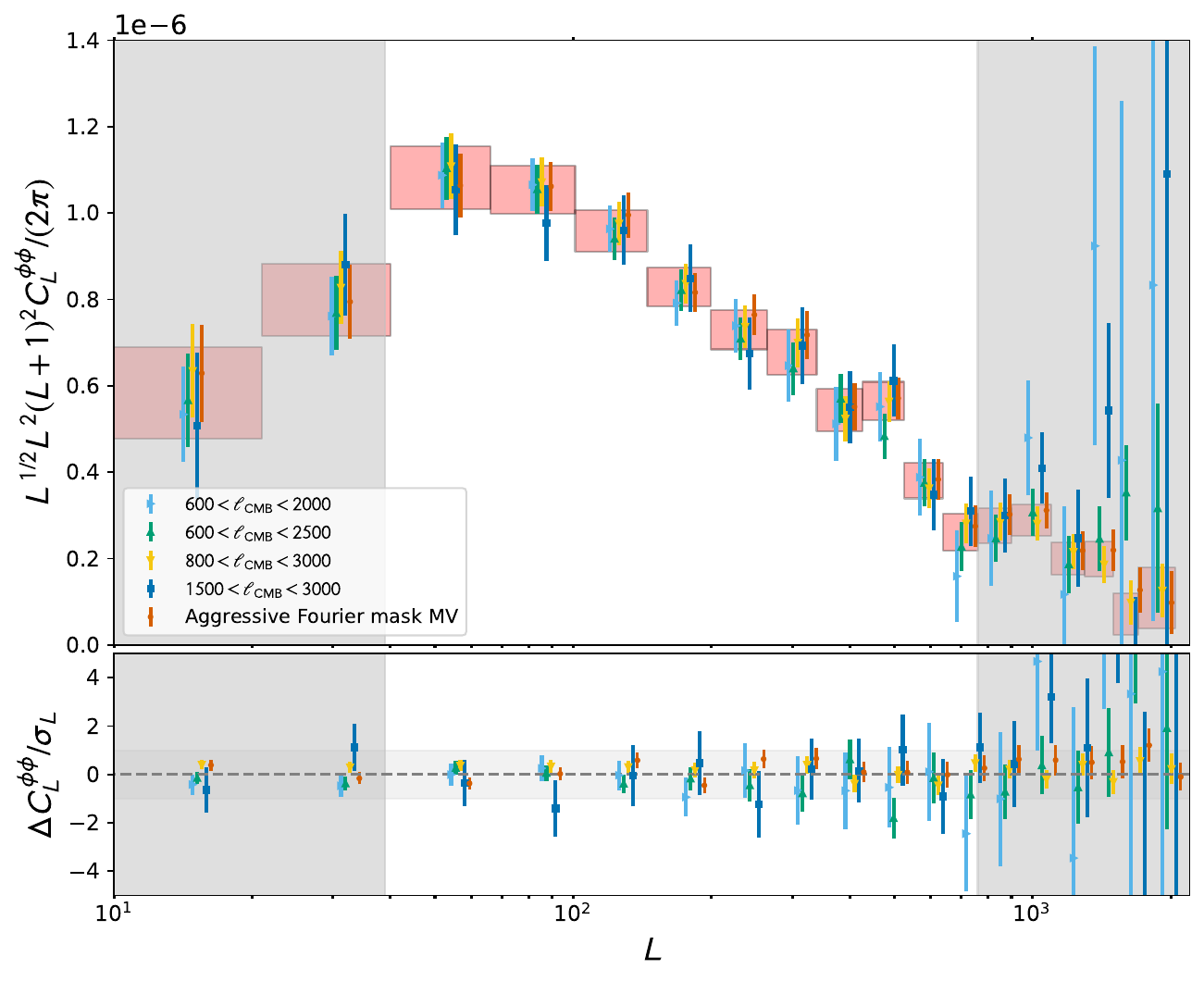}
  \caption{Lensing power spectra obtained from the MV estimator with different CMB scale cuts. These can be compared with our baseline analysis (red boxes), which corresponds to a range of CMB scales $600<\ell_{\mathrm{CMB}}<3000$ used for the measurement. The differences of the resulting lensing spectra compared with the baseline are shown in the bottom panel. We find excellent stability of our results to different CMB scale choices, with all difference spectra consistent with zero. }
  \label{Fig.multipole_cut}
\end{figure*}

The lensing bandpowers and the null bandpowers can be seen in Figure~\ref{Fig.multipole_cut}. We observe excellent consistency with the baseline bandpowers.

\subsubsection{Dust mask variation}\label{mask40}

Galactic foregrounds, such as dust, are non-Gaussian and can\abds{, in principle,} contribute to both the lensing estimator, inducing a bias to the lensing power spectrum measurement, and the curl estimator, causing a curl null-test failure. Given that we only consider small angular scales, $\ell>600$, in our lensing analysis, we expect foreground power from Galactic foregrounds to be a subdominant component in both temperature and (to a lesser extent) in polarization.

We test for any effects of dust in the sky maps by preparing apodized masks that include a more conservative $40\%$ Galactic dust mask in addition to $60\%$ (our baseline). The footprints of the baseline mask and the $40\%$ mask are shown in Figure~\ref{fig:lensing_map} in orange and red, respectively. \fk{The $40\%$ is otherwise prepared in the same way as the $60\%$ mask, as described in Section~\ref{sec.mask}.}

Figure~\ref{Fig.galbandpowers} shows the bandpower difference between baseline lensing spectra and those obtained when measuring lensing with a $40\%$ mask, which cuts out more of the Galactic emission. \fk{For the filtering operation, we use a filter with noise levels consistent with the $40\%$ mask}.
We find a passing PTE of 0.71, with no evidence of contamination from dust in the MV channel. The corresponding curl bandpowers are shown in Appendix \ref{app:null tests curl}.

\subsubsection{North Galactic vs. South Galactic}\label{northsouth}

\fk{We compare the bandpowers obtained from our observations north and south of the Galactic equator, using filters appropriate to the noise levels of each region.}  The gradient bandpowers reconstructed from each region are shown in Fig.\ref{Fig.galbandpowers}; we find that the lensing signal is consistent across these two regions with a null PTE for their difference of 0.91.  Similarly, the curl spectra in Appendix \ref{app:null tests curl}  are also consistent with zero, with PTE values of 0.14 and 0.47 for the Southern and Northern patches respectively.

\begin{mycomment}{In the early pre-blinding stages of our analysis, we used a 60\% dust mask. We subsequently decided to look at null and consistency tests with a 70\% mask, and if these passed, we intended to switch to 70\% as the baseline. Once the baseline mask was decided, we compared our bandpowers against those from more restrictive masks.}\end{mycomment}

\subsubsection{Fourier-space filter variations}\label{sec:ktest}
We vary the extent of the Fourier mask used to eliminate ground pickup and compare the resulting reconstructed bandpowers with our baseline analysis, which removes the \fk{Fourier} modes $|\ell_x|<90$ and $|\ell_y|<50$.
We introduce a more aggressive masking of $|\ell_x|<180$ and $|\ell_y|<100$, doubling the size of the excised region compared to the baseline. 
The lensing bandpowers obtained with this aggressive filter are consistent with the baseline bandpowers, as shown in Figure~\ref{Fig.multipole_cut} for the gradient reconstruction (red points); the PTE for the null difference with respect to the baseline is 0.14.

\subsubsection{Temporal null tests}\label{sec:temporal}

\fk{Finally, we consider null tests with lensing power spectra reconstructed from data taken under different observing conditions or at different times. For these tests, we use filters with noise levels consistent with each dataset to maximize the SNR of the lensing spectra obtained from each data subset.}
The first such test uses two sets of sky maps prepared specially according to the level of precipitable water vapour (PWV) in the observations. This tests for instrumental effects that depend on the level of optical loading and the impact of different levels of atmospheric noise.
Table \ref{table:nullbandpower} shows the results of the PWV high versus PWV low test, in which we compare the lensing bandpowers obtained from data with high and low PWV; see Appendix \ref{pwvdiff} for implementation details. No statistically significant difference is seen between the lensing power reconstructed from the high PWV and low PWV data, with a PTE for their difference of $0.89$.

The second test compares lensing bandpowers obtained from sky maps constructed with observations in the period 2017--2018 (Time-split 1) and 2018--2021 (Time-split 2), as well as their comparison against the baseline MV reconstruction. This tests for the impact of any drifts in the instrument characteristics with time that are not accounted for in the analysis.
These results are presented in Appendix~\ref{time-diffband}, with PTE values additionally reported in Table~\ref{table:nullbandpower}. Again, no statistically significant differences are seen in the bandpower differences.

\subsection{Null-test results summary}\label{sec.chi_dist}

\begin{figure}
    \centering
    \includegraphics[width=1.\columnwidth]{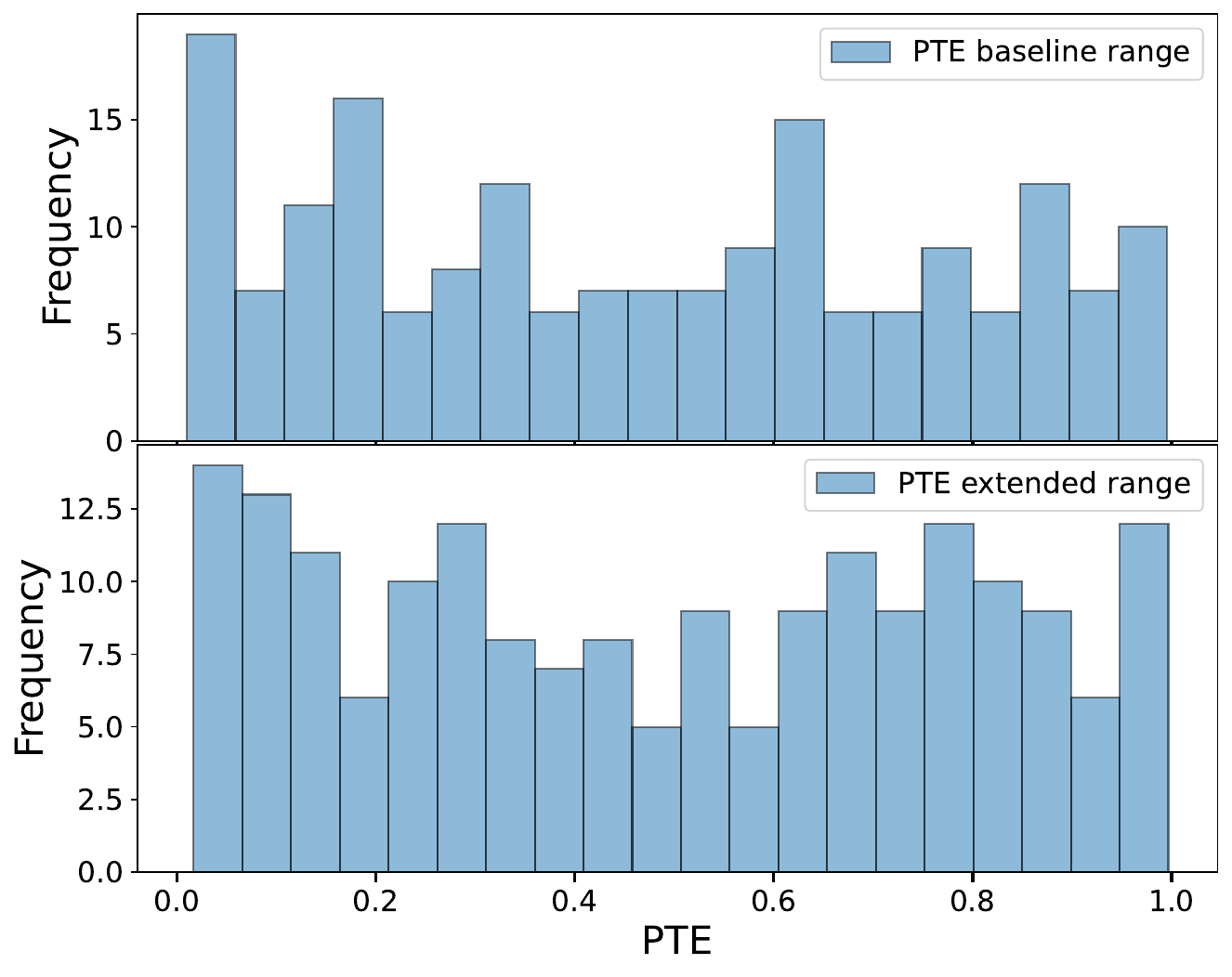}
    \caption{Histogram of PTE values for all of our null tests; the two panels show results for our baseline range of scales in the lensing power spectrum (top panel) and extended range of scales (bottom panel). Both histograms are consistent with a uniform distribution using a Kolmogorov--Smirnov test at $10\%$  and $74\%$, respectively. We argue in Section~\ref{sec.chi_dist} that the tests that have the lowest PTEs are unlikely to indicate significant problems for our baseline analysis. }  
    \label{distribution}
\end{figure}

In this section, we present an overview of all our null test results. We first summarize the results of the previous sections using the overall distribution of PTEs, shown in Figure~\ref{distribution}. Our conclusion is that the distribution of the PTEs of all the null tests is consistent with a uniform distribution for both the baseline and extended range (passing at $10\%$ and $74\%$ with the Kolmogorov--Smirnov test, respectively).

\begin{figure}
  \includegraphics[width=0.5\textwidth]{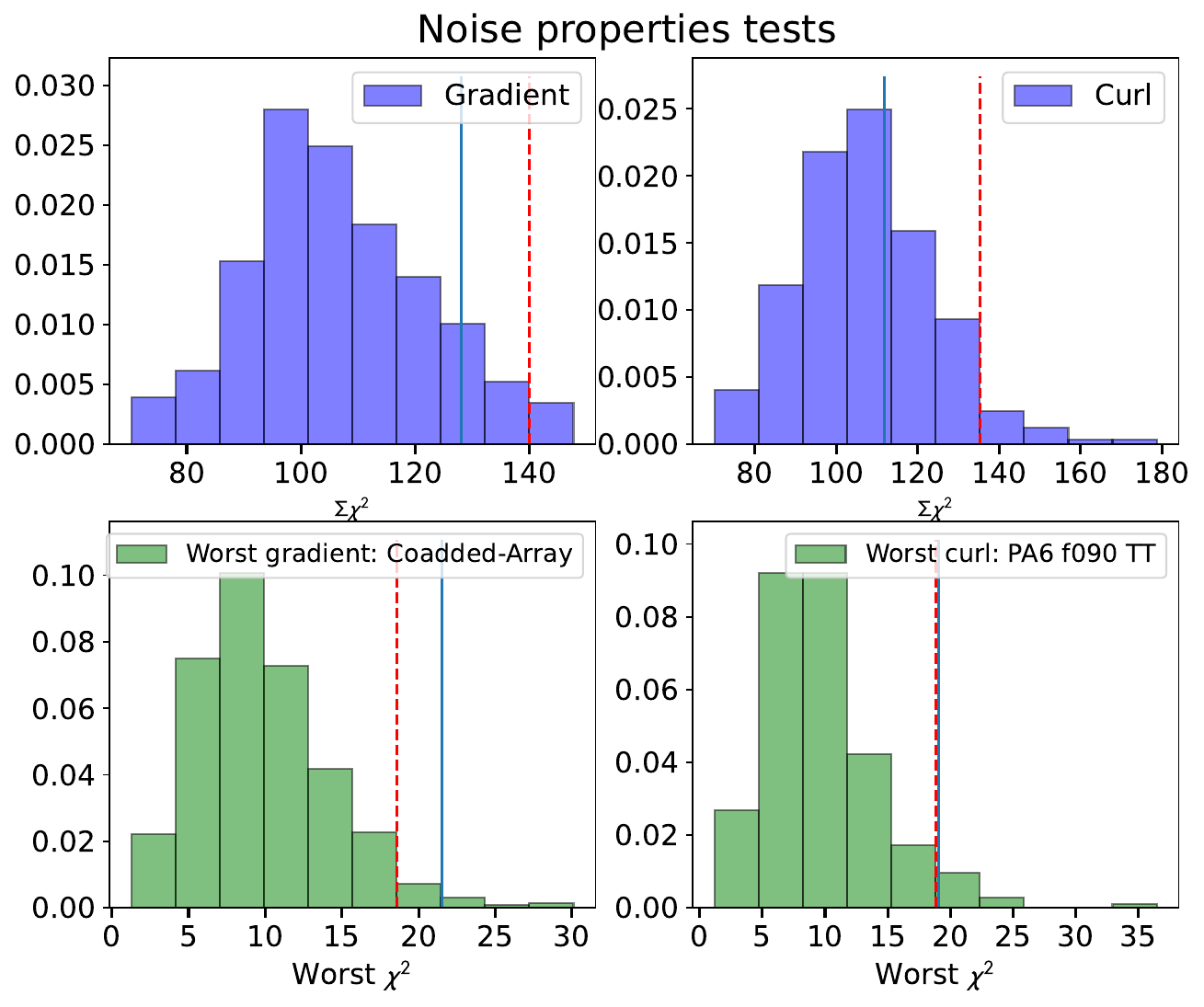}
  \caption{Results of $\chi^2$ tests applied to the bandpowers reconstructed from null (noise-only) maps constructed for individual array-frequencies. \abds{To examine an entire set of null tests of a certain type, we investigate two quantities: (i) we sum the $\chi^2$ of all the relevant tests of this type; and (ii) we select the worst $\chi^2$ of all tests of this type. We then compare the sum or worst $\chi^2$ from data to the distribution of the same summary statistic in simulations. 
  In this figure, we consider the set of noise-only null tests} \abds{described in Section \ref{sec. noise_only}.}\blake{right?}\abds{. The top row shows the $\chi^2$ sum statistic, the bottom row shows the worst $\chi^2$ statistic. The histograms are obtained from simulations, and the red dashed lines indicate the simulation-derived $2\sigma$ limits (in the sense that 95\% of values fall below these limits). The solid blue lines show the $\chi^2$ sum (or the worst $\chi^2$) obtained from the data. Although there are isolated, mild failures at the $2\sigma$ level, overall, the sum of $\chi^2$ tests do not provide evidence for significant problems in our data.} } 
  \label{Fig.noise_only_hist_test}
\end{figure}

\begin{figure}
 \centering
  \includegraphics[width=0.5\textwidth]{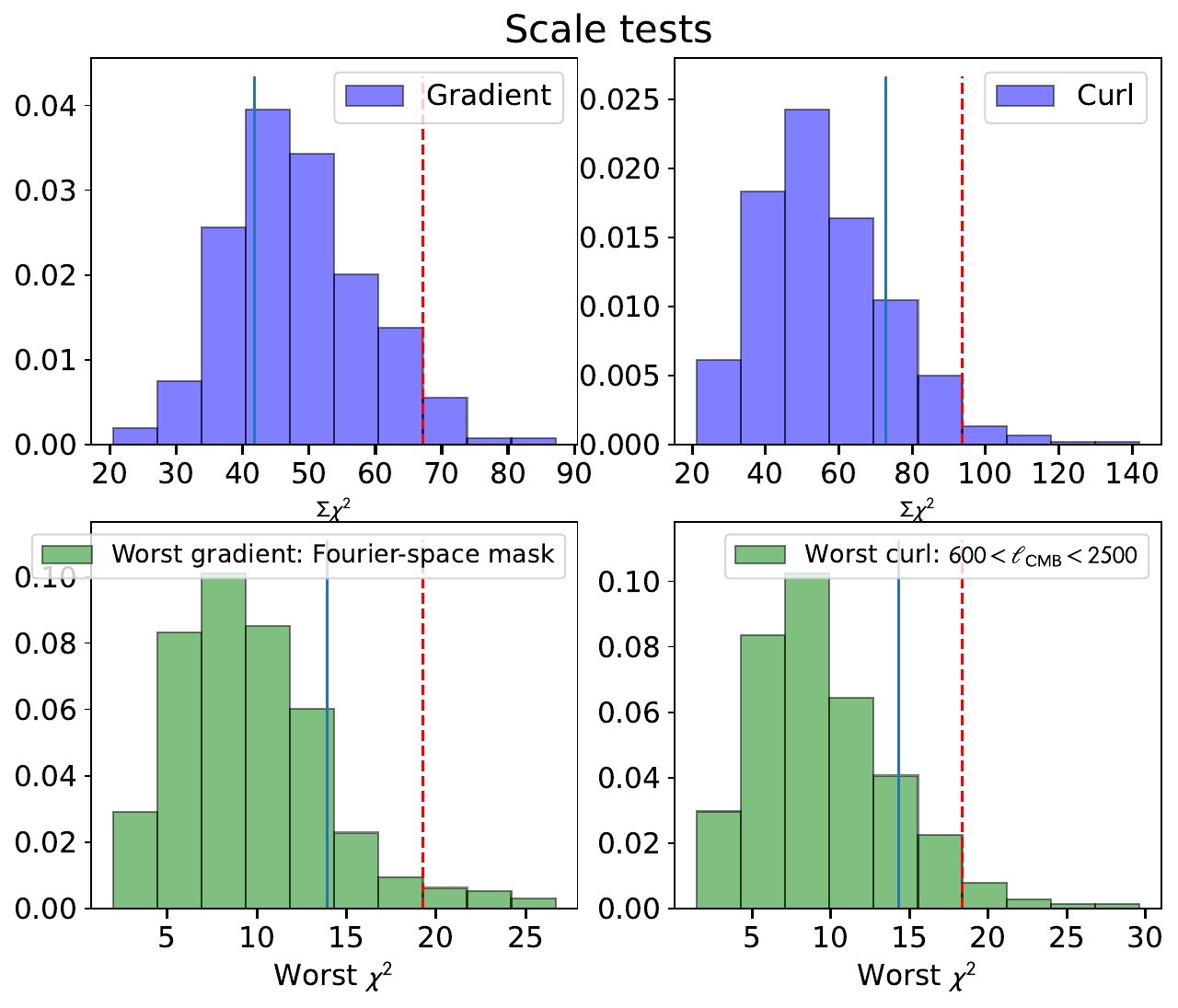}
  \caption{As for Figure \ref{Fig.noise_only_hist_test}, but for the set of scale-related null tests of Section \ref{sec:multipole}, namely the consistency of {$600<\ell<3000$ (baseline), $800<\ell<3000$, $1500<\ell<3000$, $600<\ell<2500$, $600<\ell<2000$ analysis ranges and the \abds{aggressive Fourier-space mask test described in Section \ref{sec:ktest}.}
  We find no evidence for systematic effects in our data in these tests.}} 
  \label{Fig.scale_only_hist}
\end{figure}

\begin{figure}
  \includegraphics[width=0.5\textwidth]{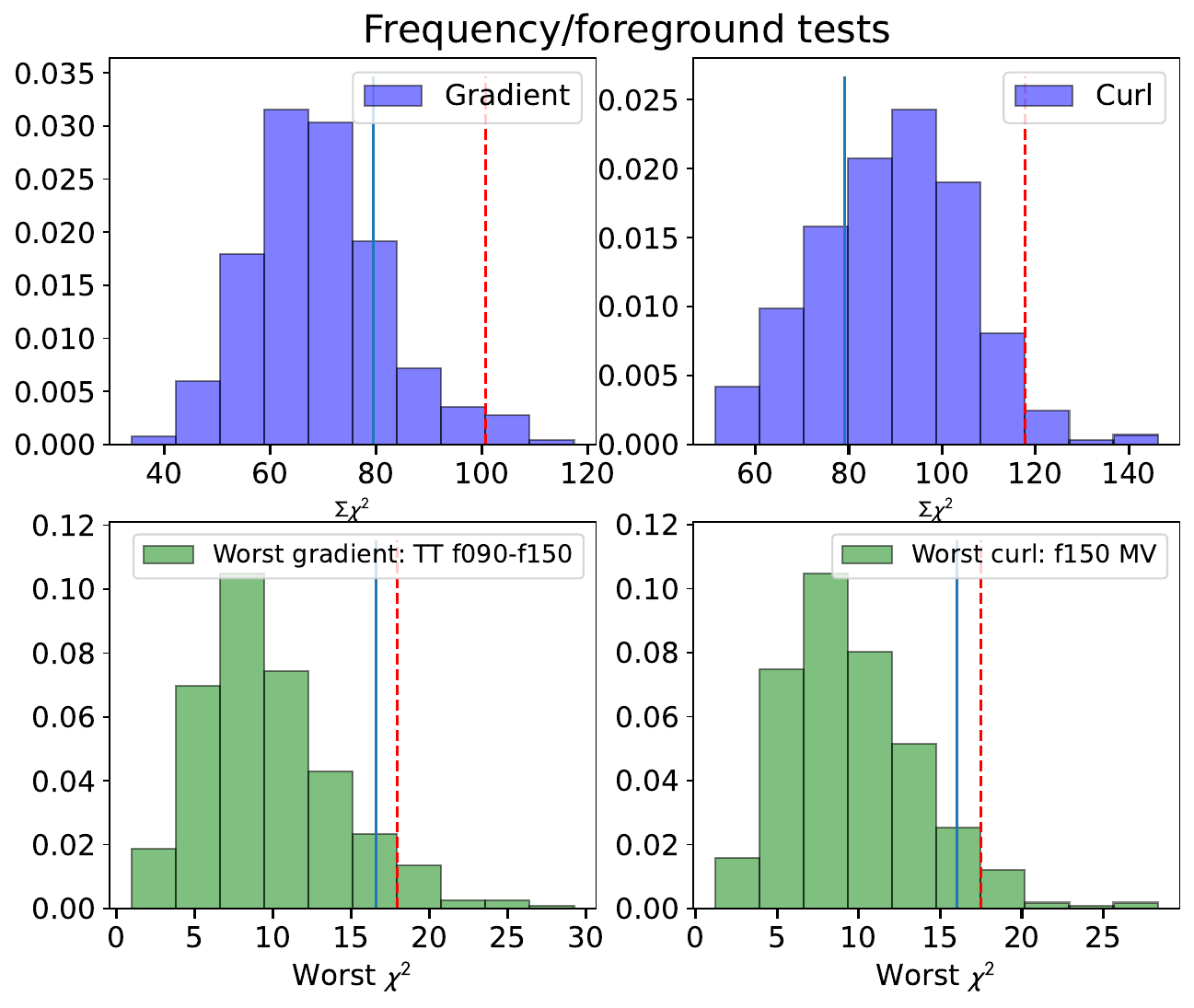}
  \caption{As for Figure \ref{Fig.noise_only_hist_test}, but for null tests related to foregrounds: \abds{map-level $\text{f090}-\text{f150}$ difference tests of Section \ref{sec:maplevelfreq}, bandpower $\text{f090}-\text{f150}$ difference tests of Section \ref{sec:bandfreqtest}, the CIB deprojection consistency test of Section \ref{sec:cibdeprojection}, the shear consistency test of Section \ref{sec:shear} and the consistency tests between different polarization combinations of Section~\ref{sec.nullpol}.}
  } 
  \label{Fig.freq_only_hist}
\end{figure}

\begin{figure}
  \includegraphics[width=0.5\textwidth]{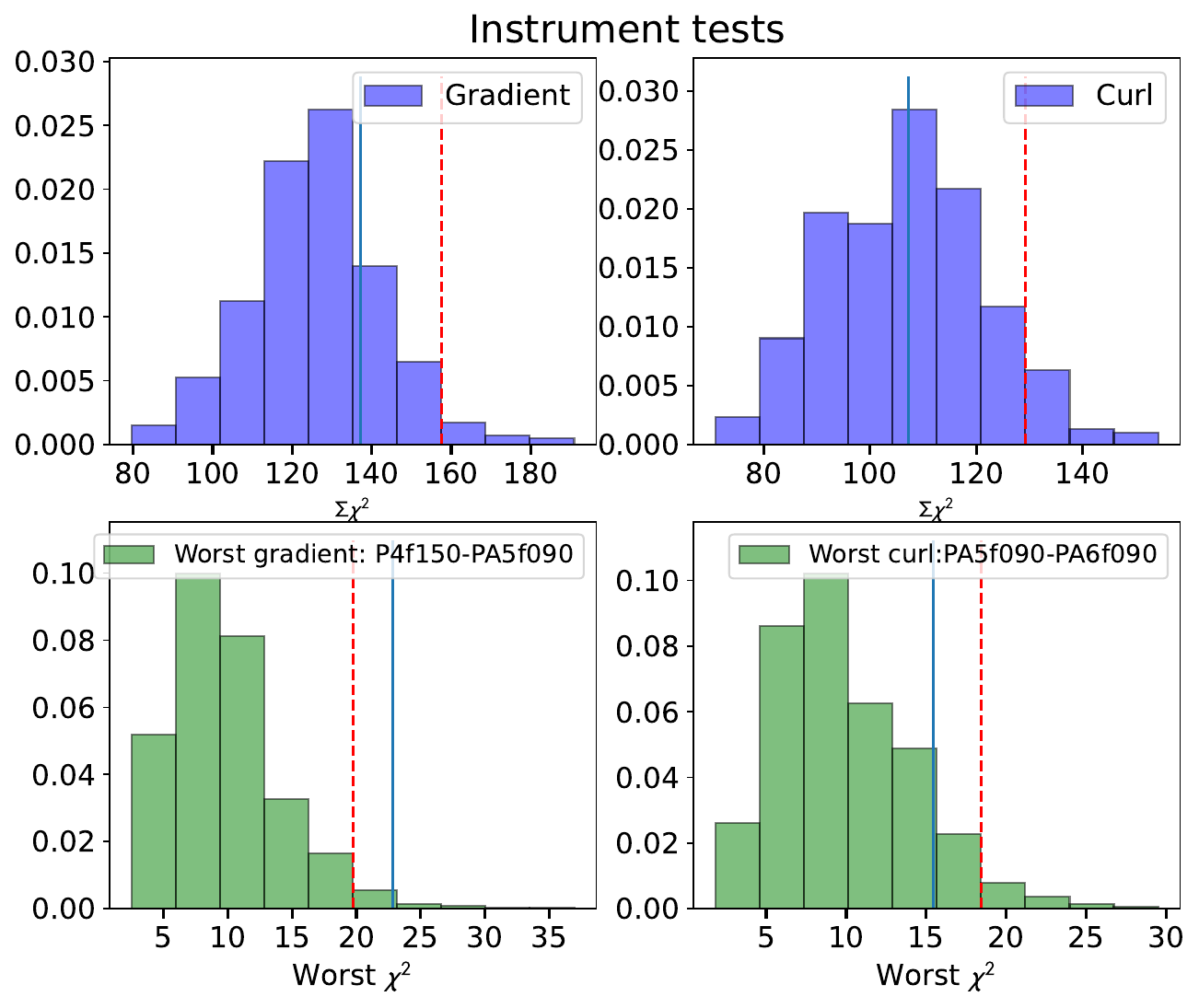}
  \caption{As for Figure \ref{Fig.noise_only_hist_test}, but for null tests related to instrument systematics.  \abds{These include the map-level array-difference tests described in Section~\ref{sec:arraydiff} and the PWV and season-difference null tests introduced in Section~\ref{sec:temporal}.} We again find no evidence for systematic effects in our data.}
  \label{Fig.instrument_only_hist}
\end{figure}

\begin{figure}
  \includegraphics[width=0.5\textwidth]{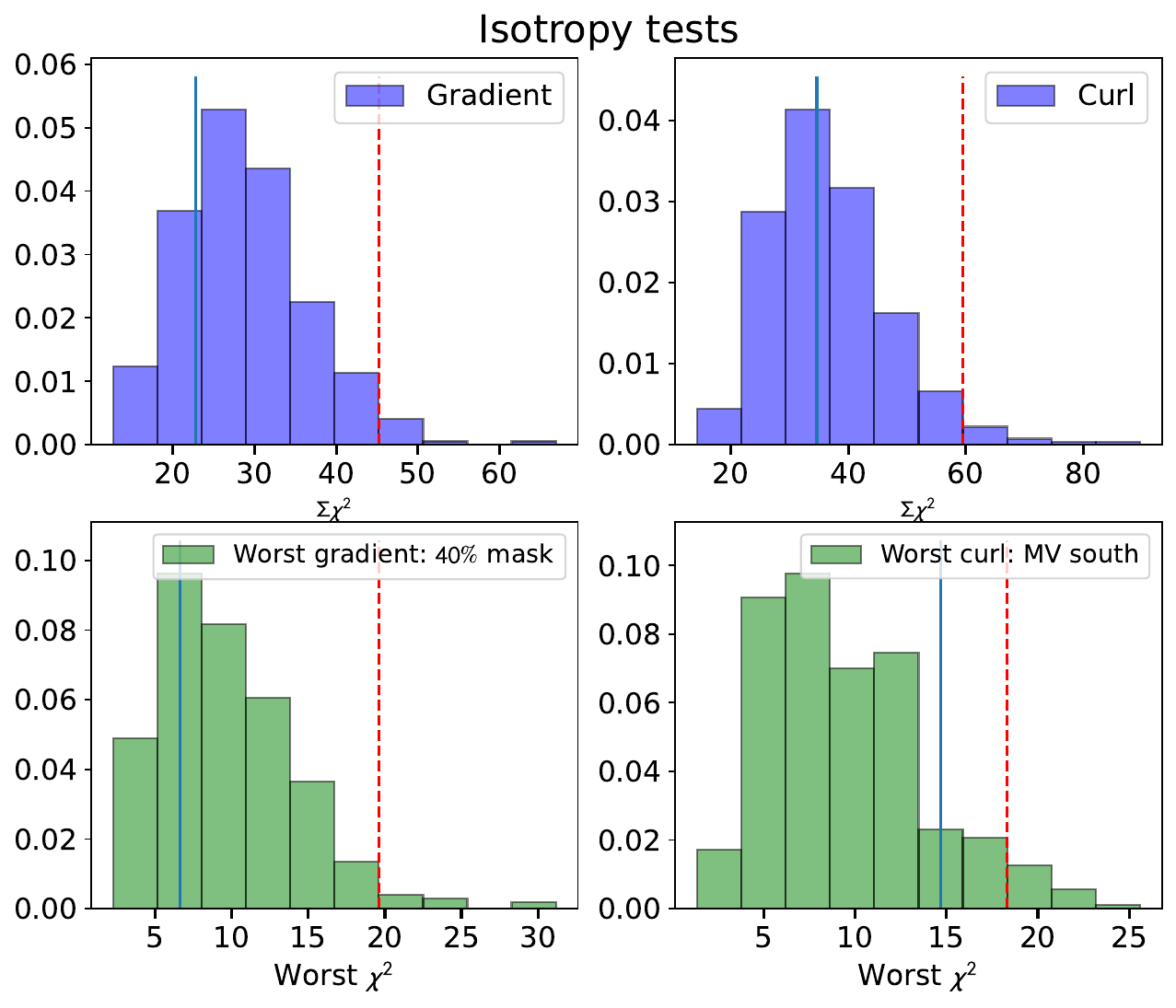}
  \caption{\fk{As for Figure \ref{Fig.noise_only_hist_test}, but for null tests related to isotropy:{ the cross-linking null test, as described in Section~\ref{sec:maskxlink}, and consistency tests between the north and south of the Galactic equator, as detailed in Section~\ref{northsouth}.  This suite of tests again appears nominal.}}} 
  \label{Fig.isotropy_only_hist}
\end{figure}

\abds{Beyond performing individual null tests, a powerful check for subdominant systematic errors is to analyze suitable groups of null tests that probe similar effects. To that end, we group all the null tests discussed previously into five categories: tests focusing on noise properties; different angular scales; different frequencies; foregrounds; isotropy; and instrument-related systematics. We then compare: (i) the sum of all $\chi^2$ values within each category and (ii) the worst (i.e., largest) $\chi^2$ within each category from the data with the distribution of such statistics obtained in simulations, as shown in Figures~\ref{Fig.noise_only_hist_test}--\ref{Fig.isotropy_only_hist}. 
For each test, we consider both gradient lensing modes and curl modes. In the figures, the blue lines show the sum or worst $\chi^2$ statistic for the data and the dotted red lines show the $2\sigma$ limits from the ensemble of simulated measurements of the relevant statistic. We do not see strong evidence of a systematic effect from the above analysis. Only the worst-$\chi^2$ statistics for the noise-only and instrument null-test sets slightly exceed the $2\sigma$ limits from simulations. These tests correspond to the coadded noise-only test (Section~\ref{sec. noise_only}) and the map-level null for the PA4 f150 and PA5 f090 array-frequencies (Section~\ref{sec:arraydiff}), respectively.}

For the following reasons, we do not consider the \fk{array-frequency} null-test failures related to PA4 f150 and PA5 f090 concerning. 
\begin{itemize}
\item The \abds{array-frequency map difference} tests are very hard to pass, as the signal-variance contribution to the error bars is absent; this implies, first, that the null test errors can be much smaller than the measurement errors, and second, that the requirements on the fidelity of the noise simulations are much more stringent than needed for the standard lensing spectrum measurement. \bds{We also note that, since we do not subtract an $N_0$ bias from these null tests, a failure could simply indicate that the CMB power spectrum of the two different maps being nulled is inconsistent. This does not necessarily imply an inconsistency in lensing, because in our lensing power spectrum analysis, the realization-dependent bias subtraction methodology should absorb small changes in the CMB power spectra.}
\item \abds{Some of these worst-performing map-level null tests involve the array-frequency PA4 f150. We further checked for possible inconsistencies of this array with the others by performing bandpower-level null tests in Section~\ref{array_bandpower}; these tests should be more sensitive to multiplicative biases affecting the different datasets than the map-level test and, furthermore, include the signal part of the covariance matrix. In these targeted tests, we found no evidence of a systematic difference in PA4 f150 compared to the other array-frequencies.}
\item In addition, the array PA4 f150 has the least weight in our coadd data; as can be seen in  Figure~\ref{fig.weights}, it only contributes less than $10\%$ at each CMB multipole to the total coadded sky map used in our analysis.
\item \abds{Foregrounds are not a likely cause of the $\text{PA4 f150} - \text{PA5 f090}$ null failure since the much more sensitive $\text{f090} - \text{f150}$ coadd map null test passes.} 

\end{itemize}


\section{Systematic Error Estimates}\label{sec.systematics}

\subsection{Foreground mitigation methodology and verification}\label{foreground}

\begin{figure*}[t]
    \centering
    \vspace{0pt}
    \includegraphics[width=0.95\columnwidth]{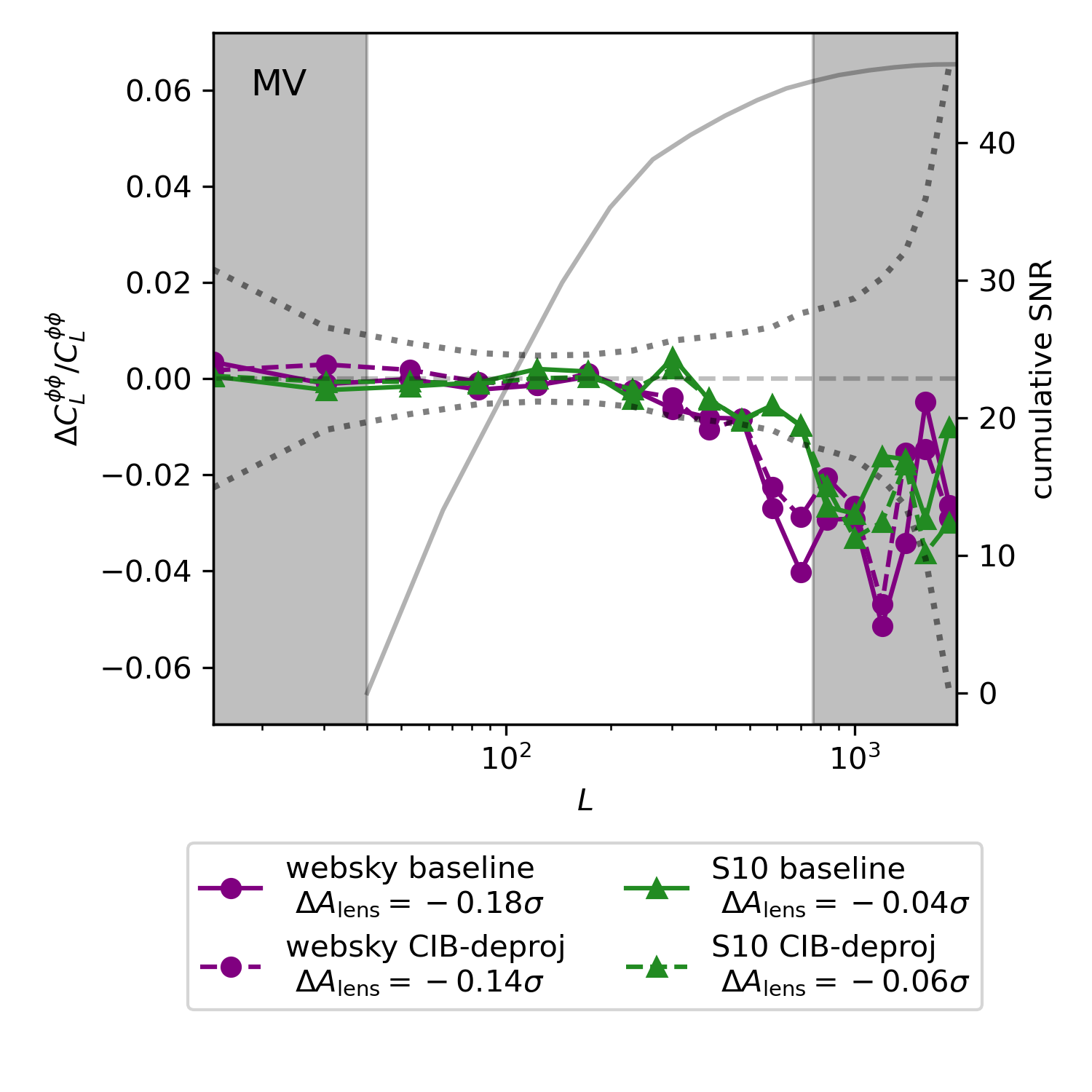} 
    \vspace{0pt}
    \raisebox{45pt}{\includegraphics[width=0.95\columnwidth]{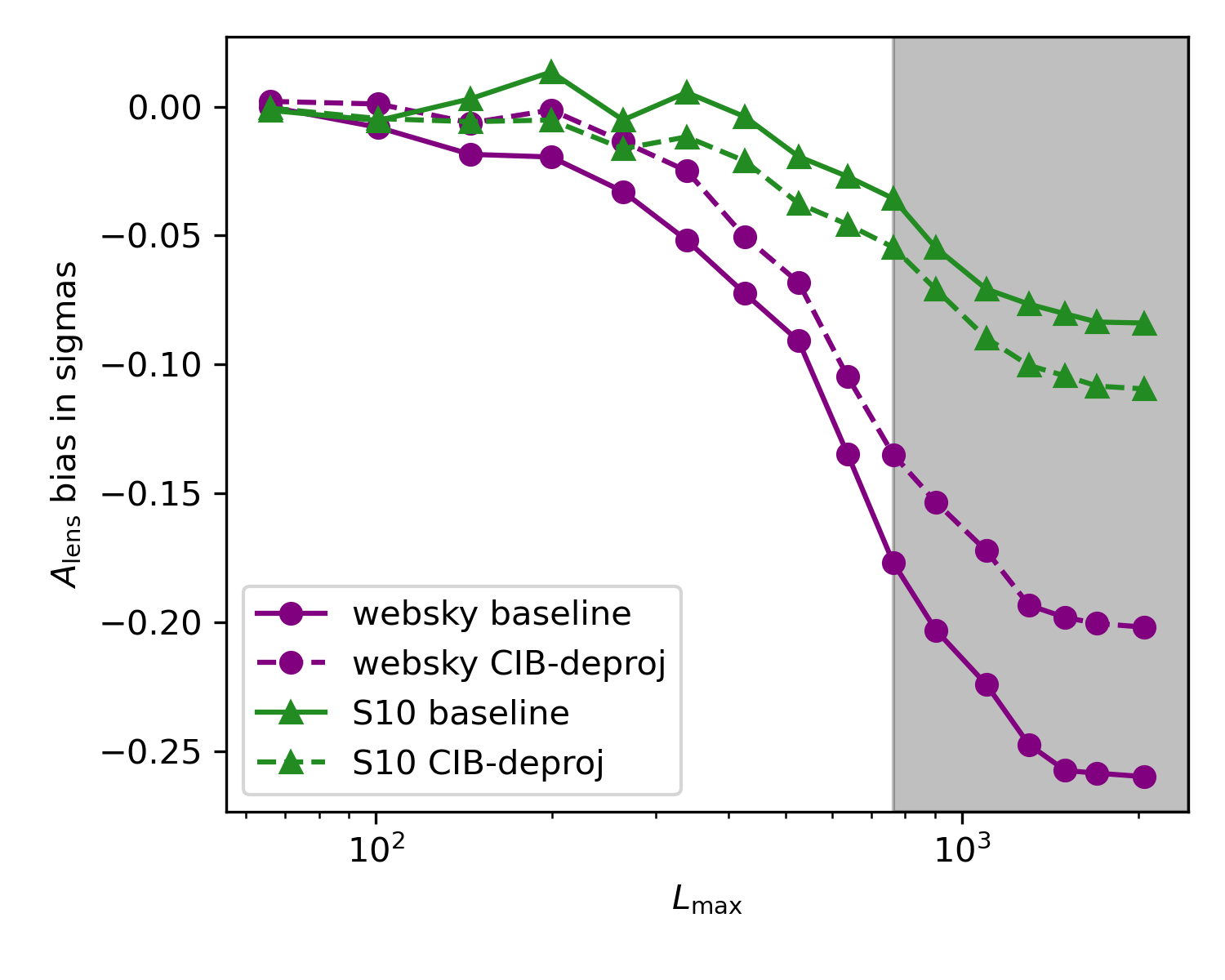}}
    \caption{\textit{Left}: Estimates of the fractional bias due to extragalactic foregrounds to the estimated CMB lensing power spectrum, for an ACT DR6-like analysis,  in the case of the \abds{baseline (MV)} estimator. Circles with solid connecting lines indicate biases predicted from the \textsc{websky} \citep{websky} simulations, triangles with solid connecting lines indicate biases estimated from the \citet{Sehgal_2010} simulations (denoted S10). The symbols with dashed connecting lines are for the CIB-deprojected analysis, which additionally uses \textit{Planck} data at $353\,\si{GHz}$ and $545\,\si{GHz}$, described in the main text. The grey dotted line indicates 10\% of the  $1\sigma$ uncertainty of the DR6 bandpower measurement, and the solid grey line indicates the signal-to-noise ratio when only scales up to $L$ are included. The grey shaded regions indicate scales not used in the baseline cosmological inference.
    \textit{Right}: Bias in inferred $\Alens$ as a function of the maximum multipole in the CMB lensing power spectrum used, $\Lmax$, in units of the $1\sigma$ uncertainty (with the uncertainty  also re-computed as a function of  $\Lmax$). Both panels indicate that the biases to the lensing power spectra are negligible in all our analyses; in particular, using the baseline range, foreground biases are generally mitigated to below $0.2\sigma$.}
    \label{fig:simbias_highL}
\end{figure*}

\fk{\citet{dr6-lensing-fgs} focuses exclusively on the systematic impact of extragalactic foregrounds; we briefly summarise that work here.} We characterise and validate the mitigation strategies employed in our lensing analysis via a three-pronged approach.
\begin{enumerate}
\item{We refine and test the mitigation strategies on two independent microwave-sky simulations: those from  \citet{Sehgal_2010} and the \textsc{websky} simulations \citep{websky}. We demonstrate that the baseline approach taken here, i.e., finding/subtracting $\text{SNR}>4$ point-sources, finding and subtracting models for $\text{SNR}>5$ clusters, and using a \abds{profile-hardened} quadratic estimator (see Section~\ref{sec:biashardening} for details), performs better or equally well to the other tested mitigation strategies, with sub-percent fractional biases to $C_L^{\phi\phi}$ on both simulations within most of the analysis range (see left panel of Figure 2 of \citealt{dr6-lensing-fgs}, reproduced here as Figure~\ref{fig:simbias_highL}). We also find that for both simulations, biases in the inferred lensing power spectrum amplitude, $\Alens$, are below $0.2\sigma$ in absolute magnitude, and below $0.25\sigma$ when extending the analysis range to $\Lmax=1300$, as shown in the right panel of Figure~\ref{fig:simbias_highL}. In addition, we demonstrate that a variation of our analysis where we include \textit{Planck} $353\,\si{GHz}$ and $545\,\si{GHz}$ channels in a harmonic ILC while deprojecting a CIB-like spectrum also performs well (see the dashed lines in Figure~\ref{fig:simbias_highL}); this motivated also running this variation on the real DR6 data to ensure our measurement is robust to the CIB.}
\item{Since the results described above to some extent rely on the realism of the extragalactic microwave-sky simulations, we also demonstrate our robustness to foregrounds using targeted null-tests that leverage the fact that the foreground contamination is frequency-dependent, while the CMB lensing signal is not. These \abds{null tests} are also presented in this paper (Figures~\ref{Fig.90150MVdiff}  and \ref{Fig.freqbandpowers}).}
\item{Finally, we demonstrate that the DR6 data bandpowers are consistent when we use either CIB-deprojected maps, which should reduce CIB contamination but to some extent increase tSZ contamination (see Section~\ref{sec:cibdeprojection}), or the shear estimator (see Section~\ref{sec:shear}).}
\end{enumerate}

\subsection{Instrumental systematics}\label{sec:instrument.summary}

We investigate the effects of various instrumental systematic factors on the lensing auto-spectrum measurement and present the summary here. These include: (i) miscalibration; (ii) beam uncertainty; (iii) temperature-to-polarization leakage; (iv) polarization efficiency; and (v) polarization angles. A comprehensive evaluation of these factors and their impact on the measurement is presented in Appendix~\ref{app:instrument.details}. 
Here, we summarise the estimated systematic biases to the lensing power spectrum in terms of $\Delta\Alens$, the typical bias in the lensing amplitude.
This is calculated from Equation~\eqref{eq.al.bias}, with $\hat{C}^{\mathrm{null}}_{L_b}$ replaced by $\Delta\hat{C}^{\phi\phi}_{L_b}$, the shift to the power spectrum due to a systematic effect.  Note that the $ \Delta\Alens$ is computed for the baseline range ($L \in [40,763]$). Table~\ref{Tab:systematics.summary} summarises the $ \Delta\Alens$ values of the systematics considered in this study. The estimated levels of instrument systematics would not produce a significant bias to our lensing measurements. We now briefly discuss these systematic effects in turn.

The lensing power spectrum can be impacted by miscalibration of the sky maps and beam uncertainties. To address this, we have estimated error budgets for both factors using exact as well as approximate methods, and found no significant bias (Appendix~\ref{app:calandbeam}). \abds{Summary statistics for $\Delta \Alens$, given the estimated uncertainties in the calibration and beam transfer functions, are reported in Table~\ref{Tab:systematics.summary}.} Furthermore, the consistency of patch-based isotropy tests in Section~\ref{iso} shows that there is no evidence of spatially varying beams at levels relevant for lensing.

\abds{As discussed in Section~\ref{subsection:beams}, measurements on Uranus provide evidence of temperature-to-polarization leakage in the ACT data.} We estimate the impact of this by adding leakage terms to the polarization maps and analyzing their response; see Appendix~\ref{app:syst.t2pleakage} for details. These results indicate only a small effect in our lensing measurement, as summarized in Table~\ref{Tab:systematics.summary}.


\fk{Some evidence has been found for a larger temperature-to-polarization leakage in the maps than specified by the nominal Uranus leakage model, based on power spectrum differences between detector arrays.} We analyse a simple model for such leakage in Appendix~\ref{app:syst.t2pleakage} and propagate this through our lensing measurement for the extreme case of the same leakage for all array-frequencies. The bias on $\Alens$ is again a small fraction of the statistical error (see Table~\ref{Tab:systematics.summary}).

The absolute polarization angle error in the ACT DR6 data set is found to be consistent with the previous ACT DR4 data set and rotating the $Q$/$U$ maps by an amount equal to the estimated DR4 angle plus its uncertainty (a total $\Phi_{\textrm{P}}= -0.16\si{deg}$) has minimal impact on the measurement (Table~\ref{Tab:systematics.summary}, with further details in Appendix~\ref{app:syst.polrot}).

The polarization efficiency correction that we apply to the polarization maps was discussed in Section~\ref{sec.polareff}; see also Appendix~\ref{app:polfit} for a full description. It is based on the mean of the measured values for each array, obtained by fitting the ACT DR6 EE spectra to the fiducal model. A test where the estimated polarization efficiency is lowered by $1\sigma$, \abds{where $\sigma$ is the mean of the per-array statistical errors from the fitting}, 
shows no significant impact (Table~\ref{Tab:systematics.summary} and Appendix~\ref{app:syst.poleff}). We note that this is a very conservative test as it assumes the errors in the fitted polarization efficiences are fully correlated across array-frequencies.


Finally, we note that a broader, simulated investigation of the impact of instrument systematics on CMB lensing was performed in \cite{PhysRevD.103.123540}; although the simulation choices are not an exact match to ACT DR6, it is reassuring that the resulting systematic biases reported there appear negligible at the levels of precision considered in our analysis.




\begin{table}
\centering
\caption{Summary of instrumental systematic error budgets. The bias to the lensing amplitude, $ \Delta\Alens$, is defined similarly to Equation~\eqref{eq.al.bias}, \fk{but with the null spectrum replaced by the bias in the lensing spectrum due to the indicated instrument systematic.}
The bias is calculated for the baseline analysis range ($L \in [40,763]$). The ``T$\rightarrow$P leakage (const.)'' and  ``Polarization eff.'' represent the values from very conservative models as described in Appendix~\ref{app:syst.t2pleakage} and Appendix~\ref{app:syst.poleff}, respectively; they should hence be understood as upper limits.}
\begin{tabular}{c c c} 
 \hline \hline
 Systematic & Estimator & $ \Delta\Alens$  \\ [0.5ex] 
 \hline
 Miscalibration & MV & $\albiasMiscalMV  \;(\biasSigmaMiscalMV)$ \\ 
 Miscalibration & MVPol & $\albiasMiscalMVPol  \;(\biasSigmaMiscalMVPol)$ \\
 Beam uncertainty & MV & $\albiasBeamMV  \;(\biasSigmaBeamMV)$  \\
 Beam uncertainty& MVPol & $\albiasBeamMVPol  \;(\biasSigmaBeamMVPol)$  \\
 T$\rightarrow$P leakage (beam) & MV & $\albiasLeakBeamMV  \;(\biasSigmaLeakBeamMV)$   \\
 T$\rightarrow$P leakage (beam) & MVPol & $\albiasLeakBeamMVPol  \;(\biasSigmaLeakBeamMVPol)$  \\
 T$\rightarrow$P leakage (const.) & MV & $\albiasLeakConstMV  \;(\biasSigmaLeakConstMV)$   \\
 T$\rightarrow$P leakage (const.) & MVPol & $\albiasLeakConstMVPol  \;(\biasSigmaLeakConstMVPol)$  \\
 Polarization eff. & MV &  $\albiasPolEffMV  \;(\biasSigmaPolEffMV)$  \\
 Polarization eff. & MVPol & $\albiasPolEffMVPol  \;(\biasSigmaPolEffMVPol)$   \\
 Polarization angle & MV & $\albiasPolAngMV  \;(\biasSigmaPolAngMV)$   \\
 Polarization angle & MVPol & $\albiasPolAngMVPol  \;(\biasSigmaPolAngMVPol)$  \\
 \hline
\end{tabular}
\label{Tab:systematics.summary}
\end{table}

\subsection{A Note on Map Versions}

\NS{This analysis uses the first science-grade version of the ACT DR6 maps, namely the \texttt{dr6.01} maps. Since these maps were generated, we have made some refinements to the map-making that improve the large-scale transfer function (discussed in Section \ref{subsection:cal}) and polarization noise levels, and include data taken in 2022. We expect to use a second version of the maps for further science analyses and for the DR6 public data release.  While we caution that we cannot, with absolute certainty, exclude changes to the lensing power spectrum with future versions of the maps, we are confident that our results are robust and stable for the reasons discussed below.}



\NS{First, as discussed in Section \ref{subsection:cal}, there are non-idealities such as transfer function and leakage effects in the analyzed version of the maps.  In principle, one could worry that the source of this transfer function problem also affects lensing measurements. However, these effects appear primarily at low multipoles below $\ell \sim 1000$. Fortunately, the lensing estimator is insensitive to effects at low multipoles, which is the reason for the negligible change found when correcting for the transfer function (as discussed in Section \ref{subsection:cal}).}

\abds{Second, after refinements to the map-making were implemented, improving the large-scale transfer functions and polarization noise levels, a small region of sky was mapped with the old and new map-making procedures to allow comparison. We measure lensing from the old and new maps and construct lensing power spectrum bandpowers and map-level null tests as described in Section~\ref{sec:null}. In the comparison of the lensing power spectra, the new maps produce a lower spectrum by 0.3$\sigma$ (of the measurement error on this region) or 3.5\%; while at the time of publication noise simulations from the new mapping procedure were not available to us to assess the significance of the difference, this variation appears consistent with random scatter.} 
\blake{How do you make this judgement without noise simulations? From determining the degree of correlation of the two CMB maps, maybe?}
\abds{In addition, residuals in the map-level null test are entirely negligible (below $C_L^{\phi \phi}/1000$), ruling out significant non-Gaussian systematics that differ between the map versions.}

\abds{Finally, 
several of our null tests should be sensitive to the known non-idealities in the maps. For example, transfer function effects appear worse in f150 than in f090; related systematics might therefore affect the $\text{f090}-\text{f150}$ frequency null tests or array-frequency null tests. Similarly, systematics affecting low or high multipoles differently should affect the multipole range stability tests shown in Figure \ref{Fig.multipole_cut}. Despite having run a suite of more than 200 null tests, no significant evidence for systematics was found; this, in particular, gives us confidence in our results.}


\section{Measurement of the Lensing Power Spectrum}\label{sec:results}

\subsection{Lensing power spectrum results}

\begin{figure}
  \includegraphics[width=\linewidth]{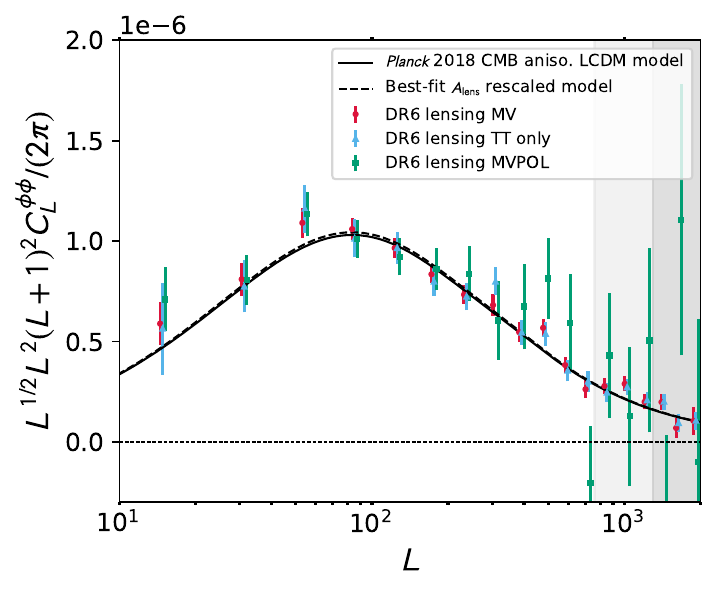}
  \caption{The ACT DR6 lensing bandpowers for the baseline analysis, combining temperature and polarization (MV) are shown in red and are in good agreement with the $\Lambda$CDM theoretical prediction based on the Planck 2018 CMB power spectrum best-fit cosmology, which is shown in the solid black line. A model based on the best-fit rescaling of this prediction with $A_\mathrm{lens} = 1.013\pm0.023$ is shown in the dashed line. We also show the polarization-only (MVPOL) and temperature-only (TT) analyses in green and blue, respectively. We find good consistency of the lensing measurements using temperature and polarization.}
  \label{Fig.polcomb_consistentt}
\end{figure}

\abds{Our baseline lensing \abds{power spectrum} measurement is shown in Figure~\ref{Fig.results_clkk} and Table \ref{tab.band}} In addition to this combined temperature and polarization (MV) measurement, we also provide temperature-only and polarization-only lensing bandpowers in Figure~\ref{Fig.polcomb_consistentt} and their respective $\Alens$ values in Figure~\ref{Fig.polcomb_consistent_alens}.  A compilation of the most recent lensing spectra made by different experiments is shown in Figure~\ref{Fig.results_compilation}. Our measurement reaches the state-of-the-art precision obtained by the latest \textit{Planck} \texttt{NPIPE} analysis \citep{Carron_2022}; compared with other ground-based measurements of the CMB lensing power spectrum, our result currently has the highest precision. \fk{Furthermore, our polarization-only lensing amplitude estimate, determined at $19.7\sigma$, is the most precise amongst measurements of its kind, surpassing the $12.7\sigma$ and $10.1\sigma$ from \texttt{NPIPE}~\citep{Carron_2022} and SPTpol~\citep{Wu_2019}, respectively.}


\begin{table}
\centering
\caption{Lensing bandpowers in the baseline range, for our minumum-variance (MV) estimator. Values are given for $10^7[L(L+1)]^2C^{\hat{\phi}\hat{\phi}}_{L}/4=10^7C^{\hat{\kappa}\hat{\kappa}}_{L}$ 
averaged over non-interlapping bins with bin edges given in the first column and band centers $L_b$ in the second.} 
\begin{tabular}{ccc}
\hline\hline
$[L_{\text{min}}\quad{L}_{\text{max}}]$ & $L_b$ & $10^7[L(L+1)]^2C^{\hat{\phi}\hat{\phi}}_{L}/4$ \\ \hline
$[40 \quad 66]$                             & 53.0  & $2.354\pm0.157$                          \\
$[67  \quad 101]$                           & 83.5  & $1.822\pm0.096$                          \\
$[102 \quad 145]$                           & 123   & $1.368\pm0.068$                          \\
$[146 \quad 199]$                           & 172   & $1.000\pm0.054$                          \\
$[200 \quad 264]$                           & 231.5 & $0.758\pm0.047$                          \\
$[265 \quad 339]$                           & 301.5 & $0.617\pm0.048$                          \\
$[340 \quad 426]$                           & 382.5 & $0.439\pm0.039$                          \\
$[427 \quad 526]$                           & 476   & $0.409\pm0.032$                          \\
$[527 \quad 638]$                           & 582   & $0.249\pm0.027$                          \\
$[639 \quad 763]$                           & 700.5 & $0.156\pm0.026$                          \\ \hline
\end{tabular}\label{tab.band}
\end{table}

\subsection{Lensing amplitude}

\begin{figure}
  \includegraphics[width=\linewidth]{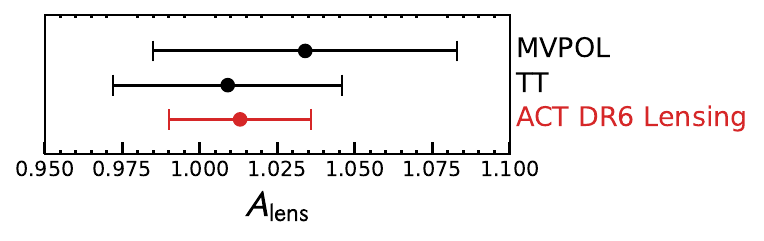}
  \caption{ACT DR6 measurements of the lensing amplitude $A_{\mathrm{lens}}$ relative to the model from the \textit{Planck} 2018 measurements of the CMB anisotropy power spectra. Results are shown for the baseline analysis (MV), which combines temperature and polarization, and for polarization-only (MVPOL) and temperature-only (TT) analyses. We find good consistency of lensing measurements using temperature and polarization.}
  \label{Fig.polcomb_consistent_alens}
\end{figure}

\begin{figure*}
\includegraphics[width=\textwidth]{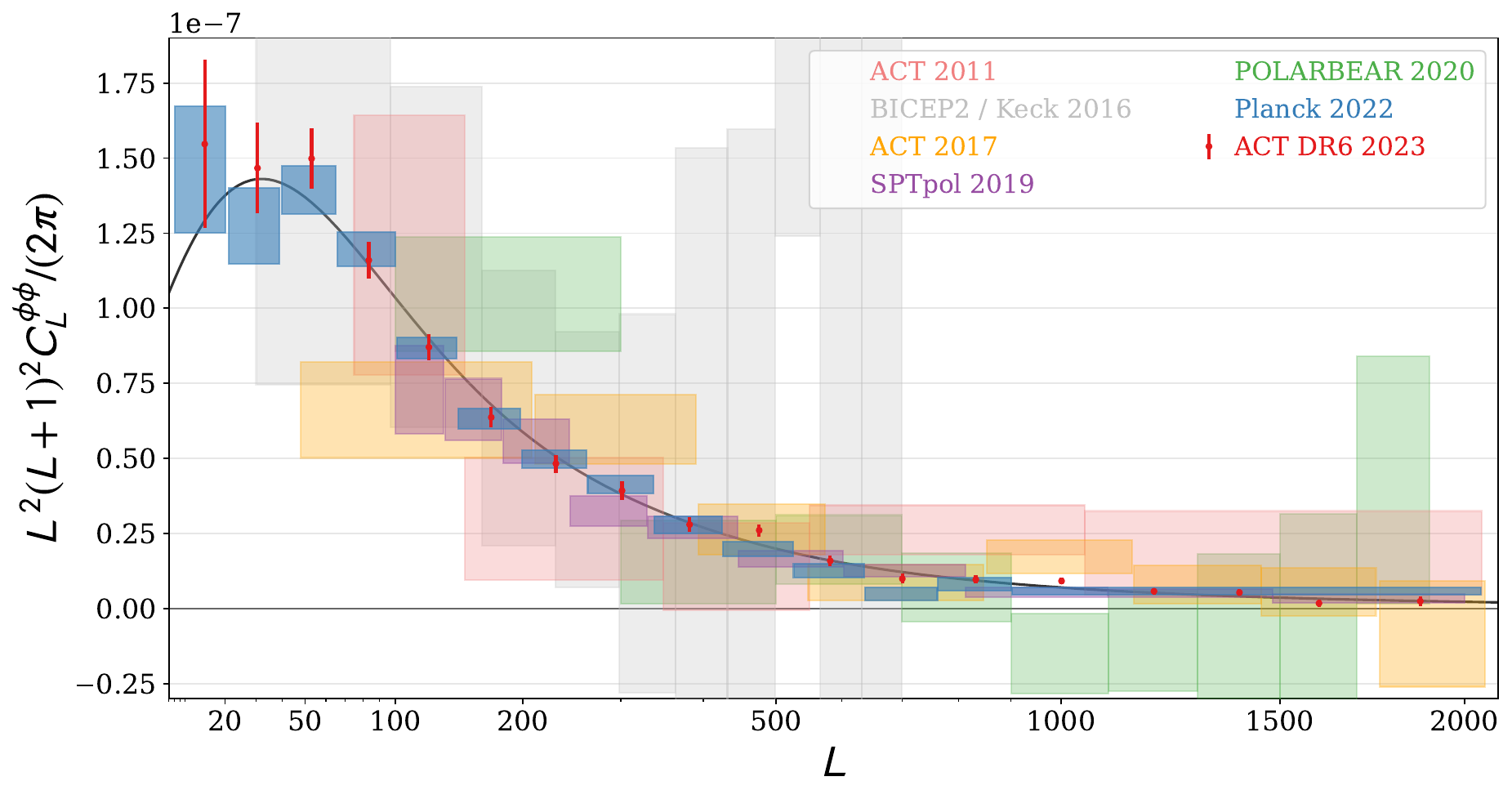}
  \caption{Compilation of CMB lensing power spectrum measurements, with our results shown as red datapoints. The CMB lensing power spectrum presented in this paper represents (along with \textit{Planck} \texttt{NPIPE}, which reaches similar precision) the highest signal-to-noise lensing spectrum measured to date.}
  \label{Fig.results_compilation}
\end{figure*}

We estimate the lensing amplitude parameter $\Alens$ by fitting our baseline bandpower measurements \NS{to a theory lensing power spectrum predicted from the best-fit $\Lambda$CDM model from the \textit{Planck} 2018  
baseline likelihood \texttt{plikHM TTTEEE lowl lowE}, allowing the amplitude of this lensing power spectrum to be a free parameter in our fit.}\footnote{\abds{Our accuracy settings, which were chosen to reproduce results obtained with the high-accuracy settings of \citet{PhysRevD.105.023517} with sufficient accuracy, while still ensuring rapid MCMC runs, are as follows: \texttt{lmax = 4000}; \texttt{lens\_potential\_accuracy = 4}; \texttt{lens\_margin = 1250}; \texttt{AccuracyBoost = 1.0}; \texttt{lSampleBoost = 1.0}; and \texttt{lAccuracyBoost = 1.0}.}} We find
\begin{equation}
{\Alens}=1.013\pm0.023 \quad(68\%\, \mathrm{limit})
\end{equation}
from the baseline multipole range $L=40$--$763$, in good agreement with the lensing spectrum predicted by the \textit{Planck} 2018 \LCDM~model ($\Alens=1$).
The scaled model is a good fit to the lensing bandpowers, with a \fk{PTE for $\chi^2$ of $13\%$}.
(The equivalent PTE for our bandpowers without rescaling the \textit{Planck} model by $A_\mathrm{lens}$ is $17\%$. Note that the PTE is higher here because we have one more degree of freedom.)
\NS{We also find}
\begin{equation}
{\Alens}=1.016\pm0.023
\end{equation}
\NS{using an extended multipole range of $40<L<1300$. The fit remains good over this extended range, \fk{with a PTE of $12\%$}.\footnote{We note that the $\Alens$ error for the extended range only improves over that from the baseline range in the next significant digit, from 0.0234 to 0.0233; the improvement is modest because the small angular scales in the lensing power spectrum are still quite noisy.}}
For comparison, the latest CMB lensing analysis from \textit{Planck} \texttt{NPIPE} obtained ${\Alens}=1.004\pm0.024$ \citep{Carron_2022}.\footnote{\fk{Using the same theory model used in our work, we obtain ${\Alens}=0.993\pm0.025$ from the NPIPE lensing bandpowers.}}
For measurements performed purely with temperature and polarization we obtain the lensing amplitudes of ${\Alens}=1.009\pm0.037$ and ${\Alens}=1.034\pm0.049$, respectively. The consistency of the $A_\mathrm{lens}$ amplitude obtained from MV, TT, and MVPOL is illustrated in Figure~\ref{Fig.polcomb_consistent_alens}.
Our measured lensing bandpowers are therefore fully consistent with a $\Lambda$CDM  cosmology fit to the \textit{Planck} 2018 CMB power spectra, with no evidence for a lower amplitude of structure. 

It is also important to emphasise that the agreement of our late-time measurements with the structure growth predicted by the CMB power spectra also holds for CMB power spectra measured by other experiments (not just \textit{Planck}). 
Our lensing measurements are also consistent with a \LCDM~model fit to independent CMB power spectra measurements from ACT DR4 + \textit{WMAP} \citep{Aiola_2020}: \NS{fitting to a rescaling of the best-fit ACT DR4 + \textit{WMAP} $\Lambda$CDM model prediction yields} 
an amplitude of lensing of $A_{\mathrm{lens}}=1.005\pm0.023$. 
We shall explore the consequences of our measurements for structure growth further in the next section.

\section{Implications for structure growth}\label{sec.discussion}
\subsection{Likelihood}\label{sec:likelihood}
\fk{We obtain cosmological constraints  by constructing a lensing likelihood function} $\mathcal{L}$ assuming Gaussian errors on $\hat{C}^{\phi\phi}_L$ obtained from the MV estimator:
 \begin{equation}
     -2\ln{\mathcal{L}}\propto\sum_{bb^\prime}\big[\hat{C}^{\phi\phi}_{L_b}-{C}^{\phi\phi}_{L_b}(\boldsymbol{\theta})\big]{\mathbb{C}}^{-1}_{bb^\prime}\big[\hat{C}^{\phi\phi}_{L_{b^\prime}}-{C}^{\phi\phi}_{L_{b^\prime}}(\boldsymbol{\theta})\big],
 \end{equation}
where $\hat{C}^{\phi\phi}_{L_b}$ is the measured baseline lensing power spectrum, ${C}^{\phi\phi}_{L_b}(\boldsymbol{\theta})$ is the theory lensing spectrum evaluated for cosmological parameters $\boldsymbol{\theta}$, and ${\mathbb{C}}_{bb^\prime}$ is the baseline covariance matrix. We discussed the construction of the covariance matrix in Section~\ref{sec.covmat}, \abds{while verification of the Gaussianity of the lensing bandpowers can be found in Appendix~\ref{app:gaussianity}.}
\abds{Further corrections to this likelihood are applied when considering joint constraints with CMB power spectra, as described in our companion paper, \citet{dr6-lensing-cosmo}. These account for the dependence of the normalization of the lensing bandpowers on the true CMB power spectra and of the $N_1$ correction on both the true CMB and lensing power spectra. For the lensing-only constraints presented in this paper, we account for uncertainty in the CMB power spectra by sampling 1000 flat-$\Lambda$CDM CMB power spectra from $\text{ACT DR4} + \text{\textit{Planck}}$ and propagating these through the lensing normalization; the scatter in the normalization leads to an additional broadening of the bandpower covariance matrix. For further details, see our earlier discussion in Section~\ref{sub:norm.corr.summary} and also Appendix~\ref{app:norm.corr.details}.}

\subsection{Constraints on the amplitude of structure from lensing alone}

We now consider constraints on the basic
 $\Lambda$CDM  parameters --- cold dark matter and baryon densities, $\Omega_ch^2$ and $\Omega_bh^2$, the Hubble constant $H_0$, the optical depth to reionization $\tau$, and the amplitude and scalar spectral index of primordial fluctuations, $A_s$ and $n_s$ --- from our lensing measurements alone. These parameters are varied with priors as summarised in Table \ref{table:priors}; these are the same priors assumed in the most recent \textit{Planck} lensing analyses~\citep{Planck:2018,Carron_2022}. Since lensing is not sensitive to the CMB optical depth, we fix this at $\tau=0.055$ \citep{planck2015}. \fk{We fix the total neutrino mass to be consistent with the normal hierarchy, assuming one massive eigenstate with a mass of $60\,\si{meV}$.}

\begin{table}
\centering
\caption{Priors used in the lensing-only cosmological analysis of this work. Uniform priors are shown in square brackets and Gaussian priors with mean $\mu$ and standard deviation $\sigma$ are denoted $\mathcal{N}(\mu,\sigma)$. The priors adopted here
 are identical to those used in the lensing power spectrum analysis performed by the \textit{Planck} team~\citep{Planck:2018}. }
\begin{tabular}{cc}
\hline\hline
Parameter       & Prior      \\ \hline
$\ln (10^{10}A_s)$ & $[2,4]$           \\ 
$H_0$           & $[40,100]$        \\ 
$n_s$           & $\mathcal{N}(0.96,0.02)$     \\ 
$\Omega_bh^2$   & $\mathcal{N}(0.0223,0.0005)$ \\ 
$\Omega_ch^2$   & $[0.005,0.99]$    \\ 
$\tau$          & $0.055$           \\ \hline
\end{tabular}

\label{table:priors}
\end{table}

Weak lensing observables in cosmology depend on both the late-time amplitude of density fluctuations in terms of $\sigma_8$ 
and the matter density $\Omega_m$; there is an additional dependence on the Hubble parameter $H_0$.  
\begin{figure}
    \centering
    \includegraphics[width=\linewidth]{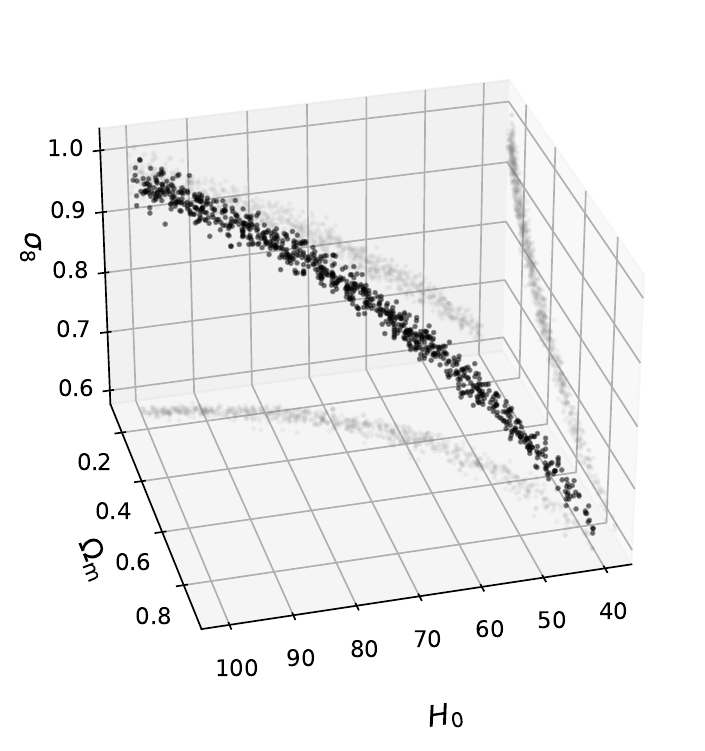}
    \caption{Posterior samples in $\sigma_8$--$H_0$--$\Omega_m$ space from our lensing-only likelihood, with the projection into 2D spaces shown with light grey points. As expected, our constraints form a degeneracy ``line'' in this parameter space. The $H_0$ units are $\si{km\,s^{-1}\,Mpc^{-1}}$. }
    \label{fig:sig8om_nobao3d}
\end{figure}
In Figure~\ref{fig:sig8om_nobao3d} we show the CMB-lensing-only constraints derived from our spectrum measurement; these follow, \abds{as in previous lensing analyses,} a narrow line in the space spanned by $\sigma_8$--$H_0$--$\Omega_m$. \abds{In our companion paper \cite{dr6-lensing-cosmo}, we argue that this line-shaped posterior arises because CMB lensing on large and small scales constrains two different combinations of $\sigma_8$--$H_0$--$\Omega_m$.
These two constraint planes intersect in a constraint ``line'', which explains the form of the posterior seen in the figure.} \fk{The DR6 lensing data provides the following constraint on this 3-dimensional $\sigma_8$--$H_0$--$\Omega_m$ parameter space:}
\begin{equation}
    \frac{\sigma_8}{0.8}\Big(\frac{\Omega_m}{0.3}\Big)^{0.23}\Big(\frac{\Omega_mh^2}{0.13}\Big)^{-0.32}=0.9938\pm0.0197
\end{equation}

\begin{figure}
  \includegraphics[width=\linewidth]{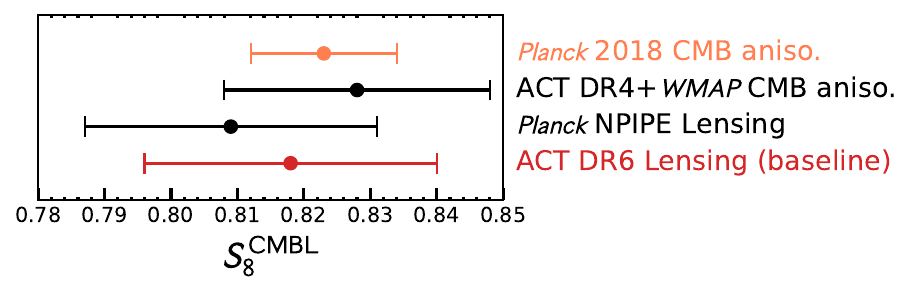}
\caption{Marginalised posteriors for $S^\mathrm{CMBL}_8$ from ACT DR6 lensing (red). We also show the constraints from \textit{Planck} \texttt{NPIPE} lensing and the early-universe extrapolation from CMB anisotropy measurements of ACT DR4 + \textit{WMAP}  and \textit{Planck} 2018.}
  \label{Fig.polcomb_posterior}
\end{figure}

This line-like degeneracy projects into constraints within a narrow region in the $\sigma_8$--$\Omega_m$ plane, as shown in Figure~\ref{fig:sig8om_nobao}. The best-constrained direction corresponds approximately to a determination of 
$\sigma_8\Omega^{0.25}_m$. Constraining this parameter combination with our data, we obtain 
\begin{equation}
\sigma_8\Omega^{0.25}_m=0.606\pm{0.016}.
\end{equation}
This translates to a constraint on the CMB-lensing-equivalent of the usual $S_8$ parameter, which we define as
\begin{equation}
    S^{\mathrm{CMBL}}_8 \equiv \sigma_8\Big(\frac{\Omega_m}{0.3}\Big)^{0.25},
\end{equation}
of
\begin{equation}
 S^{\mathrm{CMBL}}_8=0.818\pm{0.022}~~(0.830\pm{0.020}).
\end{equation}
\NS{In the constraint shown above, the result for the baseline analysis is shown first,} followed by the constraint from the extended range of scales in parentheses. These can be compared with the value expected from \textit{Planck} CMB power spectrum measurements \fk{assuming a \LCDM~cosmology}. Extrapolating the \textit{Planck} CMB anisotropy measurements to low redshifts yields a value of $ S^{\mathrm{CMBL}}_8=0.823\pm{0.011}$.
This is entirely consistent with our direct ACT DR6 lensing measurement of this parameter. As in the case of $A_\mathrm{lens}$, this agreement is not limited to comparisons with \textit{Planck}; similar levels of agreement are achieved with ACT DR4 + \textit{WMAP} CMB power spectrum measurements, which give a constraint of  $S^{\mathrm{CMBL}}_8=0.828\pm{0.020}$. Our measurement is also consistent with the direct $S^{\mathrm{CMBL}}_8$ result from \texttt{NPIPE} lensing, $S^{\mathrm{CMBL}}_8=0.809\pm0.022$. \abds{This consistency of $S^{\mathrm{CMBL}}_8$ between extrapolations from CMB power spectra, which probe primarily $z\sim 1100$, and direct measurements with CMB lensing at lower redshifts $z\sim 0.5$--$5$} can be seen in Figure~\ref{Fig.polcomb_posterior}.

\begin{figure}
  \includegraphics[width=\linewidth]{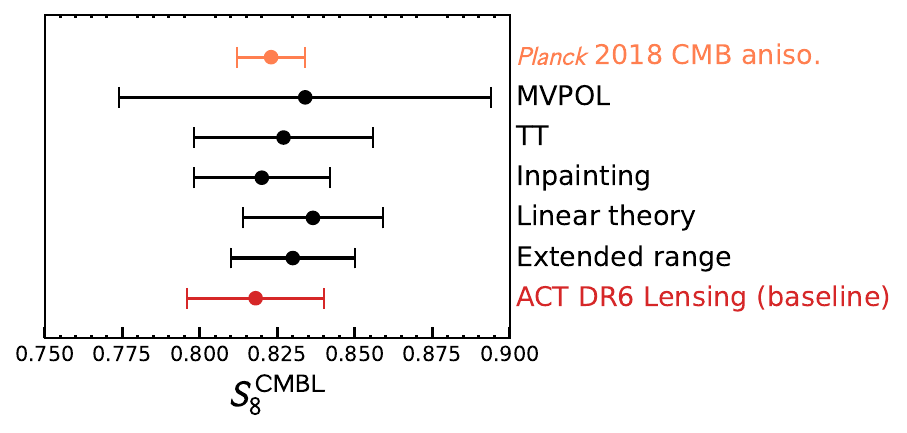}
\caption{Marginalized posteriors for $S^\mathrm{CMBL}_8$ using variations of our ACT DR6 analysis choices and datasets (black). Moving from bottom to top, we first show results obtained using the extended multipole range up to $L=1300$, and a constraint obtained using linear theory only in the model predictions. The small shift for the latter compared to the baseline (given in red) shows that we probe scales where details of non-linear physics are not so important.
Next, the result from the SZ-inpainting method, which we used before unblinding, is shown; this produces a very small shift compared to our post-unblinding method of template subtraction. Finally, variations using only temperature data (TT) and polarization-only data (MVPOL) are shown to be consistent with the baseline, further demonstrating that foregrounds are well mitigated. For comparison, we also show the early-universe extrapolation for $S^\mathrm{CMBL}_8$ from the \textit{Planck} 2018 CMB power spectrum measurements (orange).} 
  \label{Fig.s8_consistency}
\end{figure}

 \fk{Our constraint on $S^{\mathrm{CMBL}}_8$ is robust to the details of our analysis choices and datasets. In Figure~\ref{Fig.s8_consistency}, we present the marginalized posteriors of $S^{\mathrm{CMBL}}_8$ obtained using different variations of our analysis. Since levels of extragalactic foregrounds are significantly lower in polarization than in temperature, the consistency between our baseline analysis and the analyses using temperature data and polarization data alone suggests that foreground contamination is under control. While our baseline analysis incorporates the modeling of non-linear scales using the non-linear matter power spectrum prescription of \cite{Mead2016}, 
 we also present constraints that use linear theory only. The consistency of this result with our baseline shows that we are mainly sensitive to linear scales. Finally, we present constraints using our pre-unblinding method of inpainting clusters instead of masking, as discussed in Section~\ref{sec:post-unblinding}. This method only results in a shift of $0.15\sigma$ in $S^{\mathrm{CMBL}}_8$.}

\begin{figure}
    \centering
    \includegraphics[width=\linewidth]{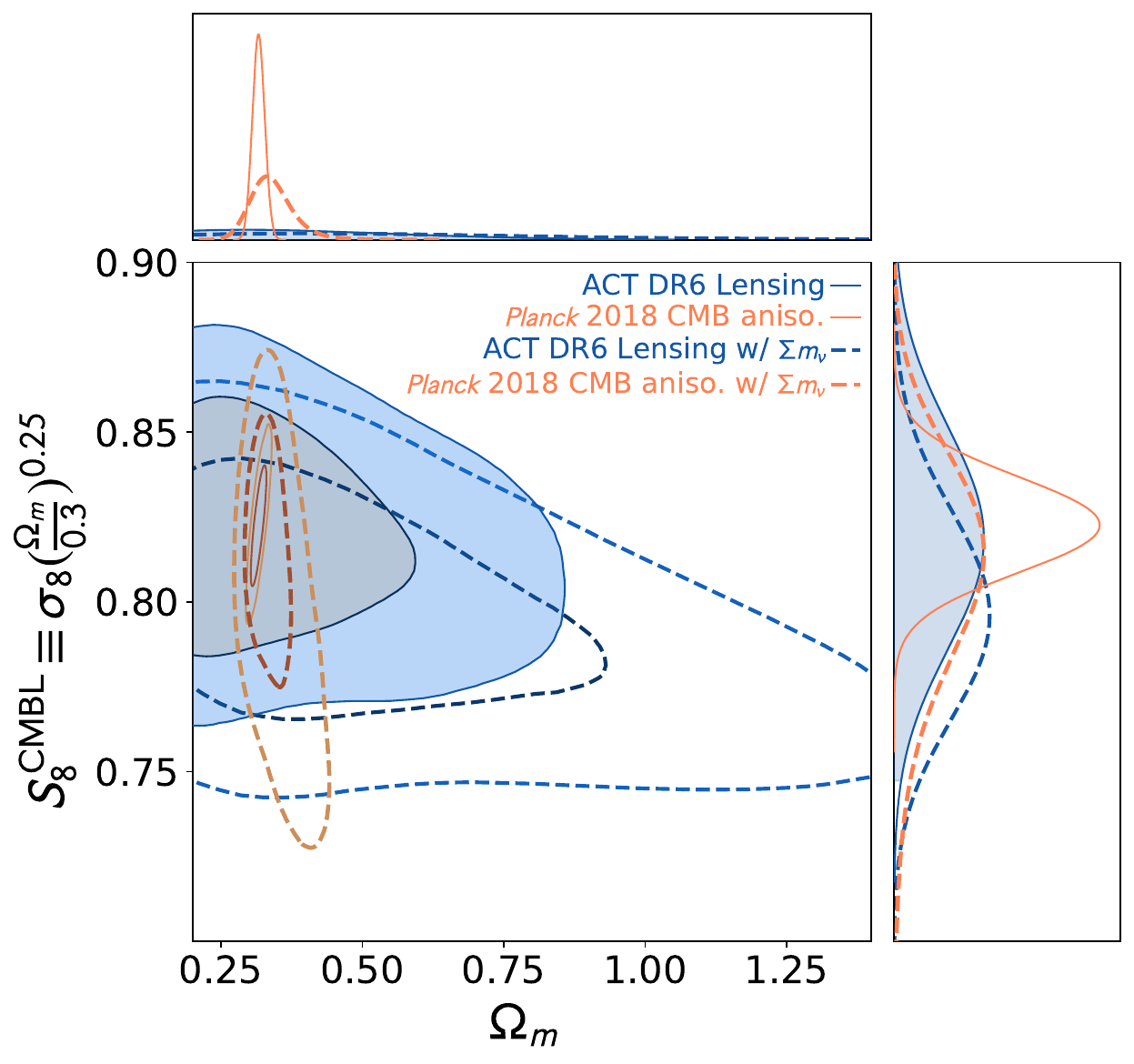}
    \caption{\Frank{Marginalised constraint on the parameters $S^{\mathrm{CMBL}}_8=\sigma_8(\Omega_m/0.3)^{0.25}$ and $\Omega_m$ from ACT DR6 CMB lensing only (blue) and \textit{Planck} 2018 CMB anisotropies (orange) with the sum of the neutrino masses allowed to vary (open, dashed contours) or fixed (filled contours). The marginalised contours show the $68\%$ and the $95\%$ confidence levels. The top and side panels show the corresponding 1D marginalized constraints.} \blake{There is a lot of white space in the top panel -- perhaps plot scaled versions of the \textit{Planck} posteriors. Also, in the 1D panels, the neutrino-freed constraints also include some shading -- I would remove the shading in this case, and for the case of fixed neutrino mass show the full posterior rather than some trucated version (68\%?). Finally, upright font for ``CMBL'' in the $y$-axis label and, possibly, horizontal ($x$-axis) and vertical ($y$-axis) numerals.}}
    \label{fig:free_mnu}
\end{figure}

\abds{In Figure~\ref{fig:free_mnu} we also show constraints on $S^{\mathrm{CMBL}}_8$ (and $\Omega_m$) with the sum of neutrino masses freed and marginalized over, instead of being fixed at the minimum allowed mass set by the normal hierarchy, $60\,\si{meV}$. \footnote{Following the arguments in \cite{LESGOURGUES_2006} and \cite{1612.00021}, here we consider a degenerate combination of three equally massive neutrinos.}
In this case, our constraint of $S^{\mathrm{CMBL}}_8=0.797\pm0.024$ become tighter than constraints from \textit{Planck} 2018 CMB power spectra,  $S^{\mathrm{CMBL}}_8=0.811\pm0.027$. The fact that our constraints are only slightly weakened when marginalizing over a free neutrino mass, whereas \textit{Planck} constraints on $S^{\mathrm{CMBL}}_8$ are significantly degraded, is expected. Since our lensing results originate from low redshifts, minimal extrapolation to $z=0$ (where $S^{\mathrm{CMBL}}_8$ is evaluated) is required, which makes our constraints comparatively insensitive to neutrino mass. In constrast, for CMB power spectrum constraints, extrapolation over a wide redshift range from $z\sim 1100$ to $z=0$ is required, which implies that the constraints have significant sensitivity to neutrino mass.}

\subsection{Constraints on the amplitude of structure from lensing alone: ACT+\textit{Planck} NPIPE}

\bds{The DR6 CMB lensing measurement from the ground contains information that is, to some extent, independent from the space-based measurement obtained with the \textit{Planck} satellite. The two measurements have different noise and different instrument-related systematics; their sky coverage and their angular scales also only have partial overlap\footnote{As previously  mentioned, ACT uses only CMB multipoles $600<\ell<3000$, whereas \textit{Planck} analyses $100<\ell<2048$.} over a sky fraction of $67\%$} \fk{used in the Planck analysis.}

\begin{figure}
    \centering
    \includegraphics[width=\linewidth]{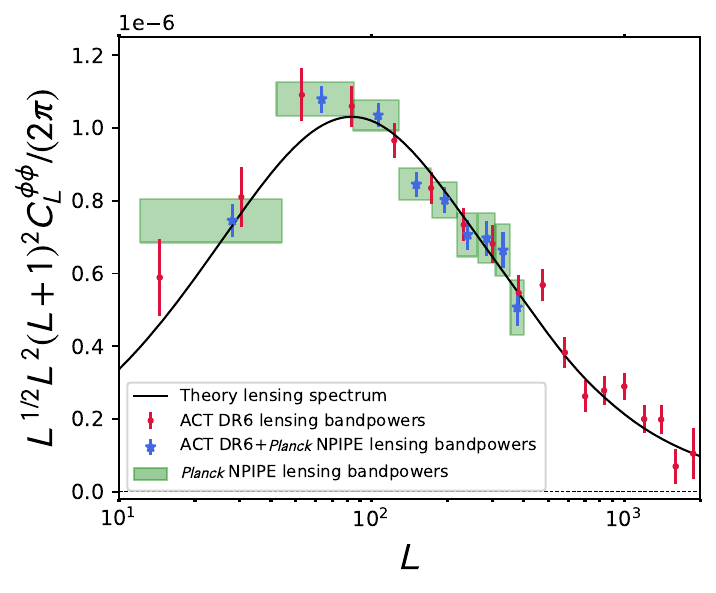}
    \caption{Lensing bandpowers for ACT DR6 (red), the \textit{Planck} \texttt{NPIPE} lensing bandpowers (green boxes) \bds{in the conservative range of scales} and the inverse-variance-weighted combination of both (blue). \bds{We note that, unlike for \textit{Planck} points, ACT bandpowers are shown over the entire observed range.} The excellent consistency between the two measurements provides motivation for combining the lensing measurements at the likelihood level and inferring joint parameter constraints.  \blake{check}}
    \label{fig:planck_bandpowers}
\end{figure}

This \abds{partial} independence motivates not just comparing the two measurements, but, given that we find good consistency between the two, combining them to obtain tighter constraints.
The excellent agreement between both measurements can be seen in Figure~\ref{fig:planck_bandpowers}, where for illustrative purposes we show also in blue the combined (with an inverse-variance weight) lensing bandpowers from ACT DR6 and \textit{Planck}.\footnote{\fk{We re-bin the ACT DR6 bandpowers using the same binning used by \texttt{NPIPE} and perform a inverse variance weighted coadd using the ACT and Planck covariance matrices. $C^{\mathrm{combined}}_L=\Big[\sum_i\mathbb{C}_i^{-1}\big]^{-1}\sum_i\mathbb{C}_i^{-1}C^{i}_L, \quad i\in\{\mathrm{ACTxACT},\mathrm{PlanckxPlanck}\}$
}} 

In this paper, we combine the two lensing spectra at the likelihood level, taking into account the small correlation between the two datasets in order to obtain further improved constraints on $S_8$.
 For the \texttt{NPIPE} lensing measurements, we use the published \texttt{NPIPE} lensing bandpowers.\footnote{\url{https://github.com/carronj/planck_PR4_lensing}} \bds{The \texttt{NPIPE} part of the covariance matrix is obtained using a set of 480 \texttt{NPIPE} lensing reconstructions.}\footnote{As a test, we replace the ACT DR6 lensing bandpowers and covariance matrix in our likelihood with those from \texttt{NPIPE}.
We recover $\sigma_8\Omega^{0.25}_m=0.590\pm0.019$, compared to the  \texttt{NPIPE} constraint of $\sigma_8\Omega^{0.25}_m=0.599\pm0.016$. The difference in errors arises mainly from the normalization marginalization step we perform (Appendix~\ref{app:norm.corr.details}) and the shift in central value is explained by the use of the measured \textit{Planck} power spectrum in the \texttt{NPIPE} normalization (see, for example, Section~3.2.1 of~\citealt{Planck:2018}).}

\subsubsection{Covariance matrix between ACT and \textit{Planck} lensing spectra}\label{sec.covplanck}

Although we expect the ACT and \textit{Planck} datasets to have substantially independent information, the sky and scale overlap are large enough that we cannot, without further investigation, neglect the correlation between these two datasets. We, therefore,  compute the joint covariance, proceeding as follows. We start with the same set of 480 full-sky FFP10 CMB simulations used by \texttt{NPIPE} to obtain the \textit{Planck} part of the covariance matrix.
We apply the appropriate ACT DR6 mask and obtain lensing reconstructions using the standard (not cross-correlation-based\footnote{\abds{This is justified because the ACT-\textit{Planck}-covariance does not depend on instrument or atmospheric noise in ACT.}}) estimator with a filter that has the same noise level as the DR6 analysis. We do not add noise to the \abds{ACT CMB maps as we expect the instrument noise of \textit{Planck} and ACT to be entirely independent; the ACT noise should hence not enter the ACT-\textit{Planck}-covariance (when using the cross-correlation-based estimator for ACT)}. We use these 480 reconstructed ACT bandpowers along with the corresponding \texttt{NPIPE} lensing bandpowers to obtain the off-diagonal ACT-\textit{Planck} elements of the joint bandpower covariance matrix for the two datasets. 

\begin{figure}
    \centering
    \includegraphics[width=\columnwidth]{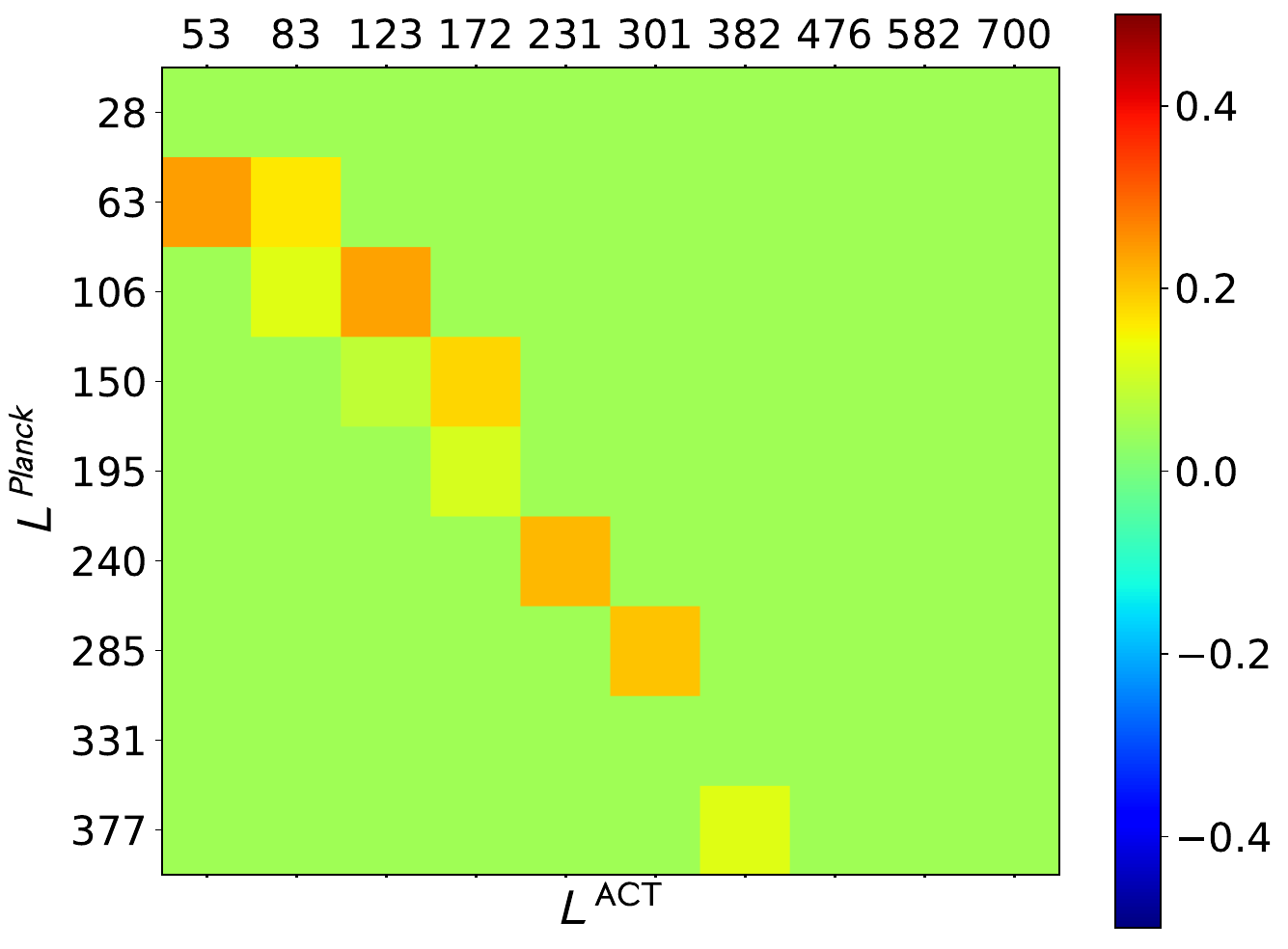}
    \caption{Visualization of the $\text{ACT lensing}\times\text{\textit{Planck} lensing}$ part of the covariance matrix for the joint dataset: we plot the correlation coefficient of the ACT bandpowers (horizontal axis) and the \textit{Planck} \texttt{NPIPE} bandpowers computed from a set of 480 FFP10 simulations. As argued in Section \ref{sec.covplanck}, correlations between disjoint bandpowers can be neglected in the $\text{ACT}\times\text{\textit{Planck}}$ covariance; they are therefore set to zero to improve convergence with a finite number of simulations.}
    \label{fig:combcov}
\end{figure}

The correlation between the two datasets is small, as we will show, but the computation of this larger covariance matrix causes challenges in ensuring convergence with a modest number of simulations. To reduce the impact of fluctuations and ensure convergence of the resulting covariance matrix, we use a criterion that allows us to simulate only the most correlated bandpowers. This is done by only simulating the terms in the covariance matrix that we expect to be significantly different from zero, and nulling all other elements. \bds{To determine which elements can be safely nulled}, we compute the covariance matrix between ACT and \textit{Planck} in an unrealistic scenario of maximum overlap in area and multipole by analysing a set of 396 ACT CMB simulations with the same ACT analysis mask and multipole range $600<\ell<3000$ (instead of \Frank{the \texttt{NPIPE} analysis mask covering $67\%$ of the sky and multipole range} $100<\ell<2048$) but with a filter, \Frank{used for the inverse-variance and Wiener filtering of the CMB maps,} with noise levels appropriate for \texttt{NPIPE}. The resulting covariance matrix only shows a significant correlation among the bins where the ACT and \texttt{NPIPE} bandpowers overlap in multipole $L$. (This is unsurprising because, even within ACT, different-$L$ bandpowers are only minimally correlated, and we expect even smaller correlations between \textit{Planck} and ACT bandpowers at a different $L$). From this maximally overlapping covariance matrix we choose the elements where the absolute value of the correlation between the bandpowers is less than $0.15$ \blake{Technical point, but when measuring the \emph{correlation} do you include noise in the simulations for the $\textit{Planck}\times \textit{Planck}$ part?} 
 and zero these elements in the final $\text{ACT}\times\text{\textit{Planck}}$ covariance matrix, as these are likely due to noise fluctuations rather than physical correlation between the bandpowers.\footnote{We verified that the central value of the posterior for $S^{\mathrm{CMBL}}_8$ changes by a negligible amount when the level of this threshold is varied.}
The resulting covariance matrix is shown in Figure~\ref{fig:combcov}. 

\abds{Finally, we broaden the covariance matrix to account for marginalization over the uncertainties in the CMB lensing power spectra, as described earlier for the ACT part in Section~\ref{sec:likelihood}. This step is applied consistently to both the ACT and the \texttt{NPIPE} parts of the covariance matrix and is described in detail in Appendix \ref{app:norm.corr.details}.}

\begin{figure*}
    \centering
    \includegraphics[width=0.8\linewidth]{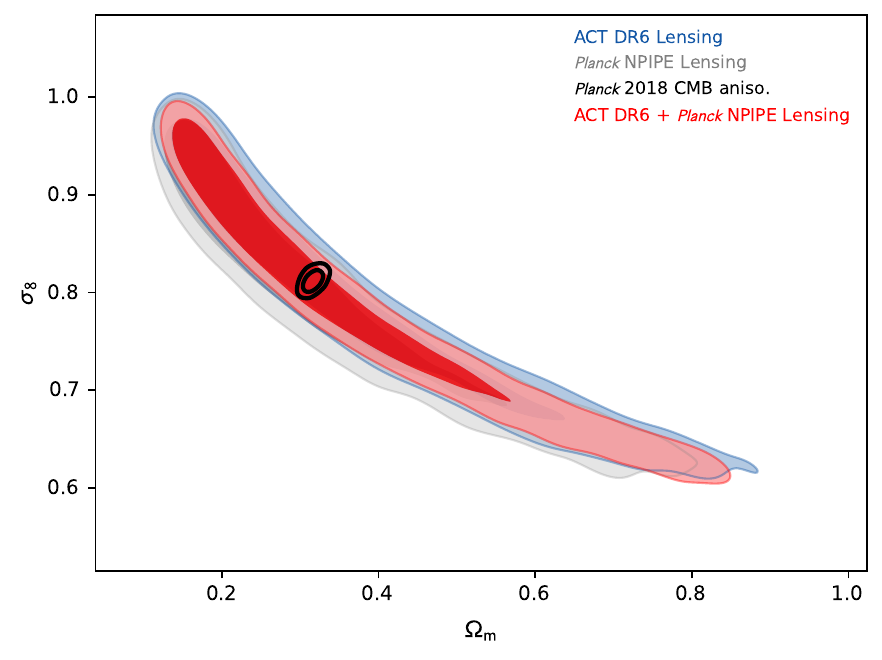}
    \caption{Constraints in the $\sigma_8$--$\Omega_m$ plane obtained by combining the ACT and \textit{Planck} CMB lensing power spectrum measurements at the likelihood level, including an appropriate correlation between the ACT and \textit{Planck} bandpowers (red). Posteriors arising from our baseline ACT DR6 lensing spectrum alone and from the \textit{Planck} \texttt{NPIPE} lensing spectrum alone are shown in blue and grey, respectively. It can again be seen that the  results are in good agreement with the prediction in the $\Lambda$CDM model from \textit{Planck} 2018 CMB power spectrum measurements (black).}
    \label{fig:sig8om_nobao_combined}
\end{figure*}

The parameter constraints on $\sigma_8$ and $\Omega_m$ derived from the combination of ACT and \textit{Planck} lensing are shown in Figure~\ref{fig:sig8om_nobao_combined}. As expected, the joint constraint in red is in good agreement with the ACT DR6-only constraint and with the \textit{Planck} CMB power spectrum prediction in black.

\begin{figure}
    \centering
    \includegraphics[width=\linewidth]{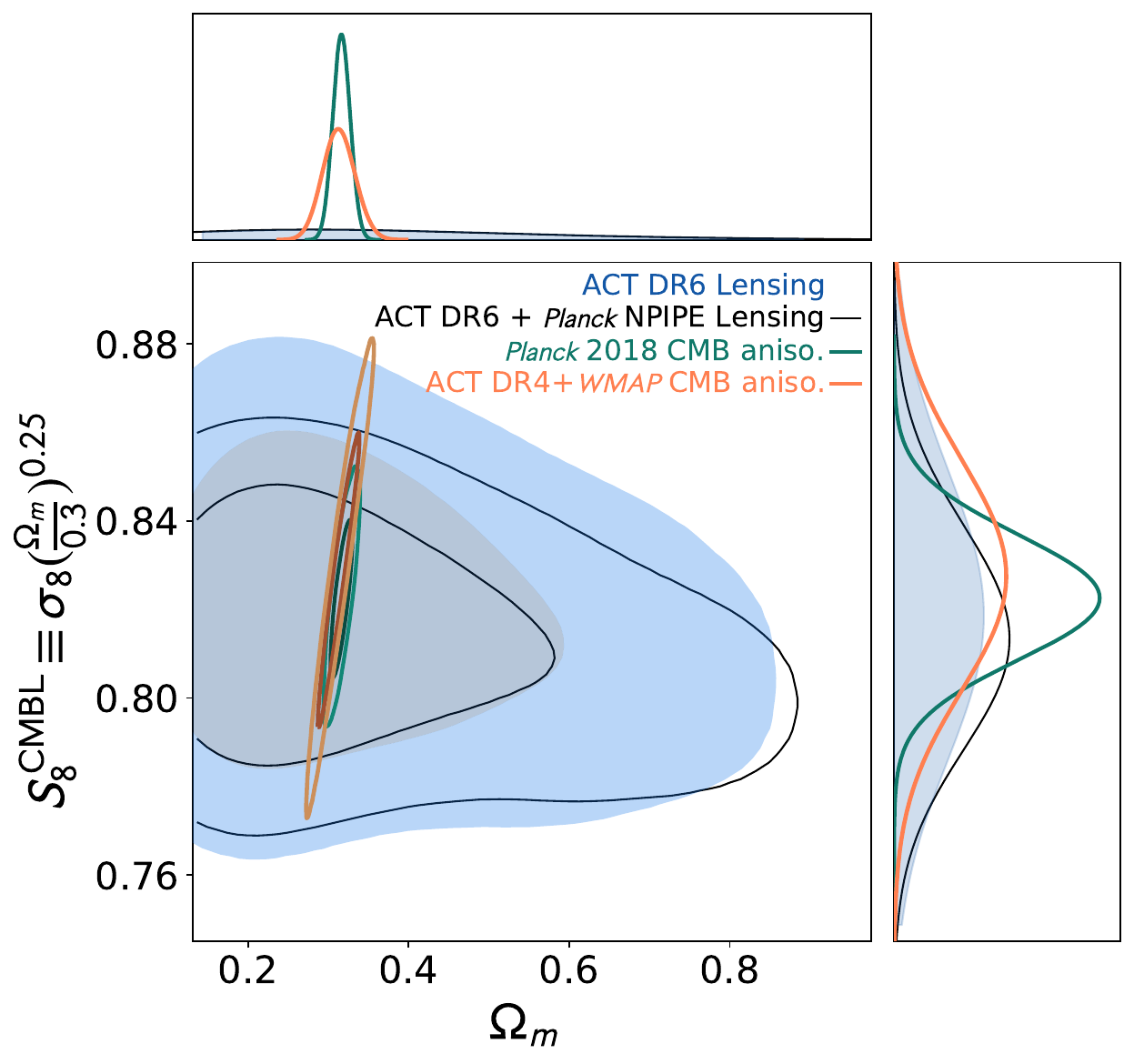}
    \caption{Marginalised constraint on the parameters $S^{\mathrm{CMBL}}_8=\sigma_8(\Omega_m/0.3)^{0.25}$ and $\Omega_m$ from ACT DR6 CMB lensing only (blue), ACT DR6 combined with \textit{Planck} \texttt{NPIPE} lensing (black),   \textit{Planck} 2018 CMB anisotropies (green) and $\text{ACT DR4}+\text{\textit{WMAP}}$ CMB anisotropies (orange). The marginalised contours show the $68\%$ and the $95\%$ confidence levels. The top and side panels show the corresponding 1D marginalized constraints. \blake{Similar comments about shading in the 1D marginals, etc.}}
    \label{fig:sig8om_nobao_combinedS8}
\end{figure}

The joint $\text{ACT DR6}+\text{\textit{Planck}}$ lensing constraints in the $S^{\mathrm{CMBL}}_8$--$\Omega_m$ plane are shown in Figure~\ref{fig:sig8om_nobao_combinedS8}.
We obtain the following constraint on $S^{\mathrm{CMBL}}_8$ from the ACT--\textit{Planck} combined dataset:
\begin{equation}
S^{\mathrm{CMBL}}_8=0.813\pm{0.018} ~~(0.822\pm0.017),
\end{equation}
where the constraint in parentheses is from the extended lensing multipole range.\footnote{\fk{The shift in posteriors from the baseline to the extended range, despite the small change in errors, could indicate that NPIPE lensing bandpowers are more consistent with the baseline ACT DR6 range. Moreover, since the posteriors of the extended range are not well-described by a Gaussian distribution, it is difficult to make a direct comparison between the baseline and extended posteriors. Refer to Appendix \ref{app.posteriors} for the full triangle plot of the posteriors.}}

\subsection{Discussion of results}

Our lensing power spectrum bandpowers are consistent with the $\Lambda$CDM prediction over a range of scales. Good consistency with $\Lambda$CDM is found even in the extended range of scales where foregrounds and non-linear structure growth could be more relevant.

As previously discussed, our results are highly relevant for the $S_8$ tension. A powerful test of structure formation is to extrapolate, assuming $\Lambda$CDM structure growth but no other free parameters, a model fit to the \textit{Planck} CMB power spectrum at (mostly) early times down to low redshifts, and then compare this extrapolation with direct measurements of $S^{\text{CMBL}}_8 \equiv \sigma_8(\Omega_m/0.3)^{0.25}$ or $\sigma_8$. Intriguingly,  some lensing and galaxy clustering measurements seem to give lower values of $S_8$ 
 or $\sigma_8$ than predicted by \textit{Planck} \fk{at the $(2$--$3)\sigma$ level}, including KiDS, DES and HSC \citep{10.1093/mnras/stt601,KiDS:2020suj,Heymans_2021,Krolewski:2021yqy,Philcox:2021kcw,PhysRevD.105.023520,Loureiro_2022,HSCY3Real,HSCY3Fourier}, although this conclusion is not universal \citep{eBOSS2021}. Our measurement of lensing and structure growth is independent of \textit{Planck} and also does not rely at all on galaxy survey data, with the associated challenges in modeling and systematics mitigation. 

As described in the previous section, we find $S^{\text{CMBL}}_8=0.818\pm{0.022}$ as our baseline result from ACT. This is in full agreement with the expectation based on the \textit{Planck} CMB-anisotropy power spectra and $\Lambda$CDM structure growth. We emphasize that the agreement between precise lensing measurements at $z\sim 0.5-5$ with predictions from CMB anisotropies probing primarily $z\sim 1100$ is a remarkable success for the $\Lambda$CDM model. \abds{From a fit of $\Lambda$CDM at the CMB last-scattering surface, our standard cosmology predicts, with no additional free parameters, cosmic structure formation over billions of years, the lensing effect this produces, and the non-Gaussian imprint of lensing in the observed CMB fluctuations; our measurements match these predictions at the $2\%$ level.}

Unlike several galaxy lensing measurements and other large-scale structure probes, we find no evidence of any suppression of the amplitude of structure.

Our lensing power spectrum and the resulting parameters also agree with the results of the \textit{Planck} lensing power spectrum measurement. Since both ACT lensing power spectra and CMB power spectra are in good agreement with \textit{Planck} measurements, this disfavors systematics in the \textit{Planck} measurement as an explanation of the $S_8$ tension. This also means that we may combine with the \textit{Planck} lensing measurement to obtain an even more constraining measurement $S_8^{\mathrm{CMBL}}=0.813\pm{0.018}$, again in agreement with the \textit{Planck} or $\text{ACT DR4} + \text{\textit{WMAP}}$ CMB power spectra.

We note that while we appear to find  a higher value of $S_8^{\mathrm{CMBL}}$ than the $S_8$ measured by several weak lensing surveys, the two quantities differ in the power of $\Omega_m$ and can hence, in principle, be brought into better agreement by a lower value of $\Omega_m<0.3$ (although the matter density is very well constrained by other probes); in any case, the discrepancy does not reach high statistical significance. Our companion paper \citep{dr6-lensing-cosmo} discusses this comparison in more detail by combining these datasets with baryon acoustic oscillation (BAO) data to constrain $\Omega_m$ to $\approx 0.3$. 

We expect the agreement of our lensing power spectrum with the \textit{Planck} CMB power spectrum extrapolation to disfavor resolutions to the $S_8$ tension that involve new physics having a substantial effect at high redshifts ($z>1$) and on linear scales, where our lensing measurement has highest sensitivity.

However, this need not imply that statistical fluctuations or systematics in other lensing and LSS measurements are the only explanation for the \Frank{reported} tensions. The possibility remains that new physics suppresses structure only at very low redshifts, $z<1$, or on 
non-linear scales, $k>0.2h\,\si{Mpc}^{-1}$, to which our CMB lensing measurements are much less sensitive than current cosmic shear, galaxy-galaxy lensing, or galaxy clustering constraints. An example of such physics could be the small-scale matter power spectrum suppression proposed in \citet{Amon_2022} or \citet{he2023s8} 
(along with systematics in CMB lensing cross-correlation analyses); modified gravity effects that become important at only very low redshifts are another example, although these need to be consistent with expansion and redshift-space distortions (RSD) measurements.

\bds{More information on possible structure growth tensions can be obtained from ACT lensing measurements not just through combination with external data as in our companion paper, but also by cross-correlations with low-redshift tracers. Such cross-correlations will allow structure growth to be tracked tomographically as a function of redshift down to $z<1$, providing for powerful tests of new physics. Several tomographic cross-correlation analyses with the lensing data presented here are currently in progress.} 

\section{Conclusions}\label{sec:conclusion}


In this paper, we report a new measurement of the CMB lensing power spectrum using ACT data collected from 2017 to 2021. Our lensing measurement spans $9400\,\si{deg}^2$  of sky and is signal-dominated up to multipoles $L=150$. 

\bds{The CMB lensing power spectrum is determined using a novel pipeline that uses profile-hardening and cross-correlation-based estimators; these methods ensure that we are insensitive to the foregrounds and complex noise structure in high-resolution, large-sky, ground-based CMB observations.} We obtain a lensing spectrum at state-of-the-art precision: the amplitude of the spectrum is measured to $2.3\%$ precision ($43\sigma$ significance) over the baseline range of scales $40<L<763$, with very similar results (\Frank{also $43\sigma$})  obtained over the extended range $40<L<1300$. We test the robustness and internal consistency of our measurement using more than 200 null and systematic tests performed at both the bandpower and map level: these include tests of consistency of different array-frequencies, sky regions, scales, and of temperature and polarization data. \fk{We find no evidence for any systematic effects that could significantly bias our measurement.}

Our CMB lensing power spectrum measurement provides constraints on the amplitude of cosmic structure that do not depend on \textit{Planck} or galaxy-survey data, thus giving independent information about large-scale structure growth and further insight into the $S_8$ tension. We find that our lensing power spectrum is well fit by a standard $\Lambda$CDM
 model, with a lensing amplitude $A_{\mathrm{lens}}=1.013\pm0.023$ relative to the \textit{Planck} 2018 best-fit model for our baseline analysis ($A_{\mathrm{lens}}=1.016\pm 0.023$ for the extended range of scales). From our baseline lensing power spectrum measurement, we derive constraints on the best-determined parameter combination $S^{\mathrm{CMBL}}_8 \equiv \sigma_8 \left({\Omega_m}/{0.3}\right)^{0.25}$ of $S^{\mathrm{CMBL}}_8= 0.818\pm0.022$ from ACT CMB lensing alone. Since our spectrum shows good consistency with the \textit{Planck} CMB lensing spectrum, we also combine ACT and \textit{Planck} CMB lensing to obtain $S^{\mathrm{CMBL}}_8= 0.813\pm0.018$.

Our results are fully consistent with the predictions based on \textit{Planck} CMB power spectrum measurements and standard $\Lambda$CDM structure growth\abds{; we find no evidence for suppression of structure at low redshifts or for other tensions. Our companion paper \citep{dr6-lensing-cosmo} combines our measurements with BAO data to study the agreement with other lensing surveys.} Since our measurement is less sensitive to new physics at low redshifts $z<1$ and on non-linear scales, further investigation of structure growth at lower redshifts and smaller scales, for example with upcoming ACT lensing cross-correlation analyses, is well motivated.

The lensing pipeline presented here provides a foundation for high-resolution, ground-based lensing measurements covering a significant portion of the sky. This framework will be used for ongoing analyses of ACT data incorporating day-time observations from 2017--2022 as well as night-time data recorded in 2022. \bds{Moreover, the analysis presented here demonstrates a preliminary pipeline that can be used for Simons Observatory \citep{1808.07445} in the near future.}

\section*{Acknowledgements}

We are grateful to
Giulio Fabbian, Antony Lewis and Kendrick Smith for useful discussions. Some of the results in this paper have been derived using the healpy~\cite{Zonca2019} and HEALPix~\cite{2005ApJ...622..759G} packages. This research made use of Astropy,\footnote{http://www.astropy.org} a community-developed core Python package for Astronomy \citep{astropy:2013, astropy:2018,astropy:2022}. We also acknowledge use of the matplotlib~\cite{Hunter:2007} package and the Python Image Library for producing plots in this paper, and use of the Boltzmann code CAMB~\cite{CAMB} for calculating theory spectra and use of the \texttt{GetDist} \citep{1910.13970}, \texttt{Cobaya} \citep{2005.05290} for sampling and likelihood analysis, ChainConsumer \citep{Hinton2016} for producing some of the cosmology plots of this paper and Cosmopower \citep{SpurioMancini2022} for exploratory cosmology runs.

Support for ACT was through the U.S.~National Science Foundation through awards AST-0408698, AST-0965625, and AST-1440226 for the ACT project, as well as awards PHY-0355328, PHY-0855887 and PHY-1214379. Funding was also provided by Princeton University, the University of Pennsylvania, and a Canada Foundation for Innovation (CFI) award to UBC. ACT operated in the Parque Astron\'omico Atacama in northern Chile under the auspices of the Agencia Nacional de Investigaci\'on y Desarrollo (ANID). The development of multichroic detectors and lenses was supported by NASA grants NNX13AE56G and NNX14AB58G. Detector research at NIST was supported by the NIST Innovations in Measurement Science program.

Computing was performed using the Princeton Research Computing resources at Princeton University, the Niagara supercomputer at the SciNet HPC Consortium and the Symmetry cluster at the Perimeter Institute. SciNet is funded by the CFI under the auspices of Compute
Canada, the Government of Ontario, the Ontario Research Fund–Research Excellence, and the University of Toronto. Research at Perimeter Institute is supported in part by the Government of Canada through the Department of Innovation, Science and Industry Canada and by the Province of Ontario through the Ministry of Colleges and Universities. This research also used resources of the National Energy Research Scientific Computing Center (NERSC), a U.S. Department of Energy Office of Science User Facility located at Lawrence Berkeley National Laboratory, operated under Contract No. DE-AC02-05CH11231 using NERSC award HEP-ERCAPmp107. This manuscript has been authored by Fermi Research Alliance, LLC under Contract No. DE-AC02-07CH11359 with the U.S. Department of Energy, Office of Science, Office of High Energy Physics. Work supported by the Fermi National Accelerator Laboratory, managed and operated by Fermi Research Alliance, LLC under Contract No. DE-AC02-07CH11359 with the U.S. Department of Energy. The U.S. Government retains and the publisher, by accepting the article for publication, acknowledges that the U.S. Government retains a non-exclusive, paid-up, irrevocable, world-wide license to publish or reproduce the published form of this manuscript, or allow others to do so, for U.S. Government purposes. This document was prepared by ACT using the resources of the Fermi National Accelerator Laboratory (Fermilab), a U.S. Department of Energy, Office of Science, HEP User Facility. Fermilab is managed by Fermi Research Alliance, LLC (FRA), acting under Contract No. DE-AC02-07CH11359.

BDS, FJQ, BB, IAC, GSF, NM, DH acknowledge support from the European Research Council (ERC) under the European Union’s Horizon 2020 research and innovation programme (Grant agreement No. 851274). BDS further acknowledges support from an STFC Ernest Rutherford Fellowship. FJQ further acknowledges the support from a Cambridge Trust international scholarship. MM, AL acknowledge support from NASA grant 21-ATP21-0145. EC, BB, IH, HTJ acknowledge support from the European Research Council (ERC) under the European Union’s Horizon 2020 research and innovation programme (Grant agreement No. 849169). JCH acknowledges support from NSF grant AST-2108536, NASA grants 21-ATP21-0129 and 22-ADAP22-0145, DOE grant DE-SC00233966, the Sloan Foundation, and the Simons Foundation. CS acknowledges support from the Agencia Nacional de Investigaci\'on y Desarrollo (ANID) through FONDECYT grant no.\ 11191125 and BASAL project FB210003. RD acknowledges support from ANID BASAL project FB210003. ADH acknowledges support from the Sutton Family Chair in Science, Christianity and Cultures and from the Faculty of Arts and Science, University of Toronto. JD, ZA and ES acknowledge support from NSF grant AST-2108126. KM acknowledges support from the National Research Foundation of South Africa. AM and NS acknowledge support from NSF award number AST-1907657. IAC acknowledges support from Fundaci\'on Mauricio y Carlota Botton. LP acknowledges support from the Misrahi and Wilkinson funds. MHi acknowledges support from the National Research Foundation of South Africa (grant no. 137975). SN acknowledges support from a grant from the Simons Foundation (CCA 918271, PBL). CHC acknowledges FONDECYT Postdoc fellowship 322025. AC acknowledges support from the STFC (grant numbers ST/N000927/1, ST/S000623/1 and ST/X006387/1). RD acknowledges support from the NSF Graduate Research Fellowship Program under Grant No.\ DGE-2039656. OD acknowledges support from SNSF Eccellenza Professorial Fellowship (No. 186879). CS acknowledges support from the Agencia Nacional de Investigaci\'on y Desarrollo (ANID) through FONDECYT grant no.\ 11191125 and BASAL project FB210003. TN acknowledges support from JSPS KAKENHI (Grant No.\ JP20H05859 and No.\ JP22K03682) and World Premier International Research Center Initiative (WPI), MEXT, Japan.

\bibliography{refs}
\bibliographystyle{aasjournal}

\appendix

\section{Self-calibration of polarization efficiencies: Fitting of Amplitude Scaling}\label{app:polfit}

In this appendix, we describe the \DW{array-frequency} fitting for a single amplitude scaling, \DW{ $p^{\textrm{A}_\textrm{f}}_{\rm eff}$}, which relates our data polarization power spectra to the fiducial model assumed for the normalization of the lensing estimator.  We define the following $\chi^{2}$:
\ba
\chi^{2}(p^{\textrm{A}_\textrm{f}}_{\rm eff}, \alpha_{\textrm{A}_\textrm{f}})=  \sum_{\ell_b = \ell^{\rm min}_b}^{ \ell^{\rm max}_b}\sum_{\ell^{'}_b = \ell^{\rm min}_b}^{ \ell^{\rm max}_b}     \Delta^{\textrm{A}_\textrm{f}}_{\ell_b}\left[\Sigma^{\textrm{A}_\textrm{f}, \rm EE}\right]^{-1}_{\ell_b, \ell_b'} \Delta^{\textrm{A}_\textrm{f}}_{\ell_b'},
\ea 
where $\Sigma^{\textrm{A}_\textrm{f}, \rm EE}_{\ell_b, \ell_b'}$ is \DW{an analytic covariance matrix including the cosmic variance and the noise variance measured from the ACT data}. 
The difference spectrum $\Delta^{\textrm{A}_\textrm{f}}_{\ell_b}$ is given by the difference of a model for the spectrum and the data:
\begin{equation}
    \Delta^{\textrm{A}_\textrm{f}}_{\ell_b}=(C^{\rm CMB, EE}_{\ell_b} + \alpha_{\textrm{A}_\textrm{f}} C^{\rm foreground, EE, \textrm{A}_\textrm{f}}_{\ell_b}) (p^{\textrm{A}_\textrm{f}}_{\rm eff})^2 -  C^{ \rm{ACT} \times \rm{ACT}, EE,  \textrm{A}_\textrm{f}}_{\ell_b}. 
\end{equation}
\abds{Here $C^{\rm CMB, EE}_{\ell_b}$ is the EE power spectrum from our fiducial \textit{Planck} 2018 model, $C^{\rm foreground, EE, \textrm{A}_\textrm{f}}_{\ell_b}$ are foreground-model templates from} \cite{Choi_2020}\abds{, and $\alpha_{\textrm{A}_\textrm{f}}$ is a parameter characterizing the foreground amplitude}. The spectrum $C^{ \rm{ACT} \times \rm{ACT}, EE,  \textrm{A}_\textrm{f}}_{\ell_b}$ is the measured ACT EE spectrum for array-frequency $\textrm{A}_\textrm{f}$. 

We use the multipole range $\ell^{\rm min}_{b}= 1000$ and $\ell^{\rm max}_{b}=1500$ to minimise the impact of temperature-to-polarization leakage, foreground contamination and effects of the transfer function. A prior on the \DW{foreground} amplitude $\alpha_{\textrm{A}_\textrm{f}}$, \DW{which is dominated by the polarized dust component, is estimated using the measurement of the $353\,\si{GHz}$ \texttt{NPIPE} EE power spectrum from \textit{Planck}. Further details about this prior will be discussed in the upcoming ACT power spectrum analysis paper. }

We determine the efficiency scaling $p^{\textrm{A}_\textrm{f}}_{\rm eff}$ by finding the maximum-posterior value from a simple Gaussian likelihood derived from this $\chi^2$, marginalized over $\alpha_{\textrm{A}_\textrm{f}}$. We also derive an error on the efficiency scaling parameter from the posterior.

\section{Further discussion of normalization correction}\label{app:norm.corr.details}


In this appendix, we provide further details of the normalization correction discussed in Section~\ref{sub:norm.corr.summary}. The reconstructed lensing field, $\hat{\phi}_{LM}$, is dependent on \Frank{the assumed fiducial CMB power spectra} through the normalization term, $\mathcal{R}^{-1}_{L}$. Setting aside the small Monte-Carlo correction for the moment, this dependence can be expressed as  
%
\begin{equation}
\hat{\phi}_{L,M}\rvert_{C_\ell^{\textrm{CMB, fid}}}
 \approx \frac{\mathcal{R}^{-1}_{L}\rvert_{C_\ell^{\textrm{CMB,fid}}}}{\mathcal{R}^{-1}_{L}\rvert_{C_\ell^{\textrm{CMB}}}} \hat{\phi}_{L,M}\rvert_{C_\ell^{\textrm{CMB}}}
\end{equation}
%
%
where $C_\ell^{\textrm{CMB, fid}}$ is the fiducial CMB power spectra used in the lensing reconstruction and $C_\ell^{\text{CMB}}$ is the true spectrum that describes the data. The reconstruction $\hat{\phi}_{L,M}\rvert_{C_\ell^{\textrm{CMB, fid}}}$ is normalized with the fiducial power spectra and will differ from the unbiased reconstruction $\hat{\phi}_{L,M}\rvert_{C_\ell^{\textrm{CMB}}}$ (i.e., normalized with the correct spectrum) if the fiducial spectrum does not match the truth.
At the time this article was written, the final ACT DR6 CMB power spectra and their covariance matrices were not yet available. As a result, the baseline normalization was calculated using the full-sky fiducial spectra, $C_\ell^{\textrm{CMB, fid}}$, based on the \textit{Planck} 2015 TTTEEE cosmology with an updated $\tau$ prior as shown in~\cite{Calabrese2017}.

While we expect this analytic normalization to be quite accurate, it is a potential concern that the lack of a final measured power spectrum might introduce a small bias: in particular, that our assumption that the CMB power spectrum is consistent with \textit{Planck} might not be correct at high precision. Fortunately, the pre-processing steps described in Section~\ref{sec:methods} help to minimize any potential bias that could arise from assuming incorrect fiducial CMB power spectra. Recall that the error in the normalization is small as long as the data power spectra are similar to the fiducial power spectra used in the analysis. By applying the calibration and polarization efficiency determination described in Section~\ref{subsection:cal}, which are defined relative to \textit{Planck}, we can ensure that this condition is met because the amplitude of the data spectra is scaled to be in good agreement with the fiducial spectrum used in the normalization. 

\begin{figure}
  \centering
  \includegraphics[width=0.5\linewidth]{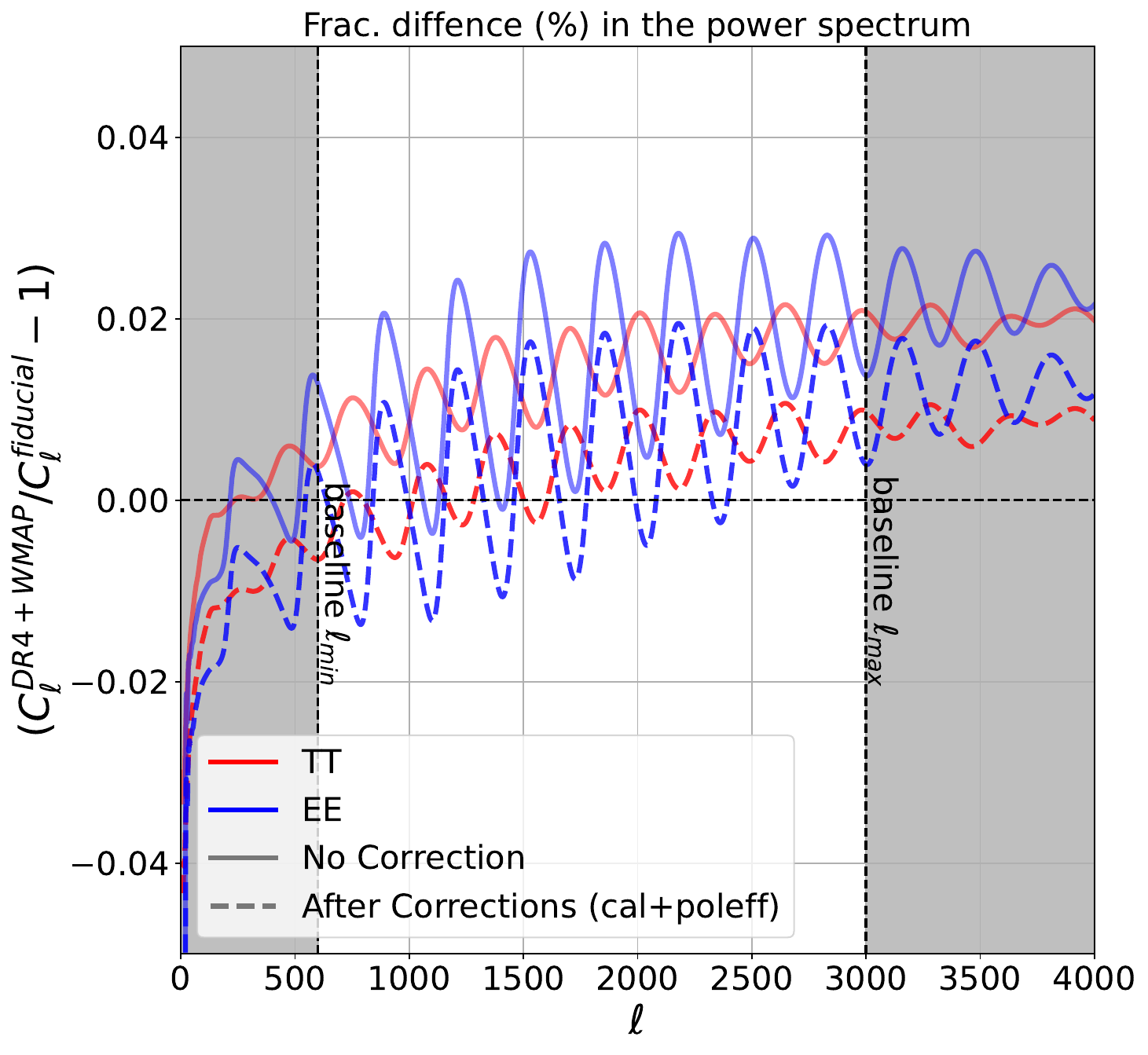}
  \caption{Differences in CMB power spectra used in the lensing normalization. Using the toy model described in Appendix \ref{app:norm.corr.details}, we illustrate the difference between the fiducial ACT DR4+\textit{WMAP} power spectra \citep[as presented in][]{Choi_2020,Aiola_2020} and $C_\ell^{\textrm{CMB, fid}}$ before and after applying calibration and polarization efficiency factors. As shown by the solid curves, the overall amplitudes of the fiducial ACT DR4+\textit{WMAP} spectra differ from the  fiducial spectra by a few percent before correction. The application of calibration and polarization efficiency factors serves to rescale these spectra and bring them more in line with the fiducial spectrua used for normalization (dashed lines).  }
  \label{Fig.2pt.with.calpoltf}
\end{figure}

To confirm quantitatively the effectiveness of our methods for reducing bias, we have tested a scenario in which the ACT DR6 data CMB power spectra match the ACT DR4+\textit{WMAP} spectra \citep[as presented in][]{Choi_2020,Aiola_2020} although we still assume $C_\ell^{\textrm{CMB, fid}}$ in our normalization. Figure~\ref{Fig.2pt.with.calpoltf} compares the fiducial ACT DR4+\textit{WMAP} spectra with $C_\ell^{\textrm{CMB, fid}}$ used in our work. In the analysis range, the ACT DR4+WMAP spectra differ by a few percent from $C_\ell^{\textrm{CMB, fid}}$. We follow the procedure described in Section~\ref{subsection:cal} and~\ref{sec.polareff} 
to compute corrections (calibration and polarization efficiency) and apply them to the ACT DR4+\textit{WMAP} spectra. The dashed lines in the figure show that the differences between the ACT+\textit{WMAP} and \DW{the fiducial} spectra are significantly reduced after these corrections are applied. 

\begin{figure*}[t]
    \centering
    \includegraphics[width=0.45\textwidth]{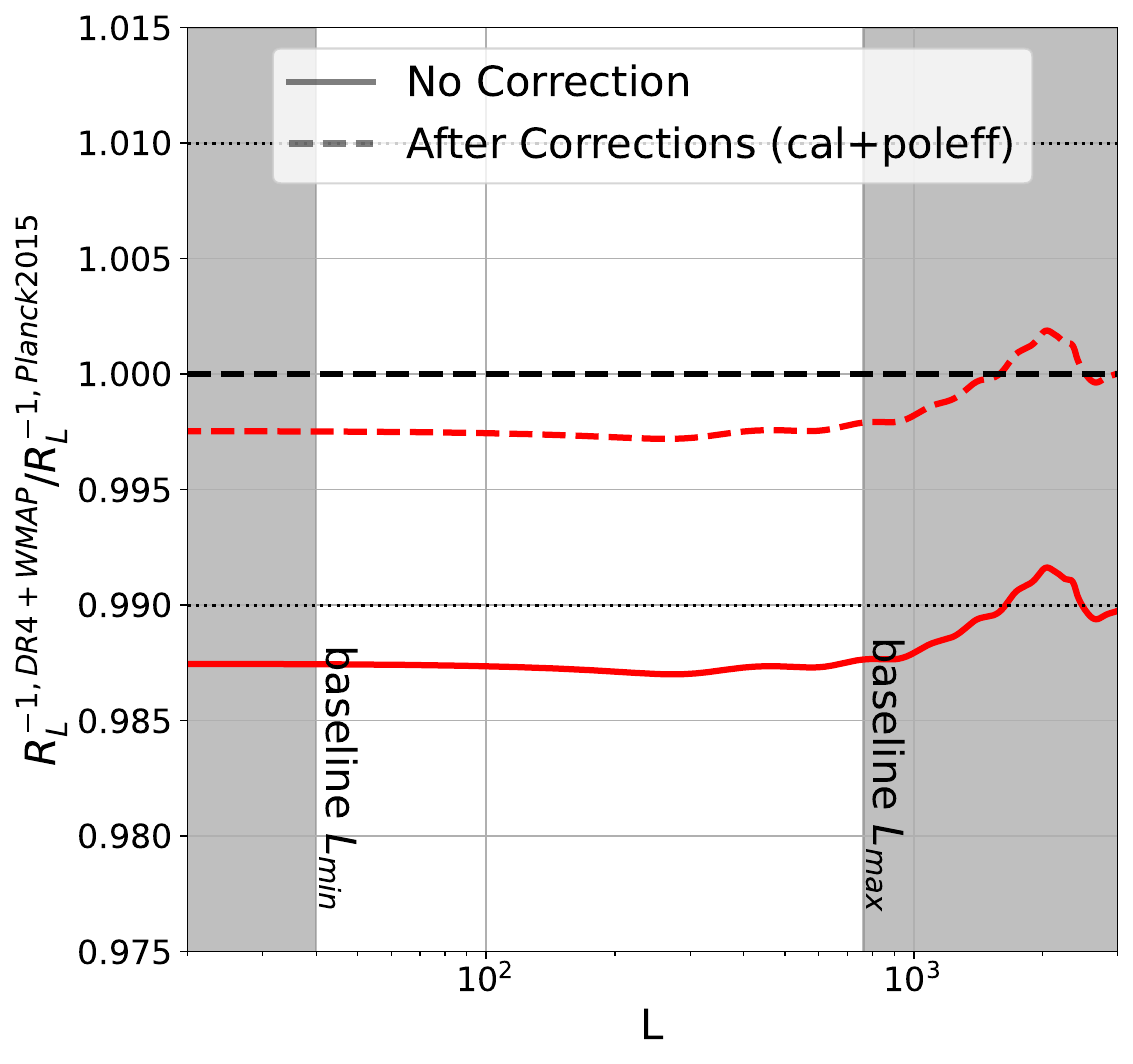} 
    \includegraphics[width=0.45\textwidth]{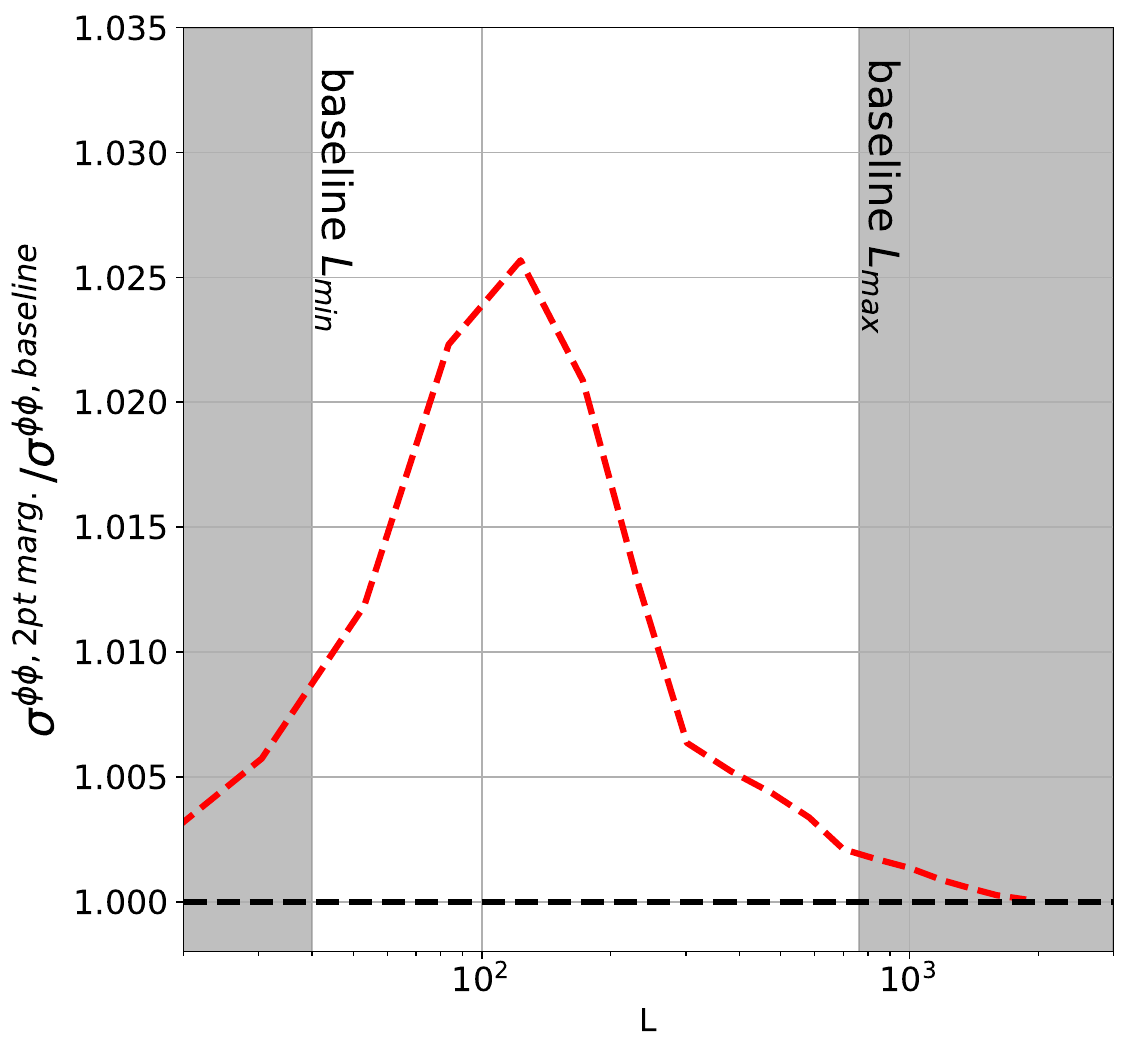}
    \caption{\textit{Left}: Similar to Figure~\ref{Fig.2pt.with.calpoltf}, but for the normalization factor $\mathcal{R}^{-1}_{L}$. Without the application of calibration and polarization efficiency factors (red solid curve), the mismatch between the power spectrum describing the data and the fiducial spectra leads to the lensing reconstruction being biased high by around $1.3\%$ corresponding to a bias of around $1.14 \sigma$ in the inferred lensing amplitude. After applying amplitude corrections to the CMB maps (red dashed curve), this bias is reduced by a factor of 5, resulting in only $0.23 \sigma$-level bias in the lensing amplitude. (If we assume that the ACT DR6 spectra fall somewhere in between \textit{Planck} and ACT DR4 spectra, any biases will be significantly less than $0.23 \sigma$.)
    \textit{Right}: Inflation of lensing bandpower error bars due to uncertainty in the true CMB power spectra, which are needed to normalize the lensing measurement correctly. As described in the text, sampling 1000 $\Lambda$CDM spectra from  ACT DR4 + \textit{Planck} chains 
    results in a less than $2.5\%$ increase in error bars and introduces small mode coupling between bandpowers. This change only has a negligible impact on cosmological parameter constraints.}
    \label{fig:Fig.al.calpoltf.2ptmarg.err.infl}
\end{figure*}

We calculated the normalization $\mathcal{R}^{-1}_{L}$ for each of these spectra and compared it to our baseline $\mathcal{R}^{-1}_{L}$. The results of this comparison are displayed in the left panel of Figure~\ref{fig:Fig.al.calpoltf.2ptmarg.err.infl}. Prior to the correction, the ACT DR4+\textit{WMAP} $\mathcal{R}^{-1}_{L}$ was approximately $1.3\%$ lower than the baseline, corresponding to a roughly $1.14 \sigma$ bias in the lensing power spectrum amplitude $\Alens$. 
After applying the correction, the difference in $\mathcal{R}^{-1}_{L}$ was reduced by a factor of around $5$, reducing the bias in $\Alens$ to $0.23 \sigma$. We anticipate that the ACT DR6 spectra will not be clearly inconsistent with the \textit{Planck} or ACT DR4+\textit{WMAP} fiducial spectra, and we thus expect that they will fall somewhere in between these two spectra. Therefore, the actual bias in the data is likely to be less than $0.23 \sigma$.

For the lensing-only likelihood analysis considered in this paper, we account for this uncertainty in the true CMB spectra as follows.
Since $C_\ell^{\textrm{CMB}}$ is well constrained, the difference between the true CMB power spectra that describe the data, $C_\ell^{\textrm{CMB}}$, and the fiducial $C_\ell^{\textrm{CMB,fid}}$ is expected to be small.  We can, therefore, Taylor expand $\mathcal{R}^{-1}_{L}\rvert_{C_\ell^{\textrm{CMB}}}$ around the fiducial spectra as
\begin{align}
\mathcal{R}^{-1}_{L}\rvert_{C_\ell^{\textrm{CMB}}} & \approx \mathcal{R}^{-1}_{L}\rvert_{C_\ell^{\textrm{CMB,fid}}} + {\frac{\partial \mathcal{R}^{-1}_{L}}{\partial C_\ell^{\textrm{CMB}}}} (C_\ell^{\textrm{CMB}}-C_\ell^{\textrm{CMB,fid}}) \nonumber \\
& \approx \mathcal{R}^{-1}_{L}\rvert_{C_\ell^{\textrm{CMB,fid}}}\left[1 + M_{L}^{\ell}\rvert_{C_\ell^{\textrm{CMB,fid}}} (C_\ell^{\textrm{CMB}}-C_\ell^{\textrm{CMB,fid}}) \right] ,
\end{align}
where $M_{L}^{\ell}\rvert_{C_\ell^{\textrm{CMB,fid}}} = \partial \ln \mathcal{R}^{-1}_L / \partial C_\ell^{\text{CMB}} \rvert_{C_\ell^{\text{CMB, fid}}}$ is the linearized normalization-correction matrix. We compute it using finite differences at each $\ell$. After correction for $N_0$, and neglecting $N_1$, this leads to the expected value of the reconstructed lensing power spectrum being related to the true spectrum by
%
\begin{align}
\langle C_L^{\hat{\phi}\hat{\phi}} \rangle &\approx \left(\frac{\mathcal{R}^{-1}_{L}\rvert_{C_\ell^{\text{CMB,fid}}}}{\mathcal{R}^{-1}_{L}\rvert_{C_\ell^{\text{CMB}}}}\right)^2 C_L^{\hat{\phi}\hat{\phi}} \nonumber \\
& \approx \left[1-2 M_{L}^{\ell}\rvert_{C_\ell^{\textrm{CMB,fid}}} (C_\ell^{\textrm{CMB}}-C_\ell^{\textrm{CMB,fid}})\right] C_L^{\hat{\phi}\hat{\phi}} .
\end{align}

For a lensing-power-spectrum-only likelihood, we need to marginalize over the uncertainty in $C_\ell^{\text{CMB}}$. In \cite{Planck:2018}, the marginalization is performed analytically using the \texttt{Plik lite} CMB bandpower covariance. At the time of writing this article, the ACT DR6 CMB covariance is not yet available. Instead, we sample 1000 $\Lambda$CDM CMB power spectra from the ACT DR4 + \textit{Planck} parameter chains used in~\citet{Aiola_2020} to generate an ensemble of smooth power spectrum curves consistent with the ACT DR4 + \textit{Planck} power spectrum measurements; we then propagate these power spectrum curves through to the lensing normalization and compute the resulting additional dispersion in the lensing bandpowers. As a result, for the lensing-only analysis, we add an additional term to the covariance matrix:
\begin{equation}
\bar{\mathbb{C}}_{bb^\prime} = \mathbb{C}_{bb^\prime} + \mathbb{C}^{\textrm{CMB}}_{bb^\prime},
\end{equation}
where now $\mathbb{C}^{\textrm{CMB}}_{bb^\prime}$ is the additional covariance induced by the CMB marginalisation. In the right panel of Figure~\ref{fig:Fig.al.calpoltf.2ptmarg.err.infl}, we demonstrate the inflation of lensing bandpower error bars resulting from the marginalization over CMB power spectra. This process leads to an increase of at most $2.5\%$ in error bars and the introduction of small correlations between bands. For comparison, we have conducted the marginalization using the ACT DR4+\textit{WMAP} parameter chains and found a slightly larger increase of up to $3.5\%$. 
In the baseline analysis range, the error bars on the lensing amplitude increase from $\sigma(\Alens) = 0.023$ to $\sigma(\Alens) = 0.0256$ when marginalizing over the ACT DR4+\textit{Planck} chains. In comparison, when marginalizing over the ACT DR4+\textit{WMAP} chains, the error bars increase to $\sigma(\Alens) = 0.0264$. Although this seems like a non-negligible change, the impact on cosmological parameters such as $S_8^{\mathrm{CMBL}}$ appears significantly smaller. \DW{In particular, the error on the $S_8^{\mathrm{CMBL}}$ only increases from $0.021$ to $0.022$ when we include the ACT DR4+\textit{Planck}-based marginalization.}
One may question the choice of using ACT DR4+\textit{Planck} chains rather than, for example, the ACT DR4 only chains, as well as the use of smooth $\Lambda$CDM curves instead of allowing the bandpowers to vary freely given the likelihood. However, it is important to note that the marginalization approach remains robust as long as the CMB bandpowers have been well-characterized, regardless of the specific CMB measurement used. The larger scales, which play an important role in determining the overall power spectrum amplitude, are already robustly measured by \textit{Planck}. Therefore, we believe it is a valid approach to incorporate this information in the marginalization procedure. With regard to the use of smooth $\Lambda$CDM curves, we explicitly test our model-dependence by using chains for an extended model (ACT DR4+\textit{Planck} $\Lambda$CDM+\DW{$\Alens^{\textrm{2pt}}$}) in the marginalization. \DW{Note that $\Alens^{\textrm{2pt}}$ is a parameter that characterizes the impact of gravitational lensing on the CMB power spectra and is distinct from the lensing amplitude parameter $\Alens$.}
We find that using the extended-model chains results in an increase of only 3\% on the error bars of the lensing amplitude, $\sigma(\Alens)$, compared to the ACT DR4+\textit{Planck} $\Lambda$CDM case. This negligible change is an illustration of the fact that this marginalization does not introduce significant model dependence into our lensing measurement.
 Lastly, it is worth highlighting that the ACT DR6 power spectra will have improved constraints on both small and large scales. As a result, future analyses will have the ability to marginalize directly over the ACT-only power spectra.


\section{Estimation of Instrumental Systematics}\label{app:instrument.details}

\subsection{Calibration and beam systematics}\label{app:calandbeam}

\DW{Both beam and calibration errors have the potential to bias lensing measurements. One way in which lensing can be affected by such errors is through their coherent rescaling of the measured power spectra across a range of scales. This rescaling then impacts the overall normalization of the measured lensing power spectrum, as discussed in detail in Appendix~\ref{app:norm.corr.details}. Hence, it is crucial to account carefully for both beam and calibration errors in any lensing analysis to avoid systematic biases.}

We run the full lensing pipeline for multiple different realizations of the calibration factor and beam transfer functions, based on their estimated means and uncertainties as discussed in Sections~\ref{subsection:beams} and~\ref{subsection:cal}. We find that the change in the amplitude of the lensing power spectrum (as opposed to a \DW{scale-dependent fractional correction}) is the dominant effect. 
In fact, in our tests, we find that the response is, to a good approximation, given by \DW{$\Delta  \ln C_L^{\phi \phi}\approx 2 \Delta  \ln \bar{C}_\ell$, where $\Delta  \ln  \bar{C}_\ell$ and $\Delta  \ln  C_L^{\phi \phi}$} are fractional changes in the CMB power spectrum averaged over $\ell \in [600,3000]$ and fractional changes in the lensing power spectrum, respectively. 
\blake{Also, comment on what you do for, e.g., the MV estimator using temperature and polarization.} Since quantifying the effect of beams and calibration errors by running the full lensing power spectrum pipeline is computationally prohibitive, we may use the approximation above to quantify their effect.

\begin{figure*}[t]
\centering
    \includegraphics[width=0.45\textwidth]{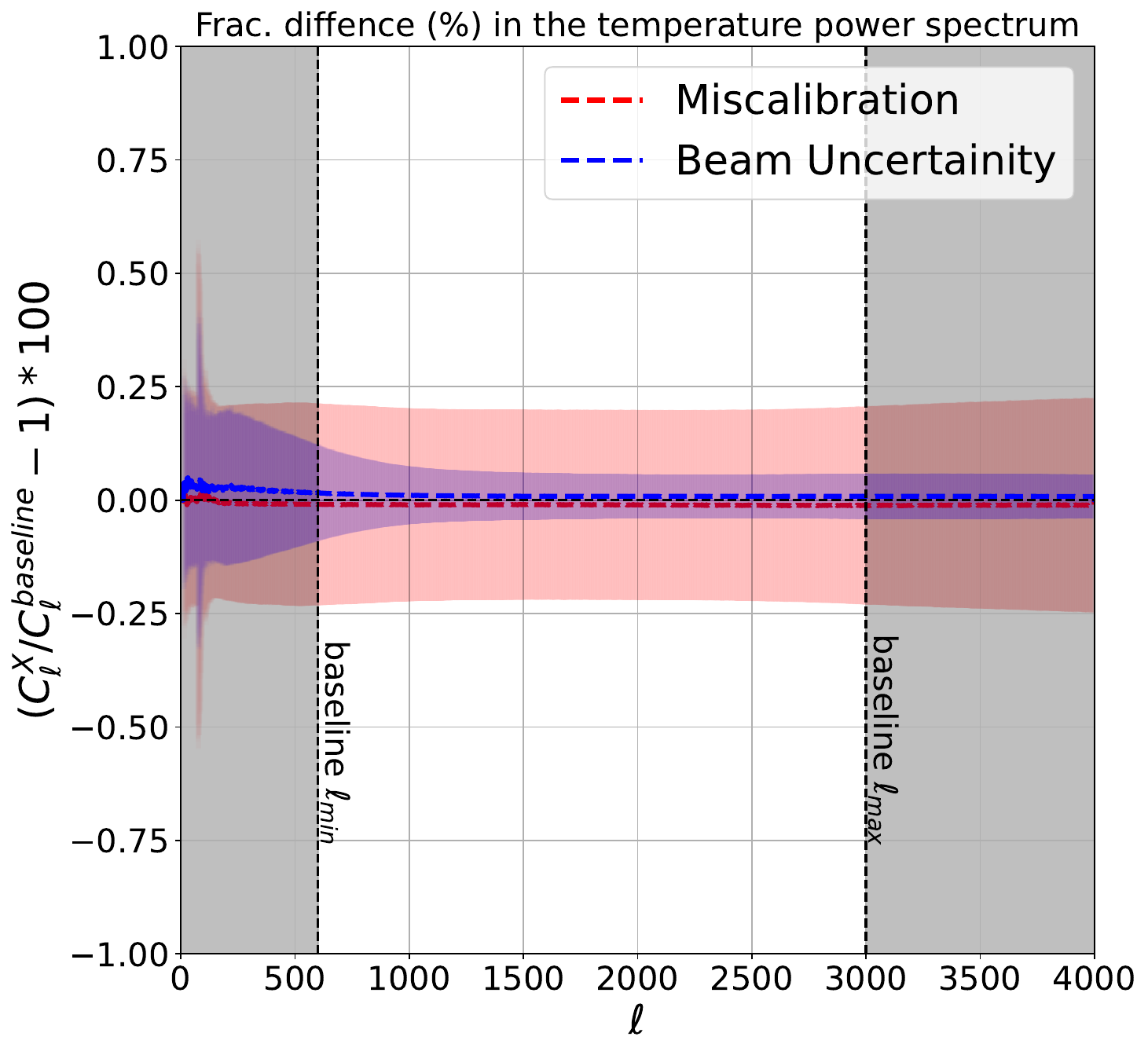} 
   \includegraphics[width=0.425\textwidth]{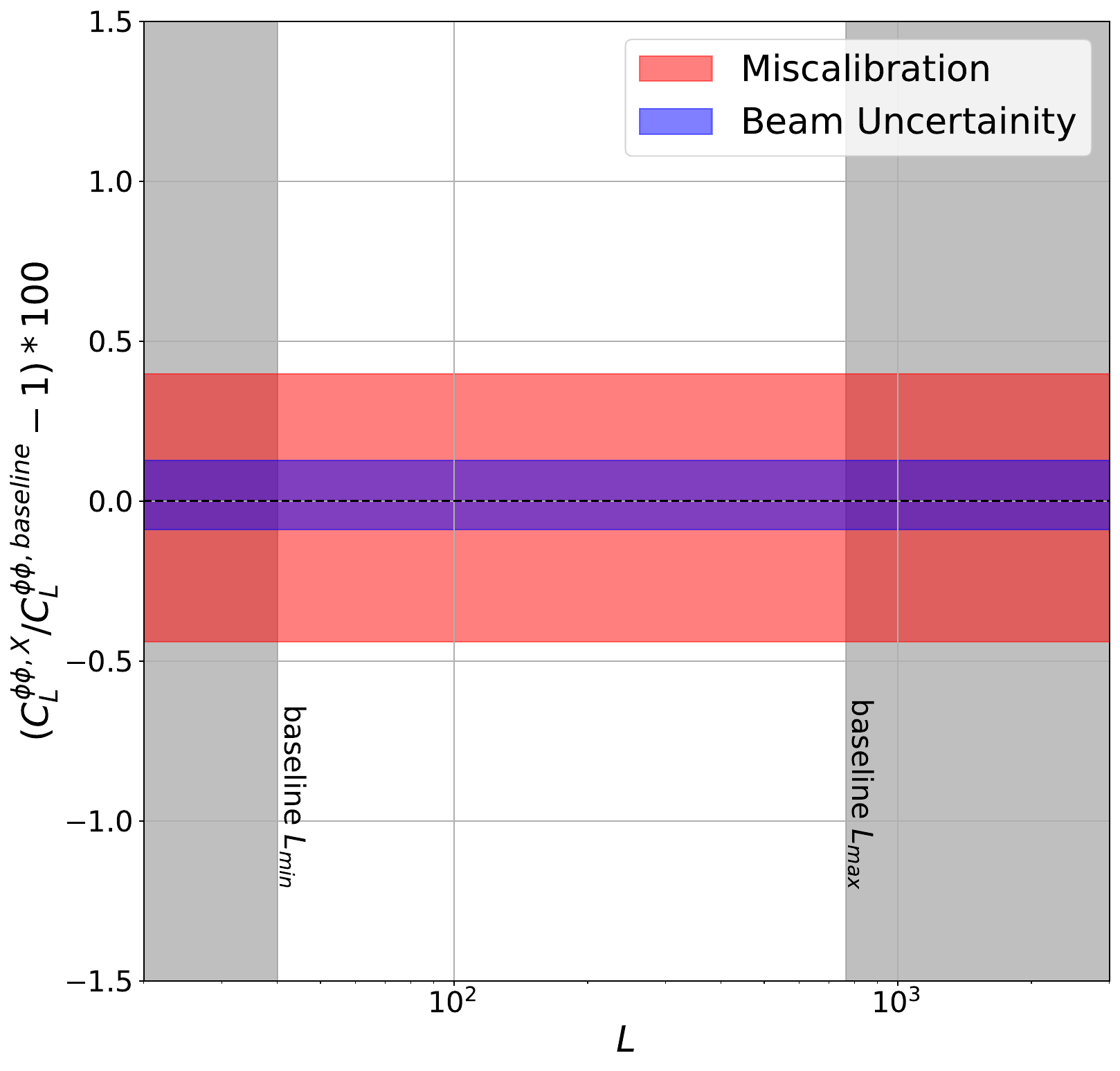}
    \caption{\textit{Left}: Mean and one-sigma confidence intervals of the reprocessed coadded temperature power spectrum compared to the baseline. The scatter is due to uncertainties in calibration factors (shown in red) and instrument beams (shown in blue). The confidence intervals are computed using the 100 Monte-Carlo samples of calibration and beams applied to raw maps processed in the same way as the baseline. The greyed-out regions are outside of the baseline analysis range.
    \textit{Right}: Similar to the left panel but for the lensing power spectrum. This is computed using the same 100 samples but with the fractional error in the CMB power spectrum propagated to the lensing power spectrum with the approximation $\Delta  \ln C_L^{\phi \phi}\approx 2 \Delta  \ln \bar{C}_\ell$ discussed in the text. The greyed-out regions are outside of the baseline analysis range.}
    \label{Fig.cal.beam.systematics.24pt}
\end{figure*}

To quantify the typical effect of beam and calibration errors on lensing, we proceed as follows. We do not attempt to incorporate beam and calibration errors into our statistical error but instead will treat the typical error as a systematic that we argue is negligible.
 First, as our calibration factors are computed with respect to the \textit{Planck} maps, they may not be independent across array-frequencies, and this correlation can increase the overall bias. To investigate this issue, we estimate their correlations by jointly sampling calibration factors from a likelihood of the following form (up to an irrelevant constant):
%
%

\begin{equation}
-2\ln \mathcal{L}(\{c_{\textrm{A}_\textrm{f}}\}|\{\Delta_\ell^{\textrm{A}_\textrm{f}}\}) = \sum_{\ell = \ell_{\rm min}}^{ \ell_{\rm max}}  \sum_{\textrm{A}_\text{f},\textrm{A}'_\textrm{f}} \Delta^{\textrm{A}_\textrm{f}}_{\ell}\left[\Sigma_\ell^{-1}\right]^{\textrm{A}_\textrm{f}\textrm{A}'_\textrm{f}}\Delta^{\textrm{A}'_\textrm{f}}_{\ell},
\end{equation}
%
where $\Delta^{\textrm{A}_\textrm{f}}_{\ell}=c_{\textrm{A}_\textrm{f}}^2 C_\ell^{\textrm{ACT}\times\textrm{ACT},\textrm{A}_\textrm{f}} - c_{\textrm{A}_\textrm{f}} C_\ell^{\textrm{ACT}\times\textrm{P}, \textrm{A}_{\textrm{f}}}$ 
is the difference between the auto-power spectrum of ACT array-frequency $\textrm{A}_\textrm{f}$ and the cross-power between ACT and the \textit{Planck} map closest in frequency, after scaling the ACT array-frequency map by the calibration factor $c_{\textrm{A}_\textrm{f}}$.
The indices $\textrm{A}_\textrm{f}$ and $\textrm{A}'_\textrm{f}$ run through the ACT array-frequencies used in our analysis, and $\Sigma_\ell$ is an approximate analytic covariance matrix for the $\Delta^{\textrm{A}_\textrm{f}}_{\ell}$ (assumed diagonal in $\ell$) computed using the fiducial CMB theory spectrum and measured noise spectrum. Our analysis shows that the 90\,GHz channels are strongly correlated at a level of 51\%, while the 150\,GHz channels exhibit lower correlations ranging from 10\% to 18\%. This difference can be attributed to the lower noise in the 90\,GHz channels. Furthermore, the correlation between the 90\,GHz and 150\,GHz channels is negligible because calibration factors are fitted at different multipole scales for the two frequency bands. We draw 100 independent Monte-Carlo samples of the calibration and beam, based on their estimated means and uncertainties  (see Sections~\ref{subsection:beams} and~\ref{subsection:cal}, and~\citealt{Lungu2022} for details), as well as accounting for the estimated correlations among calibration factors.
We then process raw data maps using these samples, following the pre-processing procedure described in Section~\ref{sec:methods}, and remeasure the CMB power spectra. Figure~\ref{Fig.cal.beam.systematics.24pt} (left) shows the fractional change in the coadded temperature power spectrum. We find that the scatter in the power spectrum is less than  $0.2\%$ in our baseline analysis range ($\ell\in [600,3000]$). 


We then convert the error in the power spectra to an error in the lensing power spectra using the approximation above. We take the RMS ($1 \sigma$) shift as an estimate of the typical systematic bias from ignoring beam and calibration uncertainty. We show the derived systematic biases for calibration and beam in the right panel of Figure~\ref{Fig.cal.beam.systematics.24pt}. We conclude that these are subdominant sources of error compared to our $2.3\%$ statistical uncertainty in our lensing power spectrum measurement.

\abds{In particular, for calibration, we find $ \Delta\Alens = \albiasMiscalMV \; (\biasSigmaMiscalMV)$ and $ \Delta\Alens = \albiasMiscalMVPol \;(\biasSigmaMiscalMVPol)$ for the MV and MVPol estimators, respectively. Similarly, for beam uncertainty, we find $ \Delta\Alens = \albiasBeamMV \;(\biasSigmaBeamMV)$ and $ \Delta\Alens = \albiasBeamMVPol \;(\biasSigmaBeamMVPol)$ for MV and MVPol, respectively. }

\subsection{Temperature-to-polarization leakage verification}\label{app:syst.t2pleakage}
As described in Section~\ref{subsection:beams}, we find evidence for some degree of temperature-to-polarization ($T \rightarrow P$) leakage in our measurements. This leakage is estimated to be small but could affect the accuracy of our polarization-based lensing estimators mainly due to its effect on normalization. To quantify its effect, we add additional leakage terms to the polarization multipoles for each array-frequency:
\begin{equation}
\begin{split}
\tilde{X}_{\ell m} &= X_{\ell m} + X^{\textrm{leakage}}_{\ell m} \\
& = X_{\ell m} + B^{T \rightarrow P}_{\ell} T_{\ell m} ,
\end{split}
\end{equation}
where $X_{\ell m}$ are either the baseline $E$ or $B$ multipoles, $B^{T \rightarrow P}_{\ell}$ is the estimated leakage beam, and $T_{\ell m}$ are the baseline $T$ multipoles The left panel of Figure~\ref{Fig.cal.beam.systematics.24pt} shows an example of the leakage beams in our analysis.
Since the baseline data, $X_{\ell m}$, already include the $T \rightarrow P$ leakage effect, our modified maps, $\tilde{X}_{\ell m}$, include twice the amount of leakage compared to a scenario with no leakage.
We run our analysis pipeline on these modified maps and find the shift in the estimated lensing power spectrum is less than $0.1\%$ compared to our MV baseline (see the right panel of Figure~\ref{Fig.cal.beam.systematics.24pt}). In terms of our $ \Delta\Alens$ metric (Equation~\ref{eq.al.bias}), we find $ \Delta\Alens = \albiasLeakBeamMV\;(\biasSigmaLeakBeamMV)$ and $ \Delta\Alens = \albiasLeakBeamMVPol\;(\biasSigmaLeakBeamMVPol)$ for MV and MVPol, respectively. 

Preliminary analysis (at the time of preparing this paper) of the ACT DR6 CMB power spectra has revealed an additional leakage for a specific array-frequency. This leakage is consistent with a constant temperature-to-polarization leakage in the form of $\hat{E}_{\ell m} = E_{\ell m} + \alpha T_{\ell m}$, where $\alpha = 0.0035$. While this feature is yet to be confirmed, we have developed a toy model to investigate its potential impact. In this toy model, we assume that temperature maps leak into polarization maps (both E and B) at all array-frequencies with a constant coefficient of $\alpha = 0.0035$. However, it should be noted that this feature has only been observed in a specific array-frequency. Thus, it is unlikely that it would affect all channels equally, and this model likely represents a worst-case scenario.  We run our analysis pipeline on this toy model and find $ \Delta\Alens = \albiasLeakConstMV\;(\biasSigmaLeakConstMV)$ and $ \Delta\Alens = \albiasLeakConstMVPol\;(\biasSigmaLeakConstMVPol)$ for MV and MVPol, respectively.

\begin{figure*}[t]
\centering
\includegraphics[width=0.45\textwidth]{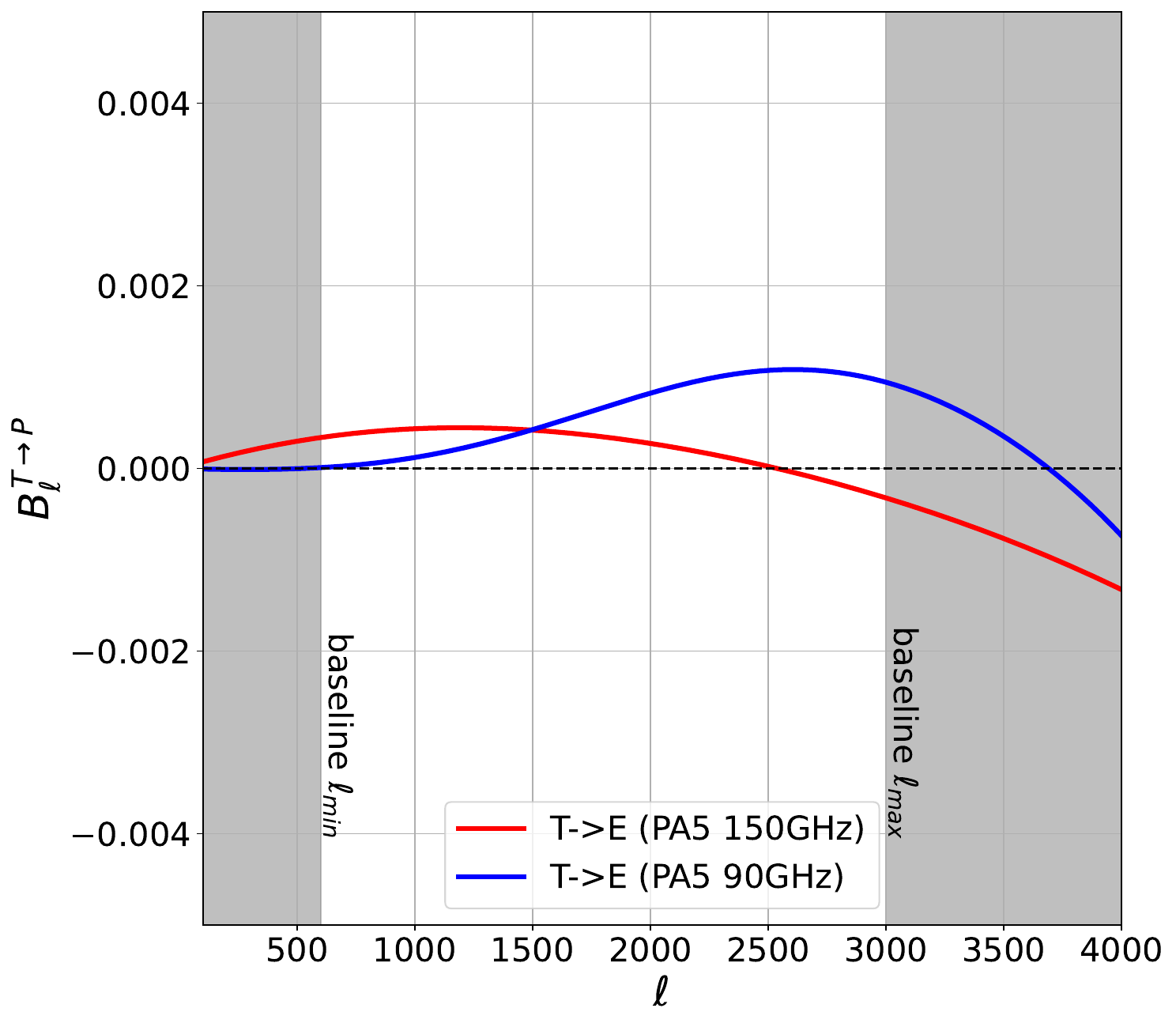}
\includegraphics[width=0.42\textwidth]{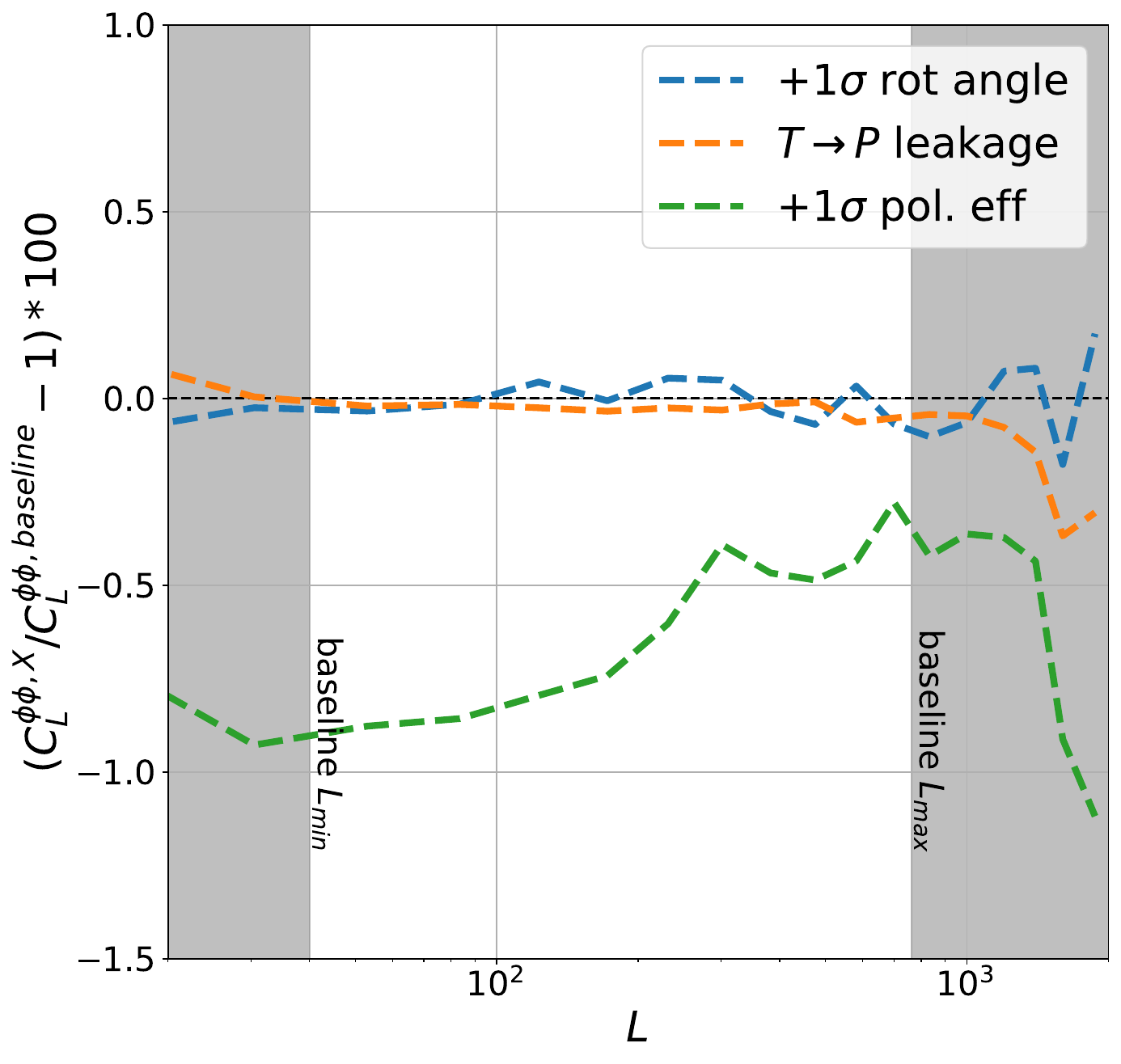}
    \caption{\textit{Left}: Example of the $T \rightarrow P$ leakage beams (discussed in Section~\ref{subsection:beams}). We calculate the impact of these leakage beams on the lensing bandpowers, following the method outlined in Appendix~\ref{app:syst.t2pleakage}, and find $ \Delta\Alens = \albiasLeakBeamMV\;(\biasSigmaLeakBeamMV)$ and $ \Delta\Alens = \albiasLeakBeamMVPol\;(\biasSigmaLeakBeamMVPol)$ for MV and MVPol, respectively. \textit{Right}: Effects of polarization angle, $T\rightarrow P$ leakage, and polarization efficiency on lensing power spectrum estimates. See Appendices~\ref{app:syst.t2pleakage},~\ref{app:syst.polrot}, and~\ref{app:syst.poleff} for further information. The effects from polarization angle and $T\rightarrow P$ leakage are neglible. The impact of polarization efficiency is larger (green dashed line), although it is important to note that the result here represents a very conservative upper limit.  The bias to the lensing amplitude for each case can be found in Table~\ref{Tab:systematics.summary} in the main text.}
    \label{Fig.leakage.restsyst}
\end{figure*}

\subsection{Polarization angle verification}\label{app:syst.polrot}
At the time of writing, a precise characterisation of the absolute polarization angle of the DR6 data set is not yet available. However, tests differencing DR4 and DR6 EB spectra have found that the EB power spectrum from the DR6 data set is consistent with that of the DR4 data set. This suggests that the DR6 polarization rotation angle should be consistent with the DR4 estimate of $\Phi_p = -0.07 \pm 0.09$ deg. In order to assess the impact of a non-zero polarization angle on our results, we artificially rotate the $Q/U$ maps of our data set by $\Phi_p = -0.16$ deg. (the mean minus one $\sigma$) and reanalyze the resulting maps using our analysis pipeline. The resulting lensing auto-spectrum estimates for these rotated maps are shown in the right panel of Figure~\ref{Fig.leakage.restsyst} as the blue dashed curve. As demonstrated in the figure, our bandpowers are not significantly affected by the rotation angle; the rotation results in minimal changes to our $ \Delta\Alens$ statistics.  We find $ \Delta\Alens = \albiasPolAngMV\;(\biasSigmaPolAngMV)$ and $ \Delta\Alens = \albiasPolAngMVPol\;(\biasSigmaPolAngMVPol)$ for MV and MVPol, respectively. 

\subsection{Polarization efficiency}\label{app:syst.poleff}
As previously discussed in Section~\ref{sec:methods}, we apply a correction to our polarization maps based on the mean of the measured polarization efficiencies. Following this correction, our baseline and polarization-only estimators show good agreement (see Figure~\ref{Fig.polcomb_consistent}). To assess the robustness of our correction method, we conduct an approximate test in which the estimated polarization efficiency was artificially lowered by  $1\sigma$. 
It is important to note that the deviations from the mean polarization efficiency are not anticipated to correlate strongly across array-frequencies, and as such, it is unlikely that all efficiencies would be biased in the same manner.
Therefore, this test represents a very conservative upper limit. The resulting lensing auto-spectrum estimates for these maps are shown in the right panel of Figure~\ref{Fig.leakage.restsyst} as the greeen dashed curve. We find $ \Delta\Alens = \albiasPolEffMV\;(\biasSigmaPolEffMV)$ and $ \Delta\Alens = \albiasPolEffMVPol\;(\biasSigmaPolEffMVPol)$ for MV and MVPol, respectively. Given that these values were obtained from conservative estimates, we anticipate that the actual data bias is much smaller than the values presented. We therefore include them in our summary table as conservative upper limits.

\section{Pipeline verification and origin of multiplicative normalization correction}\label{sec:verification}

\begin{figure}
 \centering
  \includegraphics[width=0.5\linewidth]{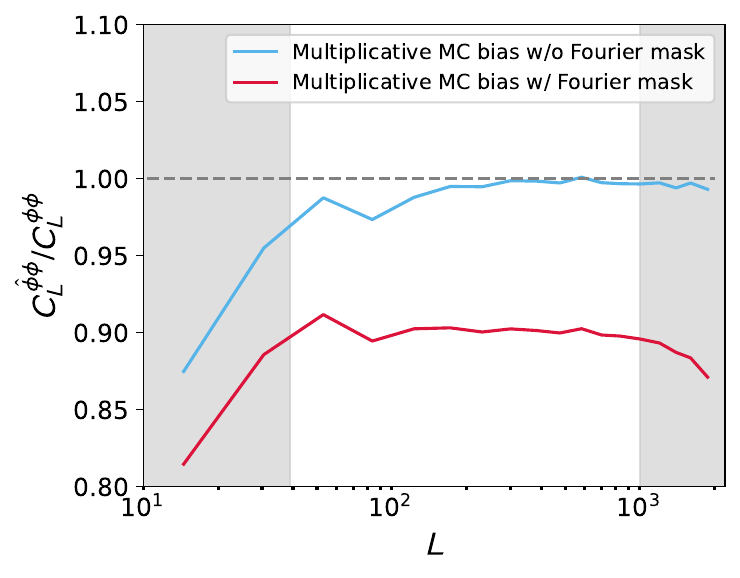}
  \caption{Cross-spectrum between the reconstructed lensing map from simulated CMB maps and the lensing map of the input simulation, $C^{\hat{\phi}\phi}_L$, divided by the input lensing auto spectrum $C^{{\phi}\phi}_L$. It can be seen in the red curve that the Fourier-space filter, which helps to remove ground pick-up, causes a multiplicative suppression of the reconstructed lensing map by $10$--$20\%$. Without this filter (blue curve), the lensing suppression is absent, except at low multipoles where mask effects become important. This well-understood multiplicative suppression due to Fourier-space filtering is simply corrected using a multiplicative factor derived from simulations. }
  \label{fig.kspace_Corr}
\end{figure}

We demonstrated the performance of our pipeline using simulations in Figure~\ref{fig:pipeline_ver}. The multiplicative MC correction constitutes a $10$--$15\%$ effect; however, it is well understood and arises almost entirely from the fact that the Fourier modes of $|\ell_x|<90$ and $|\ell_y|<50$ are filtered away. To demonstrate this, Figure \ref{fig.kspace_Corr} \abds{shows the ratio of the cross-spectrum between a lensing reconstruction and the input lensing map with the input auto-spectrum from simulations. With no Fourier-space mask applied, we find that this ratio is quite close to unity (with the departures at large scales expected from the couplings induced by the mask); in contrast, the same ratio computed on maps where the Fourier-space mask is applied is lower by $10$--$20\%$}.

\section{Lensing power spectrum biases} \label{app.biases}
This appendix describes in more detail the algorithms used to compute the lensing power spectrum biases; we focus on the computation of $N_0$, $N_1$ and additive MC biases.
\subsection{Realization-dependent $N_0$} \label{app.rdn0}
Lensing power spectrum estimation aims to probe the connected 4-point function that is induced by lensing. However, the naive lensing power spectrum estimator $\hat{C}^{\hat{\phi}\hat{\phi}}_L[\bar{X},\bar{Y},\bar{A},\bar{B}] \sim XY AB$ also contains disconnected contributions arising from Gaussian fluctuations (e.g., $\langle X A \rangle \langle Y B \rangle$), which are non-zero even in the absence of lensing. These contributions, which are typically referred to as the Gaussian or $N_0$ bias and which can be understood as a bias arising from the ``noise'' in the lensing reconstructions, must be subtracted to recover an unbiased estimator of lensing.\\
The Gaussian ($N_0$) bias is estimated using an algorithm involving different pairings of data and two sets of independent simulation maps, denoted with superscript $S$ and $S^\prime$ \citep{Namikawa2013, Planck:2018}:

\begin{align}
\Delta{C^{\text{Gauss}}_{L}} = &\langle \Bar{C}^\times_L[XY^S,AB^S]+\Bar{C}^\times_L[X^SY,AB^S]\nonumber \\
&+\Bar{C}^\times_L[X^SY,A^SB]+\Bar{C}^\times[XY^S,A^SB]\nonumber \\
&-\Bar{C}^\times_L[X^SY^{S^\prime},A^SB^{S^\prime}]-\Bar{C}^\times_L[X^SY^{S^\prime},A^{S^\prime}B^{S}]\rangle_{S,S^\prime}.
\label{eq:rdn0}
\end{align}

This estimator can be obtained from the Edgeworth expansion of the lensing likelihood; it has the useful feature that it corrects for a mismatch between two-point functions of the data and of simulations. The estimator achieves this by also using the two-point function of the data, rather than simulations alone, when calculating this Gaussian bias; the combination employed above can be shown to be insensitive to errors in the simulation two-point function, to first order in the fractional error. Furthermore, this estimator helps to reduce the correlation between different lensing bandpowers~\citep{PhysRevD.83.043005} as well as the correlation of the lensing power spectra with the primary CMB spectra~\citep{2013PhRvD..88f3012S}. Finally, the negative of the last two terms of Equation~\eqref{eq:rdn0}, obtained purely from simulations, constitutes the Monte-Carlo N0 or MCN0. This term corresponds to the Gaussian reconstruction noise bias of the lensing bandpowers averaged over many CMB and lensing reconstructions; we use this term in Equation~\eqref{app.mcbias} to estimate the size of the additive MC bias.

We use $480$ different realisations of noiseless simulation pairs $S,S^\prime$ to calculate this bias for the real measurement. Noiseless simulations can be used here because we are using the cross-correlation based estimator, and the absence of noise helps to reduce the number of simulations required to achieve convergence.  We verify that increasing the number of realizations from 240 to 480 did not substantially affect our results (with  $\Delta\Alens=5\times10^{-6}$ ). We hence conclude that 240 realizations are sufficient for convergence, though we use 480 simulations in our data and null tests to be conservative.

The same RDN0 calculation should also be carried out for each simulation used to obtain the covariance matrix,  with the `data' corresponding to the relevant simulated lensed CMB realization. However, in practice, it is computationally unfeasible to estimate this RDN0 bias for each of the 792 simulations used for the covariance matrix, due to the very large number of reconstructions required for each RDN0 computation. Therefore, we resort to an approximate version of the realization dependent $N_0$, referred to as the semi-analytic $N_0$; we discussed this in detail in Section \ref{subsection:diagonal}.


\subsection{$N_1$ bias}\label{app.n1}
In addition to the Gaussian bias, another, smaller, bias term arises from ``accidental'' correlations of lensing modes that are not targeted by the quadratic estimator, as described in detail in \cite{PhysRevD.67.123507}. Since this bias is linear in $C^{\phi\phi}_L$, it is denoted the $N_1$ bias.


We calculate the $N_1$ bias by using 90  pairs of simulations with different  CMB realizations, but common lensing potential maps denoted $(S_\phi,S^\prime_\phi)$ and 90 pairs of simulations with different CMB and lensing potential $(S,S^\prime)$. (We have carried out convergence tests as in Section \ref{app.rdn0} to ensure that 90 simulations are sufficient.)

\begin{align}
\Delta{C^{N_1}_{L}} = &\langle \Bar{C}^\times_L[X^{S_\phi}Y^{S^\prime_\phi},A^{S_\phi}B^{S^\prime_\phi}]\nonumber\\
&+\Bar{C}^\times_L[X^{S_\phi}Y^{S^\prime_\phi},A^{S^\prime_\phi}B^{S_\phi}]\nonumber \\
&-\Bar{C}^\times_L[X^{S}Y^{S^\prime},A^{S}B^{S^\prime}]\nonumber\\
&-\Bar{C}^\times_L[X^{S}Y^{S^\prime},A^{S^\prime}B^{S}]\rangle_{S,S^\prime,S_\phi,S^\prime_\phi}.
\label{eq:mcn1}
\end{align}

We do not include biases at orders higher than $N_1$, since biases such as $N^{(3/2)}$ due to non-linear large scale structure growth and post-Born lensing effects are still negligible at our levels of sensitivity \citep{gfb}.

\subsection{Additive MC Correction}\label{app.mcbias}
Although subtraction of the Gaussian and $N_1$ biases results in a nearly unbiased lensing power spectrum, we calculate an additive Monte-Carlo bias $C_L^{\text{MC}}[XY,AB]$ with simulations to absorb any residual arising from non-idealities that have only been approximately captured, such as the effects of masking. 
As can be seen in Figure \ref{fig:pipeline_ver}, this bias term is very small. This justifies why we may model this bias term as additive rather than multiplicative: any true functional form of the Monte-Carlo correction could be simply Taylor expanded to give an additive term. 
This additive MC bias is given by taking to be the difference of the fiducial lensing spectrum $C^{\phi\phi}_L$ and the  average of 480 (mean-field subtracted) cross-only lensing power spectra obtained from simulations, i.e.,
\begin{equation}
    \Delta{C}^{\mathrm{MC}}_L=C^{\phi\phi}_L-(\Delta{A}^{\text{MC,mul}}_L\langle{C^{\times}_L}\rangle_\text{sim}-\mathrm{MCN0}-\Delta{C}^{\mathrm{N}_1}_L)
\end{equation}
\DW{Here, $\Delta{A}^{\text{MC,mul}}_L$ is the multiplicative bias term defined in Section~\ref{sec:qe}.}

\section{Cross-correlation based estimator: Motivation from noise-only null tests}\label{app:crossnoise}

A novelty in our lensing measurement presented here compared to past results is the use of the cross correlation-based lensing estimator, where we estimate the lensing bandpowers from four independent data splits as outlined in Sec.~\ref{sec:qe}. At a cost of modestly reducing the signal to noise by avoiding using the data with non-independent noise, we are completely immune to the assumptions made in modelling or simulating the instrument noise.
Accurate noise modelling is important for the lensing analysis for calculating the Gaussian disconnected bias (RDN0), which receives a contribution from both the CMB signal and foregrounds, as well as the instrument noise. Our original plan was to use the standard quadratic estimator as our simulated results agree with the data within $5\%$ at the CMB two-point power spectrum level and the RDN0 algorithm itself is designed to self-correct for small differences between data and simulations.
However, using this standard estimator, we were unable to pass the noise-only null test which directly tests the robustness of our pipeline and the accuracy of our noise simulations. This noise-only null test consists of the following:
For each data array, we prepare four splits of data map $\{X_0,X_1, X_2,X_3\}$ from the original 8 splits by weighting them with their respective inverse variance maps. We then form the following signal-nulled data map:
$[(X_0+X_1)-(X_3+X_4)]/2$. The same processing is done for a set of 300 simulations used to calculate the mean-field and the RDN0. We do not need to estimate the $N_1$ here as this term is zero due to the absence of the lensing signal.

\begin{figure*}[t]
\centering
\includegraphics[width=0.45\textwidth]{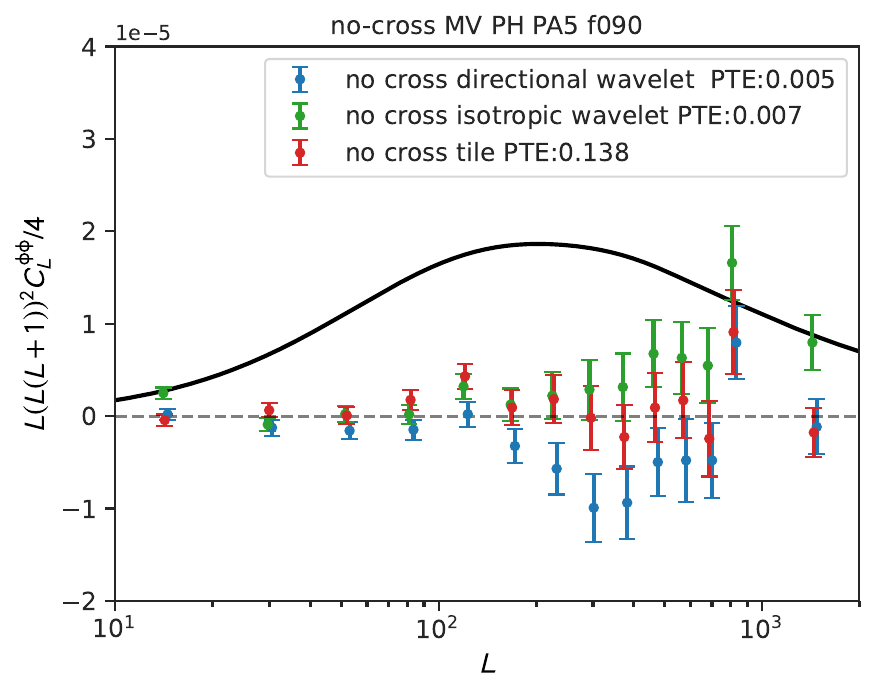}
\includegraphics[width=0.45\textwidth]{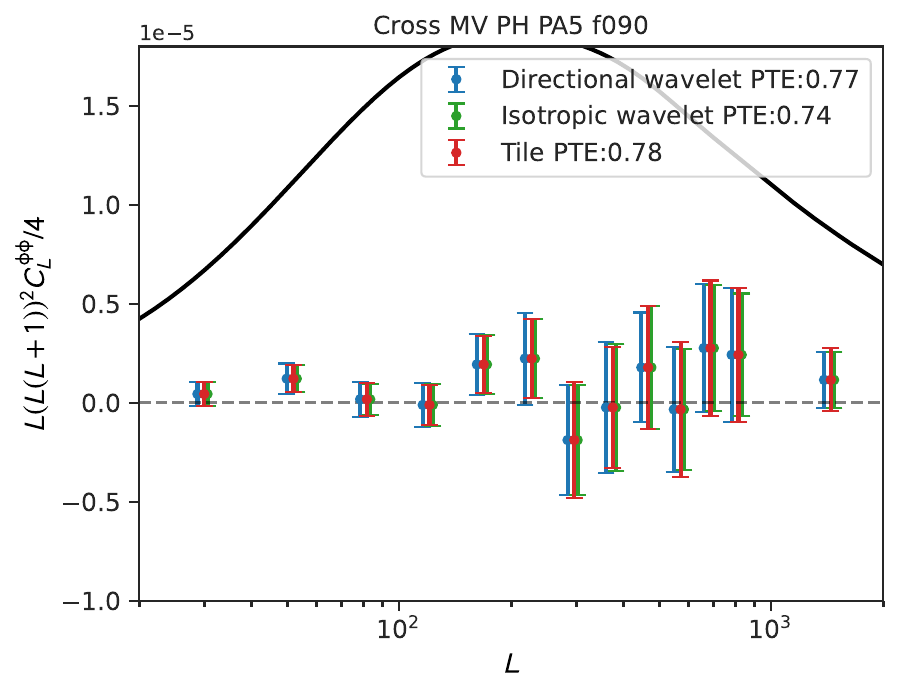}
    \caption{\textit{Left}: Noise-only null test using the standard quadratic estimator for the three different type of noise simulations (directional wavelet, isotropic wavelet and tiled) for the array PA5 f090. \blake{either write with phi or note that clkk is equivalent to the lensing potential with the following conversion}. \textit{Right}: Noise-only null test using the cross-correlation based estimator for the three different type of noise simulations (directional wavelet, isotropic wavelet and tiled) for the array PA5 f090. It can be seen that the cross-correlation based estimator avoids biases arising from noise mis-modeling. We also note that, with the cross-correlation based estimator, the errors and PTEs are similar for all noise simulations, indicating that the covariance matrix is stable for this estimator.}
    \label{fig.null.non.cross}
\end{figure*}

The left panel of Figure \ref{fig.null.non.cross} shows the resulting null test for the array PA5 f090 obtained with noise simulation drawn from three different types of noise models: directional wavelet, isotropic wavelet and tiled while using the same reconstruction filter and normalization. We see that the ‘null’ bandpowers are  dependent on the details of the noise modelling, and in this particular case, only the tile model passes the test. This scatter of results suggests that an accurate noise model required for lensing is hard to construct for ground-based CMB surveys covering a significant portion of the sky, with the main challenges coming from the combination of the atmosphere and the ACT scanning strategy. Space-based surveys like \textit{Planck} avoid these complications from the atmosphere, and many previous ground-based measurements cover a smaller region at lower precision, which allows the standard lensing power spectrum estimator to be used. \fk{It is worth noting that we are investigating improvements to the noise modeling on the basis of this null-test result.}

Switching to the four cross estimator, and performing the same null test as in \ref{sec. noise_only} means that this noise-only null does not depend on noise for the mean-field and RDN0, and hence the null test is independent of the details of the noise modelling. This can be seen in the null bandpowers in the right panel of Figure \ref{fig.null.non.cross}, obtained from the same noise sims. The null bandpowers are now consistent with zero. We also obtain nearly the same PTE values independent of the noise simulation type used, indicating that our covariance matrix is stable.

\section{Semi-analytic $N_0$}\label{app:dumb}
The semi-analytic $N_0$  bias can be computed as follows:

\begin{align}
       &\Delta{C^{\text{Gauss,diag}}_{L}}[\bar{X},\bar{Y},\bar{W},\bar{Z}] =\nonumber\\& \frac{{\mathcal{R}^{XY}_{L}}^{-1}{\mathcal{R}^{WZ}_{L}}^{-1}}{\Delta^{XY}\Delta^{ ZW}}\left[-\zeta^{XY,WZ}_L[C_\ell-\widehat{C}_\ell] + \zeta^{XY,ZW}_L[\widehat{C}_\ell]\right],
\end{align}

where $\Delta^{\rm XX}=2$, $\Delta^{\rm EB}=\Delta^{\rm TB}=1$. Here, $\zeta$ is equal to the integral of the lensing normalization if $X=Y=W=Z$ and it depends on  ${C}^{\text{fid}}_\ell$, the theoretical power spectra, and $\hat{C}_\ell$,  the realization dependent CMB spectrum. Writing this out explicitly it has the form

\begin{equation}
    \zeta^{XY,ZW}_{\l}[C] \equiv\int\frac{{d}^2{\bold{L}}}{(2\pi)^2} f^{\rm XY}_{\bold{\ell},\bold{L}}f^{\rm ZW}_{\bold{\ell},\bold{L} }F^X_LF^Z_L F^Y_{L'}F^W_{L'}({C}^{XZ}_L{C}^{YW}_{L'}+{C}^{XW}_L{C}^{YZ}_{L'})   ,
\end{equation}
where $F^{X}_\ell=1/(C^{X}_\ell+N^{X}_\ell)$ is the total diagonal power spectrum used for the filter.\footnote{For simplicity we showed the flat sky expression here but in the pipeline this is done using the curvedsky formalism.}

\section{Gaussianity of the bandpowers}\label{app:gaussianity}
\begin{figure}
 \centering
  \includegraphics[width=0.5\linewidth, clip, trim=0cm 1cm 0cm 0cm]{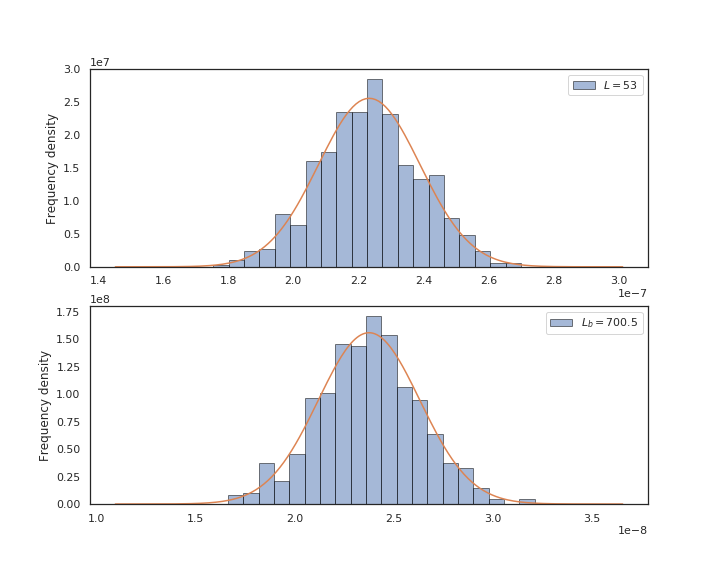}
  \caption{To test that a Gaussian likelihood for the lensing power spectrum bandpowers is sufficient, we plot histograms of the distribution of $C^{\hat{\phi}\hat{\phi}}_L$ bandpowers obtained from simulations for a bin with $L_B=53$ (top panel) and $L_B=700.5$ (bottom panel). In both cases, the distribution of the bandpowers closely follows a Gaussian distribution with the expected properties (i.e., a mean and width given by the bandpower mean and standard deviation). No evidence for departures from Gaussianity is found. \blake{label should be $L_B$ to be consistent}}
  \label{Fig.gauss}
\end{figure}

The likelihood we use in the cosmological analysis is built with the assumption that our bandpowers are Gaussian distributed. A priori, this assumption should not hold at arbitrary precision given the fact that even our lensing reconstruction map arises from a quadratic function of the (nearly Gaussian) observed CMB fields. However, the large number of effectively independent modes in our bandpowers suggests that the central limit theorem will drive the distribution of our bandpowers towards a Gaussian. We test this assumption using a set of 400 simulated lensing reconstructions by investigating the distribution of the resulting bandpowers. We choose the lowest and highest bins as representative examples; the distributions of these bandpowers is shown Fig. \ref{Fig.gauss}. The Kolmogorov–Smirnov statistic \citep{KS} shows that both distributions are well described by Gaussians, with PTEs of 0.6 and 0.9 respectively.

\section{Additional null tests}\label{app.null}

In the main text, for brevity we focused on summary statistics characterizing our ensemble of null tests and on some of the most crucial individual tests. In this Appendix, we explain and show results from the additional null tests we have performed with our data.

\subsubsection{Coadded split differences}\label{app.coadd_array}

A stringent noise-only null test, in the sense that it has  small errors, is obtained by coadding in harmonic space  the eight splits of each array with the same weights used to combine the data and forming 4 null maps by taking $X^{i,\mathrm{null}}=\mathrm{split}_{i}-\mathrm{split}_{i+4}$ .
The filter and normalization used for this test are the same used for the baseline lensing analysis.

\begin{figure*}[t]
\centering
 \includegraphics[width=0.45\textwidth]{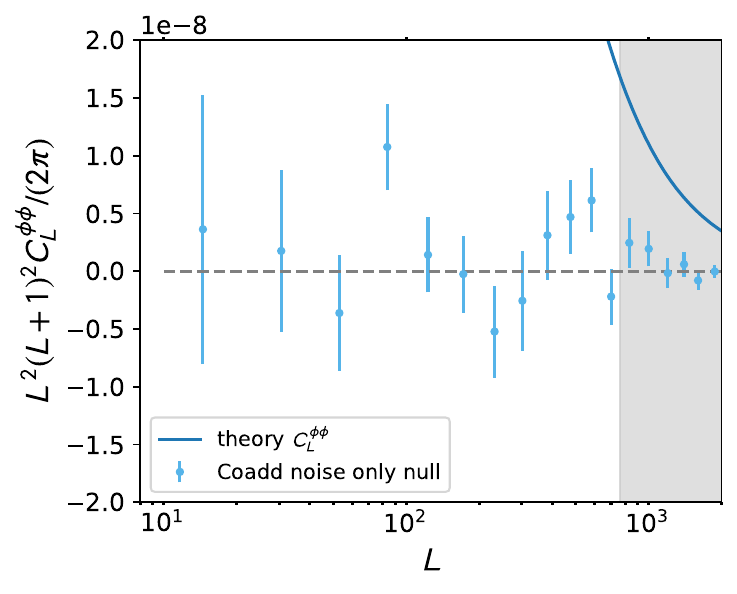}
 \includegraphics[width=0.45\textwidth]{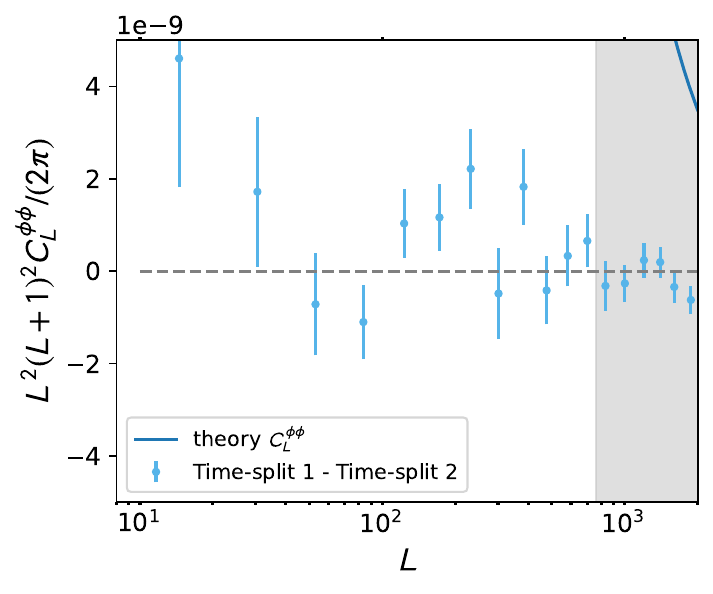}
    \caption{\textit{Left}: All array-frequencies coadd difference map noise-only bandpowers. Noise-only test obtained by coadding first the individual array-frequencies and taking the difference of the coadded splits. The low PTE of $0.02$ is not concerning as the null bandpowers do not exhibit a systematic trend and the size of the residual is small compared to the lensing signal. \textit{Right}: Null bandpowers obtained from differencing the Time-split 1 and Time-split 2 maps. \blake{It is consistent with zero with the PTE of XX}}
    \label{Fig.chisqnoise.tdiffmapgrad}
\end{figure*}

 The gradient null is shown in the left panel of Figure \ref{Fig.chisqnoise.tdiffmapgrad} and the curl is shown in the 5th panel of Fig. \ref{Fig.curl}. While the curl passes with a PTE of 0.85, the PTE for the gradient test is low, $0.02$; however, given the large number of tests we have run, the fact that the failure does not exhibit any obvious systematic trend, and the fact that magnitude of the residuals is negligibly small compared to the signal, we do not consider this 
 concerning.

 \begin{figure*}[hp!]
 \centering
 \includegraphics[width=\textwidth,height=23cm,keepaspectratio]{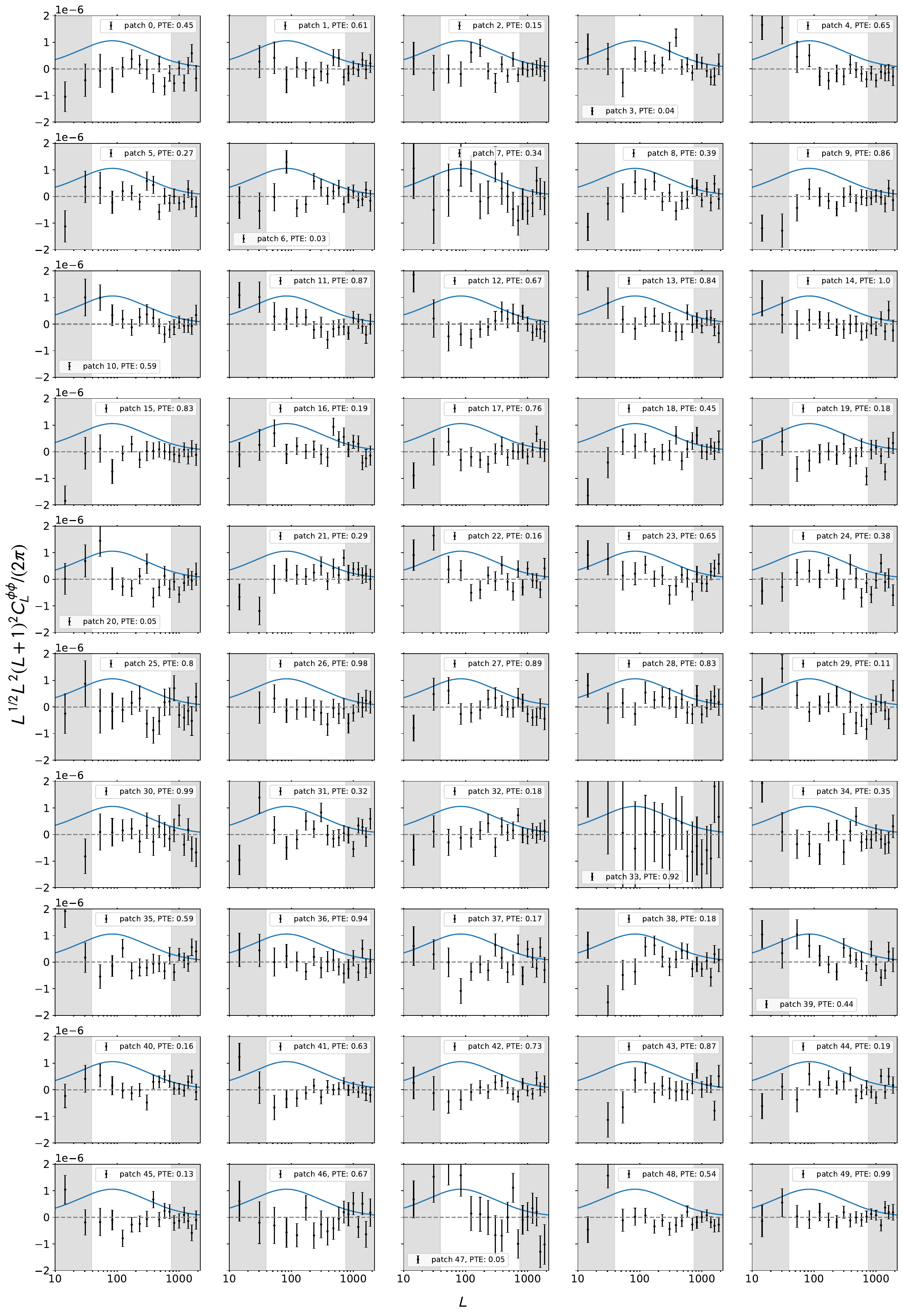}
  \caption{Null tests differencing the lensing power spectrum measured in 50 different patches (regions)  of the sky from the baseline lensing power spectrum measured on $9400~\si{deg}^2$. 
  No significant evidence is found for deviations from statistical isotropy of the lensing signal. This test also shows that there is no evidence for spatially varying beams at levels significant for our lensing measurement. Each patch spans $10 ^{\circ}\times25^{\circ}$ with patch 0 starting at the bottom right corner and going from right to left in steps of 25 degrees.}
  \label{Fig.diffband}
\end{figure*}
 
\subsubsection{Isotropy}\label{iso}

We test for consistency of our lensing measurements across different regions of the sky by dividing our sky into 50 patches $10 ^{\circ}\times25^{\circ}$ non-overlapping patches \footnote{The whole footprint is divided into $10^\circ \times 25^\circ$ rectangular patches but some patches at the edge of the map include masked, zeroed regions outside our analysis footprint. This is why  $50\times250 = 12500\,\si{deg}^2$  does not correspond to the exact area of $9400\,\si{deg}^2$; the fact that some regions (e.g., patch 33) only include a small unmasked area also explains why such regions’ spectra have larger errors.}. We compare the lensing power spectrum obtained in each patch, debiased and with errors estimated from simulations using our baseline procedure described in Sec. \ref{sec:reconstruction}. The tests show that our bandpowers are consistent with the assumption of statistical isotropy in the lensing signal and lensing power spectrum. This test is particularly targeted towards identifying spurious residual point sources in the map; higher spatial resolution and smaller patches are helpful for this. It complements well other tests of isotropy with larger regions: for example, the comparison of north vs south patches where we similarly did not find any evidence for anisotropy.
Furthermore, effects such as beam asymmetry and small time variations can all lead to spatially varying beams which could induce mode coupling similar to lensing. 

The bandpower consistency across the different regions also provides evidence that there are no problematic beam variations across the analysis footprint.

 \begin{figure}[h!]
 \centering
\includegraphics[width=0.4\textwidth,keepaspectratio]{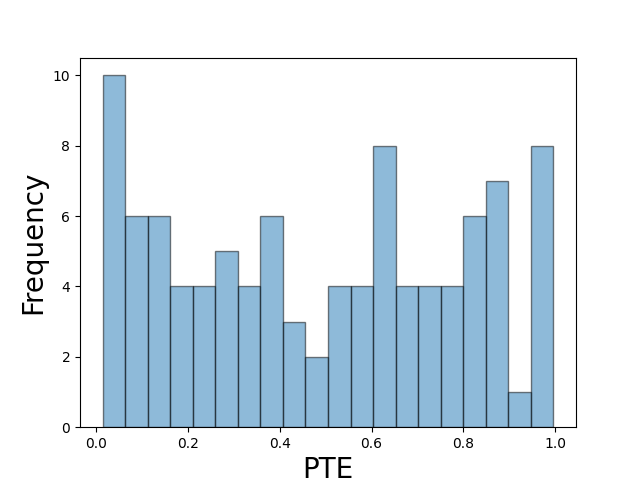}
  \caption{Distribution of the PTE's for the 100 isotropy bandpower difference and curl tests. The distribution is consistent with a uniform distribution, passing the K-S statistic with a PTE of 0.66.}
  \label{fig.distisotropy}
\end{figure}

\abds{The ensemble of 50 isotropy lensing is shown in  Fig. \ref{Fig.diffband}. These have PTE values consistent with a uniform distribution; the distribution of the combined 100 lensing and curl null bandpowers passes the K-S statistic with a PTE of 0.66 as shown in Fig. \ref{fig.distisotropy}.}

\subsubsection{Time-split differences Map level}\label{timediff_map}

We difference maps made from observations during the period 2017-2018 from those constructed from the period 2018-2021.
This test targets potential instrument systematics which may be different between these two phases and also makes sure that the beam, calibrations and transfer functions used are consistent across different observation periods.

The gradient null bandpowers in the right panel of Figure \ref{Fig.chisqnoise.tdiffmapgrad} are marginally consistent with zero with a PTE of 0.05; the curl in Fig. \ref{Fig.curl} is also consistent with zero with a PTE of 0.48.
To obtain accurate PTE's and debiasing in the bandpower test described in Sec. \ref{time-diffband} below, we produce special noise simulations capturing the characteristics of the two time-splits using $\texttt{mnms}$ \citep{dr6-noise}. 

\subsubsection{Time splits bandpower level test}\label{time-diffband}
\begin{figure*}[t]
\centering
\includegraphics[width=0.45\linewidth]{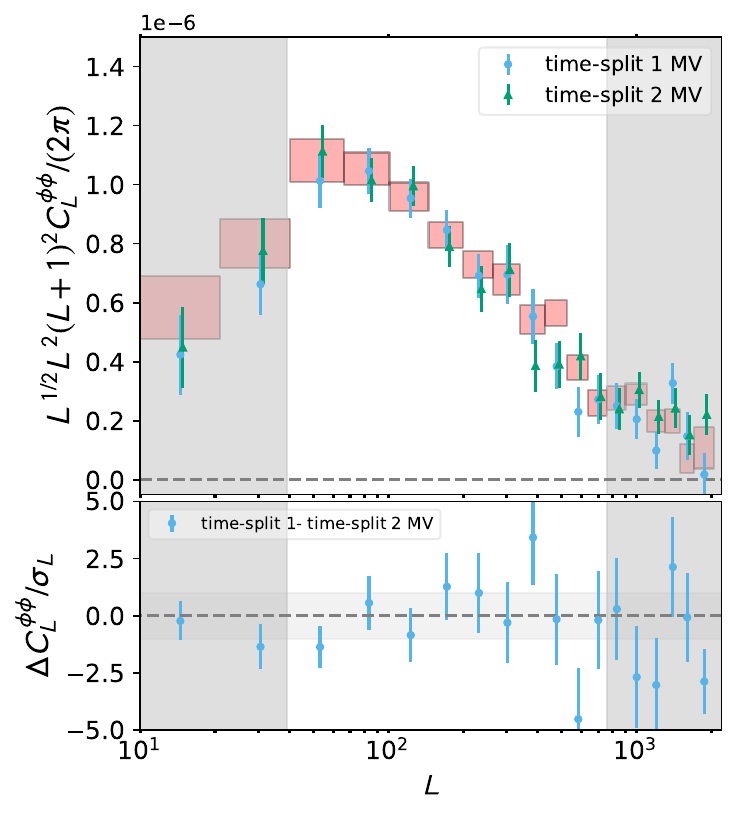}
\includegraphics[width=0.45\linewidth]{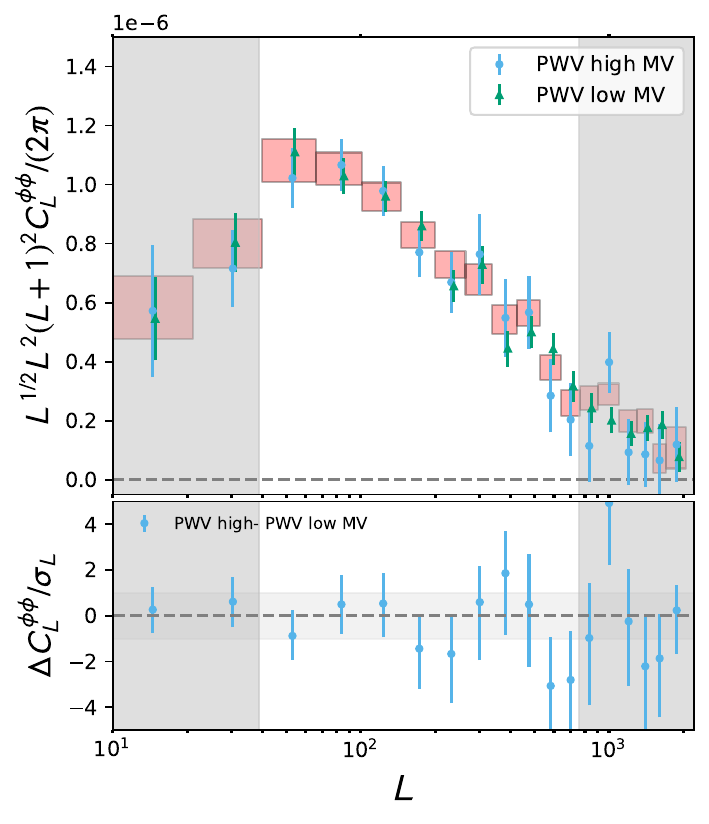}    
    \caption{\textit{Left}: Comparison of lensing spectra from time-split 1 and time-split 2 maps. \textit{Right}: The lensing power spectrum measured with PWV high and PWV low maps. No significant difference between the two measurements is found.}
    \label{Fig.season.pwvb}
\end{figure*}

We difference lensing spectra made from 2017-2018 observations from the spectra obtained from data taken during the period 2018-2021. This test targets systematic variations of beam, calibration and transfer functions with observing time. We apply the same filter and normalization used for our baseline lensing analysis and find no evidence of systematic differences between the two time-splits. The bandpower difference is shown in the left panel of Figure \ref{Fig.season.pwvb}; it has a passing PTE of 0.33.

\subsubsection{PWV split differences}\label{pwvdiff}
We obtain lensing spectra  from two sets of sky maps that are made to contain observations at high and low precipitable water vapour (PWV), respectively; a null difference of these two spectra can test for instrument systematics, e.g. systematics that have any dependence on the detector optical loading and to the level of atmospheric fluctuations. The noise levels of the high-PWV and low-PWV maps are noticeably different; hence, for the Wiener and inverse variance filtering of the maps, the filter is built using  a power spectrum whose noise is given by the average noise power of both maps. The null test results are shown in the right panel of Figure \ref{Fig.season.pwvb} with a PTE of 0.89.  To obtain accurate PTE's and debiasing in this null test, we draw special noise simulations capturing the characteristics of the two PWV-splits using $\texttt{mnms}$.

\subsection{Marginalized posteriors for cosmology runs}\label{app.posteriors}
\begin{figure}
  \centering
  \includegraphics[width=0.45\textwidth]{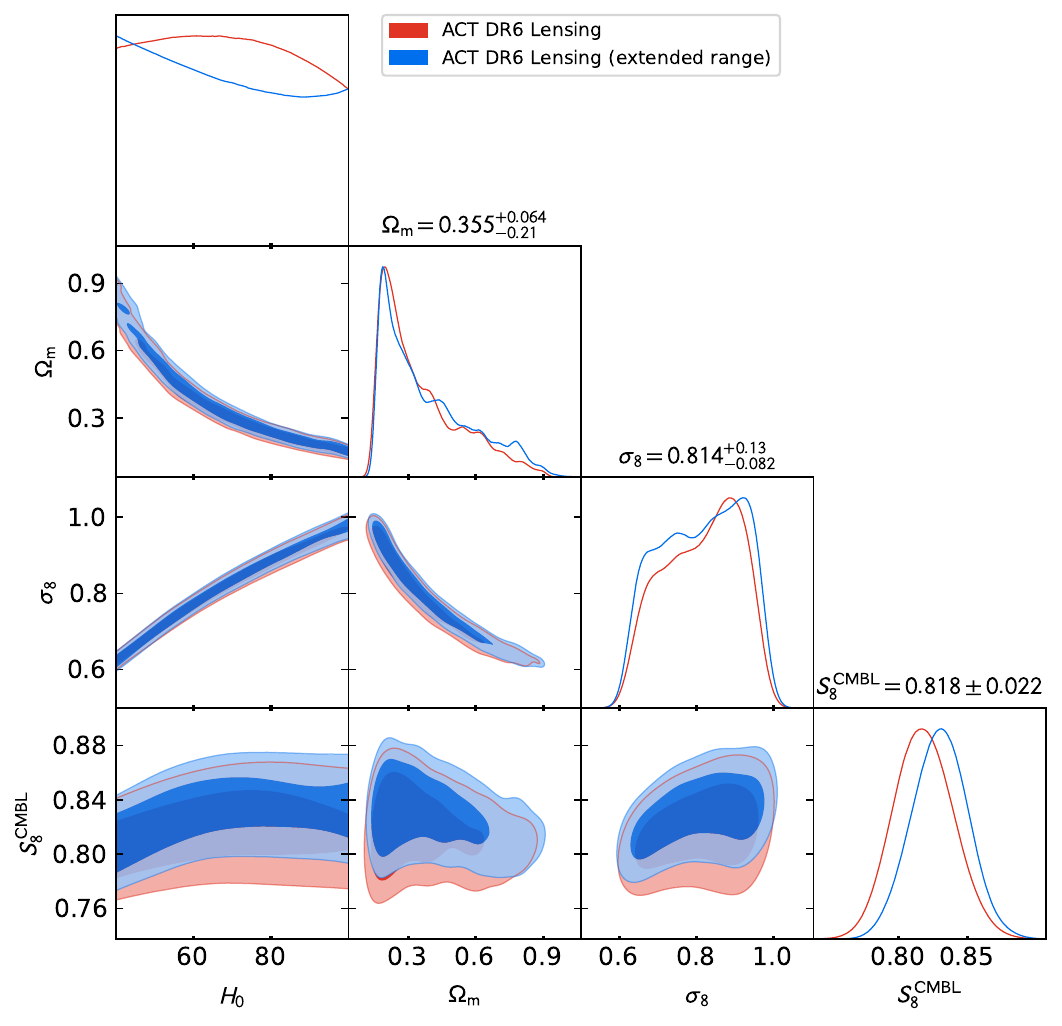}
   \includegraphics[width=0.45\textwidth]{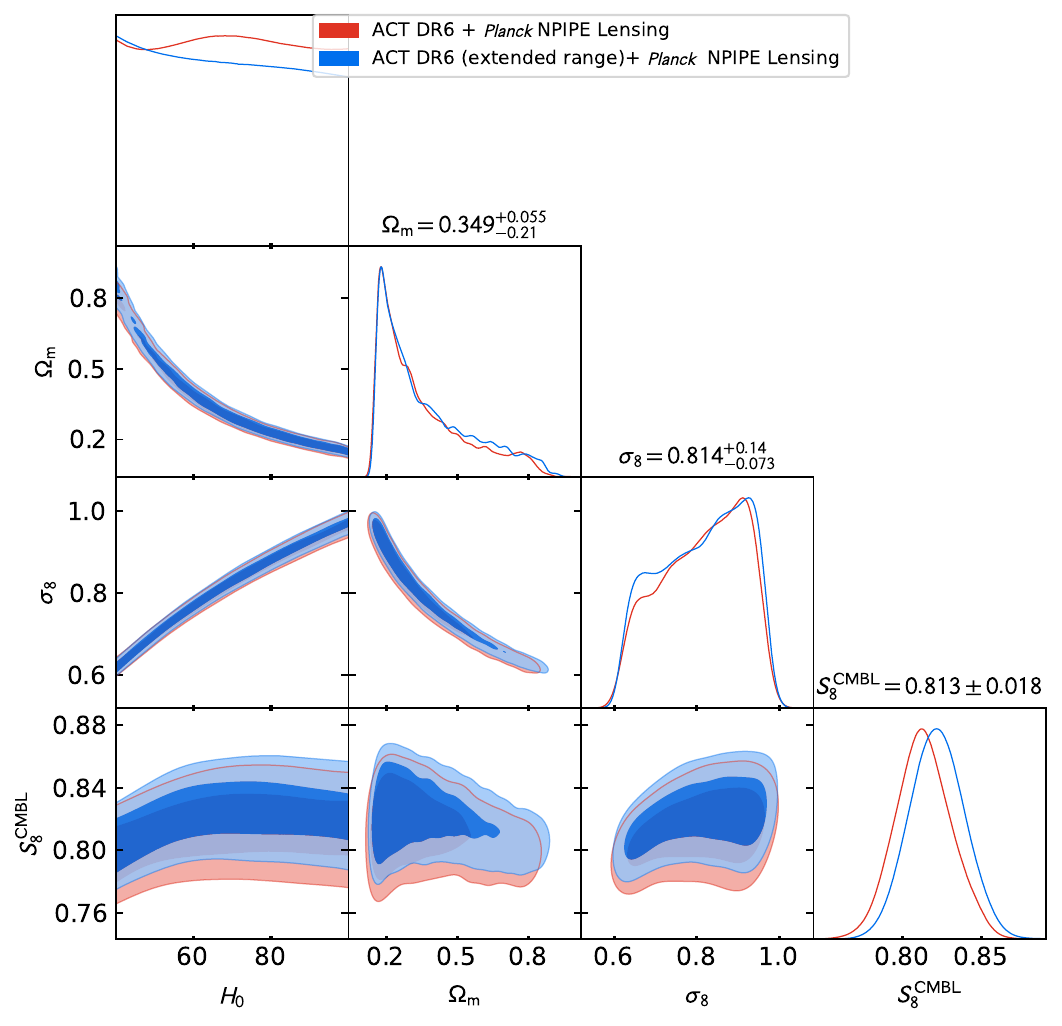}
  \caption{\textit{Left}: Marginalized posteriors for the combined ACT DR6 lensing constraints in the baseline range (red) and the constraints obtained using the extended range of scales (blue). \textit{Right}: Marginalized posteriors for the combined ACT DR6  and \textit{Planck} \texttt{NPIPE} Lensing constraints.}
  \label{fig.lensing.only.dr6.planck}
\end{figure}

\begin{figure*}
 \centering
 \includegraphics[width=\textwidth,height=23cm,keepaspectratio]{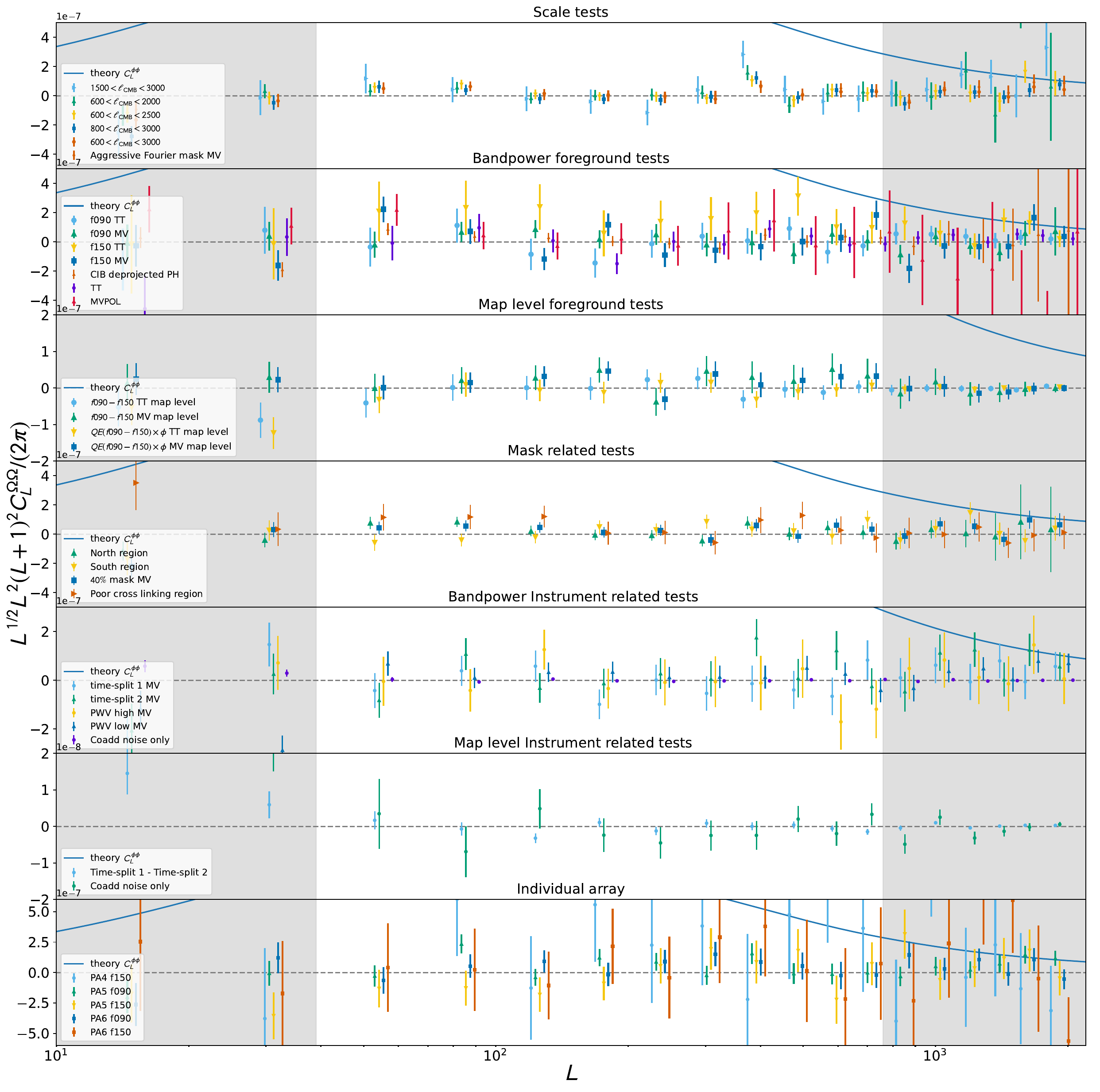}
  \caption{Compilation of curl null test bandpowers. No significant evidence for any curl signal is found. These results not only test for systematic contamination, but also provide additional validation of our covariance matrix estimates.}
  \label{Fig.curl}
\end{figure*}

\fk{Here, we provide a detailed presentation of the complete posterior contours for the cosmology runs that are presented in this paper. The constraints obtained from our DR6 lensing baseline and extended range are depicted in the left panel of Figure \ref{fig.lensing.only.dr6.planck} in blue and red, respectively. Likewise, the posterior plots for the baseline and extended range  when incorporating \textit{Planck} \texttt{NPIPE} lensing bandpowers, are shown in the right panel of Figure  \ref{fig.lensing.only.dr6.planck}.}

\subsection{Sensitivity of lensing-only constraints to $\Omega_k$}


Extensions to the flat-$\Lambda$CDM cosmology with parameters such as the spatial curvature $\Omega_k$ are considered in our companion paper \citep{dr6-lensing-cosmo}. 
In principle, the lensing-only constraints could be significantly affected by fully freeing curvature since the CMB lensing power spectrum amplitude can be affected by curvature; however, typical curvature values preferred by current constraints, such as those in our companion paper, are sufficiently close to flatness that the lensing-only constraints are minimally affected by assuming such curvature values. For example, assuming the maximum posterior value for the curvature from \cite{dr6-lensing-cosmo} only minimally shifts our lensing-only constraints to ~$S^\mathrm{CMBL}_8=0.814\pm0.022$, consistent with the baseline constraints $S^\mathrm{CMBL}_8=0.818\pm0.022$.

\subsection{Curl results summary}\label{app:null tests curl}
\DW{We present a more comprehensive summary of the curl results for the different null test, including a compilation of the PTEs and $\chi^2$ values in Table \ref{table:curl}. Our analysis reveals no significant evidence of systematic biases in our measurements.}

\begin{table}
    \centering
    \caption{Summary of results for the curl versions of our null tests.}
    \begin{tabular}{c c c}
     \hline\hline 
     Curl null test & $\chi^2$ & (PTE) \\ [0.5ex] 
     \hline
     PA4 f150 noise-only  & 5.2 & (0.88) \\ 
     PA5 f090 noise-only  & 13.9  & (0.18)\\ 
     PA5 f150 noise-only  & 8.1 & (0.62) \\ 
     PA6 f090 noise-only  & 11.6 & (0.31) \\ 
     PA6 f150 noise-only  & 19.1 & (0.04) \\ 
     Coadded noise & 5.4 & (0.85) \\
     $600<\ell_{\mathrm{CMB}}<2000$ & 14.2 & (0.16)  \\ 
     $600<\ell_{\mathrm{CMB}}<2500$ & 15.1 & (0.13)  \\ 
     $800<\ell_{\mathrm{CMB}}<3000$ & 13.1 & (0.22) \\ 
     $1500<\ell_{\mathrm{CMB}}<3000$ & 13.8 & (0.18)  \\
     CIB deprojection  & 8.6 & (0.57) \\
     $\text{PA4 f150}-\text{PA5 f090}$ & 14 & (0.16) \\
     $\text{PA4 f150}-\text{PA5 f150}$ & 9.5 & (0.48) \\
     $\text{PA4 f150}-\text{PA6 f090}$ & 8.0 & (0.63) \\
     $\text{PA4 f150}-\text{PA6 f150}$ & 9.6 & (0.48) \\
     $\text{PA5 f090}-\text{PA5 f150}$ & 15.3 & (0.12) \\
     $\text{PA5 f090}-\text{PA6 f090}$ & 5.2 & (0.88) \\
     $\text{PA5 f090}-\text{PA6 f150}$ & 6.7 & (0.75) \\
     $\text{PA5 f150}-\text{PA6 f090}$ & 14.7 & (0.14) \\
     $\text{PA5 f150}-\text{PA6 f150}$ & 12.0 & (0.29) \\
     $\text{PA6 f090}-\text{PA6 f150}$ & 11.7 & (0.30) \\
     $\text{f090}-\text{f150}$ map MV & 8.0 & (0.63)  \\
     $\text{f090}-\text{f150}$ map TT & 4.4 & (0.92)  \\
     TT f150 & 13.5 & (0.19) \\
     TT f090 & 6.0 & (0.82) \\
     MV f150 & 18.0 & (0.05) \\
     MV f090 & 4.3 & (0.93) \\
     MV coadd & 11.7 & (0.37) \\
     TT coadd & 6.8 & (0.75) \\
     MVPOL coadd & 13.2 & (0.21) \\
     North region & 9.6 & (0.47) \\
     South region & 14.7 & (0.14) \\
     $40\%$ mask & 8.3 & (0.41) \\
     Aggressive ground pick up & 10.8 & (0.37) \\
     Poor cross-linking region &9.0 &(0.54)\\ 
     \hline
    \end{tabular}
    \label{table:curl}
\end{table}

\section{Post-unblinding change: from inpainting to subtracting clusters}

\begin{figure}[t]
\centering\includegraphics[width=\textwidth]{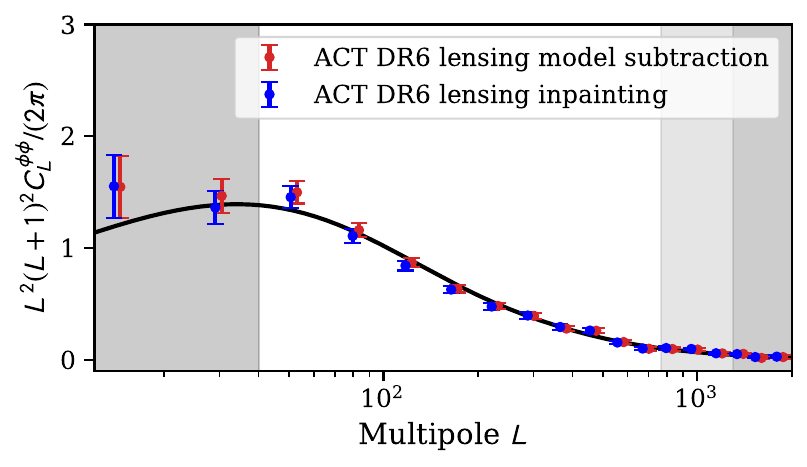}
  \caption{DR6 lensing bandpowers compared: bandpowers from cluster mitigation with inpainting (green) can be compared with bandpowers from model subtraction (red). The difference only leads to negligible shifts in cosmological parameters. }
  \label{Fig.compare}
\end{figure}

As described in Section \ref{sec:post-unblinding}, although our initial procedure for removing clusters involved inpainting a region around the cluster, we decided, based on simulation results, that subtracting a model template was more stable; we, therefore, switched to this procedure after unblinding. The two versions of the lensing spectra, which use inpainting and model subtraction, respectively, is seen in Figure \ref{Fig.compare}. As discussed in Section \ref{sec:post-unblinding}, the resulting change to $S_8^{\mathrm{CMBL}}$ are found to be negligible.

\section{Lensing power spectrum redshift kernel}\label{app.wkk}

\begin{figure}
    \centering
\includegraphics[scale=0.45]{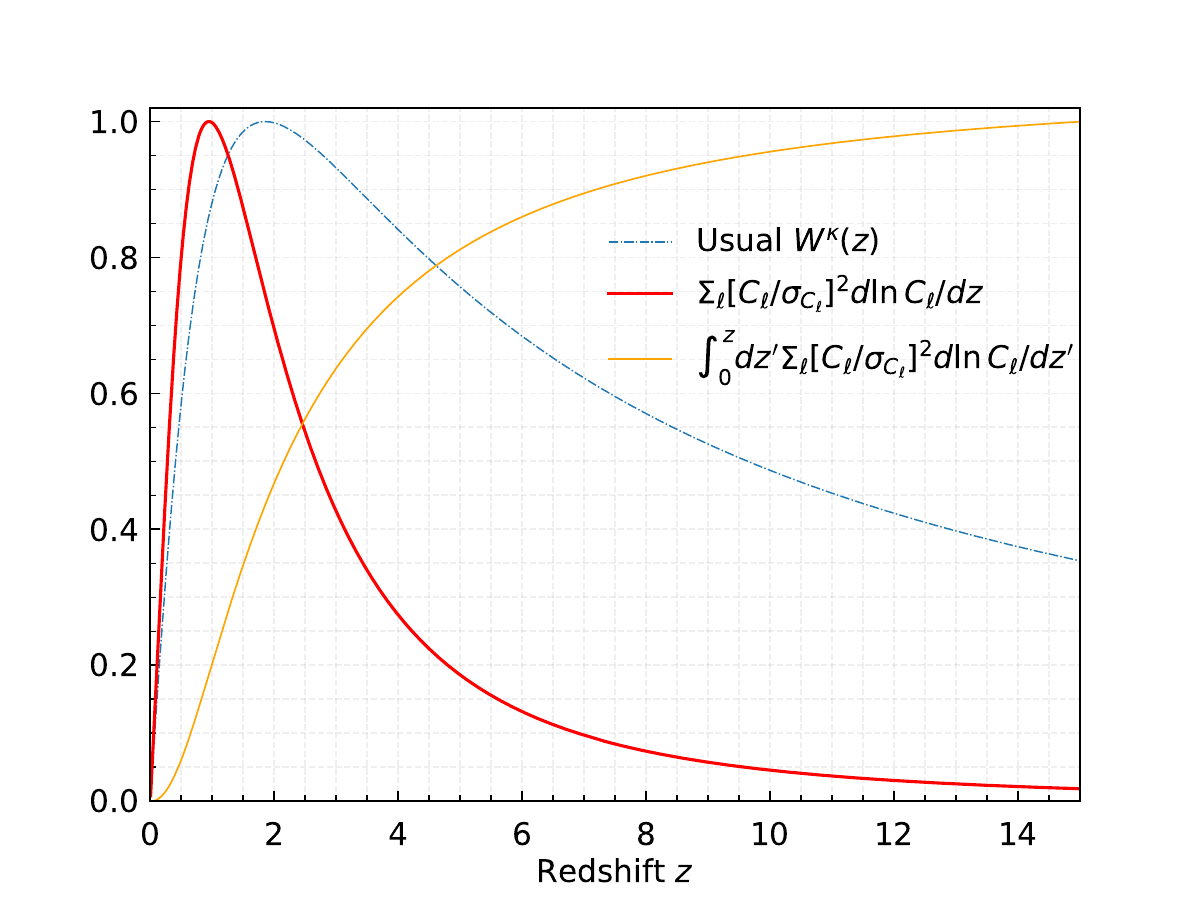}
    \caption{Comparison of the lensing kernel and weighted $C_\ell^{\phi\phi}$ integrand.}
    \label{fig:1dkernel}
\end{figure}

The lensing kernel that is usually presented represents how sensitive we are to structure at a given redshift. It is given by the formula:
\begin{equation}
    W^\kappa(z) =  \frac{3}{2}\Omega_{m}H_0^2  \frac{(1+z)}{H(z)} \frac{\chi(z)}{c} \\ 
    \left [ \frac{\chi(z_\star)-\chi(z)}{\chi(z_\star)} \right ].
    \label{eqn:cmb_kernel}
\end{equation}
It enters the lensing convergence equation
\begin{equation}
    \kappa(\mathrm{\bf{\hat{n}}}) = \int_0^{\infty} dz W^\kappa(z) \\
                \delta_{\rm m}(\chi(z)\mathrm{\bf{\hat{n}}}, z),
\end{equation}
and is shown as the dashed blue curve on Figure \ref{fig:1dkernel}.

\begin{figure}
    \centering
    \includegraphics[scale=0.35]{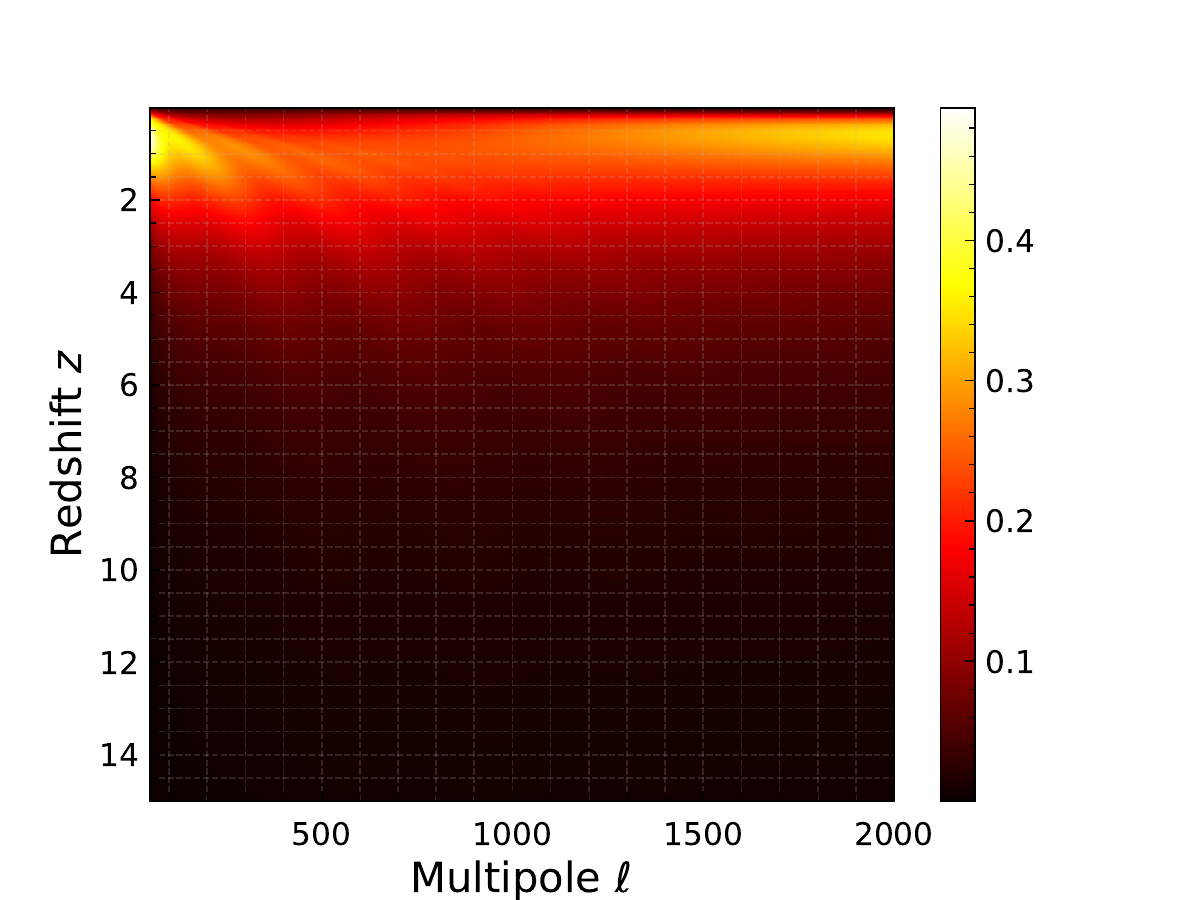}\includegraphics[scale=0.35]{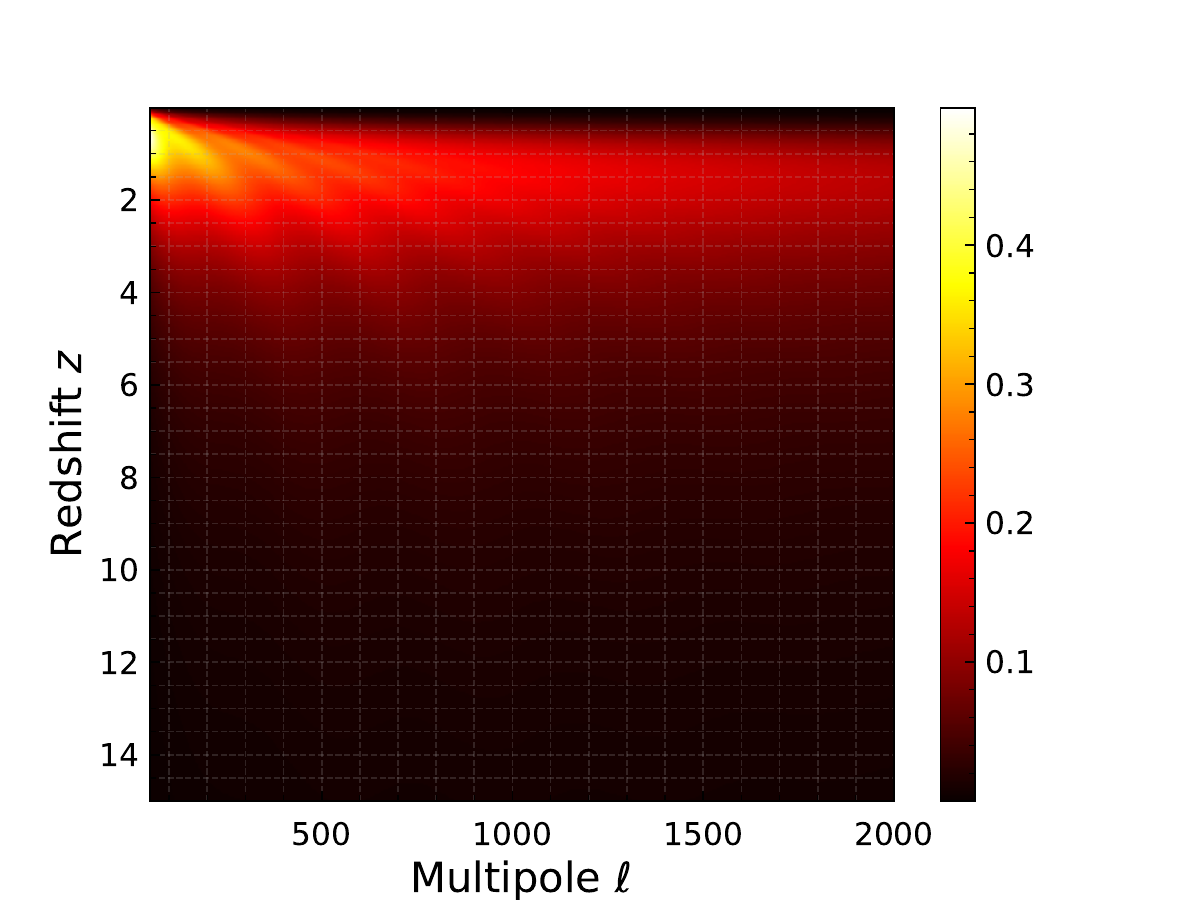}
    \caption{Lensing power spectrum integrand, impact of nonlinear evolution (left) versus linear (right).}.
    \label{fig:2dkernelpks}
\end{figure}

However, this kernel does not tell us where  lensing we probe in the measured power spectrum comes from. To answer this question, we first recall that the lensing power spectrum can be approximated by the redshift integral
\begin{equation}
    C^{\phi\phi}_\ell  =  \int d z W(z) P_{NL}(\ell/\chi(z)).
    \label{eqn:cmb_kk}
\end{equation}
where $W(z)=[H/c\chi^2]W^\kappa(z)^2$ and where $P_{NL}$ is the non-linear matter power spectrum, typically evaluated with \textsc{halofit} \citep{Takahashi:2012em}. At a each multipole, the relative contribution of a given redshift to the power spectrum amplitude is the logarithmic derivative, namely: 
\begin{equation}
    \frac{d\ln C^{\phi\phi}_\ell}{dz}  =  W(z) P_{NL}(\ell/\chi(z)).
    \label{eqn:lncmb_kk}
\end{equation}
Note that having the log in the sum is correct because we  estimate $A_\mathrm{lens}$ for each bandpower, which is $C_\ell/C_\ell^\mathrm{fiducial}$, and then average that Alens estimate over bandpowers, so it makes sense when thinking about contributions to $C_\ell$ to divide by $C_\ell^\mathrm{fiducial}$, \bds{i.e. effectively average $ \frac{d\ln C^{\phi\phi}_\ell}{dz}$ over multipoles.}
This quantity depends on both redshift and multipole, and is shown on a 2d grid on the left panels of Figure \ref{fig:2dkernelpks}. We used a \textsc{class} \citep{2011arXiv1104.2932L,Lesgourgues_2011_CLASSIII,classII} to compute this quantity efficiently. 
Our lensing power spectrum measurements are such that we are more or less sensitive to some $\ell$-range. So to give an accurate picture of where the lensing that measure in the power spectrum comes from, we want to weight \ref{eqn:lncmb_kk} in an $\ell$-dependent way. We chose to weight it by $[C_\ell^{\phi\phi}/\sigma_\ell]^2$, so the smaller the errors are, the bigger the weight is. 
Finally, we obtain the \textit{error weighted lensing kernel} by taking the average over all $\ell$'s at each $z$, computing 
    $\Sigma_\ell [C_\ell^{\phi\phi}/\sigma_\ell]^2 {d\ln C^{\phi\phi}_\ell}/{dz}$.
This quantity is shown as the red line in Figure \ref{fig:1dkernel}. And the cumulative is the orange line. 

From the cumulative we can read off several interesting redshifts: 50\% of signal is from $1.16<z<4.09$, 67\% of signal is from $0.88<z<5.44$, and 95\% of signal comes from $z<9.6$. While mean redshift from which our lensing comes is $z=2.16$, the peak of the red curve — i.e. the redshift that contributes most strongly to our $C_\ell^{\phi\phi}$ -- is slightly below 1, at $z=0.96$.

Other noticeable features are the visibility of the BAO in the integrand, which are washed out in the Limber integral, and the importance of the nonlinear evolution which gives more weight to low redshift at high multipole. This is shown in Figure \ref{fig:2dkernelpks}.

\end{document}

%% file: authors_qu.tex

\author{Frank~J.~Qu}\affiliation{DAMTP, Centre for Mathematical Sciences, University of Cambridge, Wilberforce Road, Cambridge CB3 OWA, UK}\affiliation{Kavli Institute for Cosmology Cambridge, Madingley Road, Cambridge CB3 0HA, UK}
\author{Blake~D.~Sherwin}\affiliation{DAMTP, Centre for Mathematical Sciences, University of Cambridge, Wilberforce Road, Cambridge CB3 OWA, UK}\affiliation{Kavli Institute for Cosmology Cambridge, Madingley Road, Cambridge CB3 0HA, UK}
\author{Mathew~S.~Madhavacheril}\affiliation{Department of Physics and Astronomy, University of
Pennsylvania, 209 South 33rd Street, Philadelphia, PA, USA 19104}\affiliation{Perimeter Institute for Theoretical Physics, Waterloo, Ontario, N2L 2Y5, Canada}
\author{Dongwon~Han}\affiliation{DAMTP, Centre for Mathematical Sciences, University of Cambridge, Wilberforce Road, Cambridge CB3 OWA, UK}
\author{Kevin~T.~Crowley}\affiliation{Department of Physics, University of California, Berkeley, CA, USA 94720}
\author{Irene~Abril-Cabezas}\affiliation{DAMTP, Centre for Mathematical Sciences, University of Cambridge, Wilberforce Road, Cambridge CB3 OWA, UK}
\author{Peter~A.~R.~Ade}\affiliation{School of Physics and Astronomy, Cardiff University, The Parade, 
Cardiff, Wales, UK CF24 3AA}
\author{Simone~Aiola}\affiliation{Center for Computational Astrophysics, Flatiron Institute, 162 5th Avenue, New York, NY 10010 USA}\affiliation{Joseph Henry Laboratories of Physics, Jadwin Hall,
Princeton University, Princeton, NJ, USA 08544}
\author{Tommy~Alford}\affiliation{Department of Physics, University of Chicago, Chicago, IL 60637, USA}
\author{Mandana~Amiri}\affiliation{Department of Physics and Astronomy, University of
British Columbia, Vancouver, BC, Canada V6T 1Z4}
% ...
